\documentclass{article}

\usepackage{a4wide}

\usepackage{hyperref}

\usepackage{amssymb}
\usepackage{amsmath}
\usepackage{microtype}
\usepackage{url}
\usepackage{proof}

\usepackage{amsfonts}
\usepackage[utf8]{inputenc}

\newcommand{\LJT}{\ensuremath{\mathit{LJT}}}

\newcommand{\solletter}{\mathcal S}
\newcommand{\solfunction}[1]{{\solletter}(#1)}
\newcommand{\sol}[2]{{\solletter}(\seq{#1}{#2})}
\newcommand{\Sol}[2]{{\solletter}\bigl(\seq{#1}{#2}\bigr)} 

\newcommand{\vect}[1]{\overrightarrow{#1}} 

\newcommand{\impl}{\supset}
\newcommand{\cosign}{\textit{co}}
\newcommand{\lambdabar}{\overline{\lambda}}
\newcommand{\cool}{\lambda^\cosign}
\newcommand{\fix}{{\nu}}  
\newcommand{\gfpsymb}{\mathsf{gfp}}
\newcommand{\gfp}{\gfpsymb\kern0.1em}
\newcommand{\oo}{\mathbb{O}}
\newcommand{\lb}{\lambda}

\newcommand{\bigasssymb}{\mathcal{A}}
\newcommand{\bigass}[2]{\bigasssymb_{#1}(#2)}

\newcommand{\RIntro}{\textit{RIntro}}

\newcommand{\LVecIntro}{\textit{LVecIntro}}
\newcommand{\Alts}{\textit{Alts}}

\newcommand{\forsomej}{}
\newcommand{\ol}{\lambda}
\newcommand{\coolfs}{\lambda^\cosign_{\Sigma}}
\newcommand{\coolfsE}{E\lambda^\cosign_{\Sigma}}
\newcommand{\olfsfix}{\lambda^\gfpsymb_{\Sigma}}
\newcommand{\seqt}[3]{#1\vdash #2:#3}
\newcommand{\seqn}[4]{#2\vdash_{#1} #3:#4}
\newcommand{\seq}[2]{#1\Rightarrow #2}
\newcommand{\judge}[4]{#1\,\rfloor\,#2\,\vdash\,#3:#4}
\newcommand{\tuple}[1]{\langle #1 \rangle}

\newcommand{\fl}[2]{\langle #2\rangle_{#1}}

\newcommand{\bs}[2]{#1+#2} 
\newcommand{\ns}[2]{#1+\cdots+#2} 
\newcommand{\s}[2]{\sum\limits_{#1}^{}{#2}} 

\newcommand{\fs}[2]{\s{#1}{#2}}

\newcommand{\interp}[2]{[\![#1]\!]^g_{#2}}
\newcommand{\interpwe}[1]{[\![#1]\!]^g}
\newcommand{\interps}[1]{[\![#1]\!]^s}

\newcommand{\finrepsymb}{{\mathcal F}}
\newcommand{\finrep}[2]{\finrepsymb(#1;#2)}
\newcommand{\finrepempty}[1]{\finrepsymb(#1)}

\newcommand{\colbase}{{\sf mem}}
\newcommand{\colr}[2]{\colbase(#1,#2)}
\newcommand{\colebase}{\colbase}
\newcommand{\colra}[2]{\colebase(#1,#2)}

\newcommand{\allinfsymb}{{\sf nofin}}
\newcommand{\allinf}[1]{\allinfsymb(#1)}

\newcommand{\bisim}{=}
\newcommand{\bisimp}{=}

\newcommand{\ndinfsymb}{\not\kern-0.152em\diamond_{\infty}}

\newcommand{\FPV}{\mathit{FPV}}
\newcommand{\dom}{\mathit{dom}}
\newcommand{\size}{\mathit{size}}

\newcounter{thm}
\newcounter{lem}
\newcounter{prop}
\newcounter{cor}
\newcounter{def}
\newcounter{ex}
\newtheorem{theorem}[thm]{Theorem}
\newtheorem{lemma}[lem]{Lemma}
\newtheorem{proposition}[prop]{Proposition}
\newtheorem{corollary}[cor]{Corollary}
\newtheorem{definition}[def]{Definition}
\newtheorem{example}[ex]{Example}

\newcommand{\finf}{\mathit{it}^\infty}
\newcommand{\Churchforest}{\mathit{Church}}

\newcommand{\Boolet}{\mathsf{BOOLE}}
\newcommand{\Inftyt}{\mathsf{INFTY}}
\newcommand{\Churcht}{\mathsf{CHURCH}}
\newcommand{\Peircet}{\mathsf{PEIRCE}}
\newcommand{\DNPeircet}{\mathsf{DNPEIRCE}}
\newcommand{\Threet}{\mathsf{THREE}}

\newcommand{\binfsymb}{\mbox{${\sf NEF}$}}
\newcommand{\binfpsymb}[1]{\binfsymb\kern-0.152em_{#1}} 
\newcommand{\binfp}[2]{\binfpsymb{#1} (#2)}
\newcommand{\binfcanindex}{\star}
\newcommand{\ionindex}{\dagger}
\newcommand{\binfcansymb}{\binfpsymb\binfcanindex}
\newcommand{\binfcan}[1]{\binfp\binfcanindex{#1}}

\newcommand{\nbinfsymb}{\mathsf{EF}}    
\newcommand{\nbinfpsymb}[1]{\nbinfsymb\kern-0.152em_{#1}} 
\newcommand{\nbinfp}[2]{\nbinfpsymb{#1} (#2)}
\newcommand{\nbinfcansymb}{\nbinfpsymb\binfcanindex}
\newcommand{\nbinfcan}[1]{\nbinfp\binfcanindex{#1}}

\newcommand{\NFFsymb}{\mbox{$\kern-0.152em{\sf NFF}$}}

\newcommand{\NFFpsymb}[1]{\NFFsymb\kern-0.152em_{#1}}
\newcommand{\NFFp}[2]{\NFFpsymb{#1}(#2)}
\newcommand{\NFFcansymb}{\NFFpsymb\binfcanindex}

\newcommand{\NFFionsymb}{\NFFpsymb\ionindex}
\newcommand{\NFFion}[1]{\NFFp\ionindex{#1}}

\newcommand{\inhabs}[1]{{\mathcal I}(#1)}

\newcommand{\finextsymb}{{\mathcal{E}_\mathrm{fin}}} 
\newcommand{\finext}[1]{\finextsymb(#1)}

\newcommand{\memsymb}{{\sf mem}} 
\newcommand{\mem}[2]{\memsymb(#1,#2)} 

\newcommand{\exfinsymb}{{\sf exfin}}
\newcommand{\exfin}[1]{\exfinsymb(#1)}

\newcommand{\finfinsymb}{\mathsf{finfin}}
\newcommand{\finfin}[1]{\finfinsymb(#1)}
\newcommand{\inffinsymb}{\mathsf{inffin}}

\newcommand{\FFsymb}{{\sf FF}}

\newcommand{\FFpsymb}[1]{\FFsymb\kern-0.152em_{#1}}
\newcommand{\FFp}[2]{\FFpsymb{#1}(#2)}
\newcommand{\FFcansymb}{\FFpsymb\binfcanindex}

\newcommand{\cntsymb}{\#}
\newcommand{\cnt}[1]{\cntsymb(#1)}

\newcommand{\fingsymb}[1]{\mathsf{fin}^{#1}} 
\newcommand{\fing}[2]{\fingsymb{#1}(#2)}

\newcommand{\emptysolsymb}{{\sf nosol}}
\newcommand{\emptysol}[1]{\emptysolsymb(#1)}

\newcommand{\lhs}{\mathit{LHS}}
\newcommand{\rhs}{\mathit{RHS}}

\begin{document}
\title{A Coinductive Approach to Proof Search through Typed Lambda-Calculi}
\author{Jos\'{e} Esp\'{\i}rito Santo (Centre of Mathematics, University of Minho),\\ Ralph Matthes (CNRS, Institut de Recherche en Informatique de Toulouse),\\ Lu\'{\i}s Pinto (Centre of Mathematics, University of Minho)}

\maketitle
\begin{abstract}
In reductive proof search, proofs are naturally generalized by solutions, comprising all (possibly infinite) structures generated by locally correct, bottom-up application of inference rules. We propose a rather natural extension of the Curry-Howard paradigm of representation, from proofs to solutions: to represent solutions by (possibly infinite) terms of the coinductive variant of the typed lambda-calculus that represents proofs. We take this as a starting point for a new, comprehensive approach to proof search; our case study is proof search in the sequent calculus $LJT$ for intuitionistic implication logic. A second, finitary representation is proposed,
comprising a syntax of lambda-terms extended with a formal greatest fixed point, and a type system that can be seen as a logic of coinductive proofs. In the
finitary system, fixed-point variables enjoy a relaxed form of binding that allows the detection of cycles through the type system. Formal sums are used in both representations to express alternatives in the search process, so that not only individual solutions but actually solution spaces are expressed. Moreover, formal sums are used in the coinductive syntax to define ``decontraction'' (contraction bottom-up)---an operation whose theory we initiate in this paper. A semantics is defined assigning a coinductive lambda-term to each finitary term, making use of decontraction as a semantical match to the relaxed form of binding of fixed-point variables present in the finitary system. The main result is the existence of an equivalent finitary representation for any full solution space expressed coinductively. This result is the main ingredient in the proof that our logic of coinductive proofs is sound and complete with respect to the coinductive semantics. These results are the foundation for an original approach to proof search, where the search builds the finitary representation of the full solution space, and the \emph{a posteriori} analysis typically consisting in applying a syntax-directed procedure or function. The paper illustrates the potential of the methodology to the study of proof search and inhabitation problems in the simply-typed lambda-calculus, reviewing results detailed elsewhere, and including new results that obtain extensive generalizations of the so-called monatomic theorem.
\end{abstract}

\markboth{J. Esp\'{\i}rito Santo, R. Matthes and L. Pinto}{A Coinductive Approach to Proof Search through Typed Lambda-Calculi}

\section{Introduction}\label{sec:intro}

Proof theory starts with the observation that a proof is more than
just the truth value of a theorem. A valid theorem can have many
proofs, and several of them can be interesting. In this paper, we
somehow extend this to the limit and study all proofs of a given
proposition. Of course, who studies proofs can also study any of them
(or count them, if there are only finitely many possible proofs, or
try to enumerate them in the countable case). But we do this study somehow
simultaneously: we introduce a language to express the ``full solution
space'' of proof search. And since we focus on the generative aspects
of proof search, it would seem awkward to filter out failed proof
attempts from the outset. This does not mean that we pursue impossible
paths in the proof search (which would hardly make sense) but that we
allow to follow infinite paths. An infinite path does not correspond
to a successful proof, but it is a structure of locally correct proof
steps, generated by the bottom-up application of inference rules (the perspective of \emph{reductive} proof theory). In other words, we use coinductive syntax to model \emph{all}
locally correct proof figures. This gives rise to a not necessarily
wellfounded search tree. However, to keep the technical effort
simpler, we have chosen a logic where this tree is finitely branching,
namely the implicational fragment of intuitionistic propositional
logic 
with a proof system given by the cut-free fragment of the sequent calculus $\LJT$,  introduced in  \cite{HerbelinCSL94} as the typed calculus
$\lambdabar$. Actually, we will consider the variant of $\LJT$ where axioms are restricted to atomic formulas, and, since we do not consider the cut rule,
the system is isomorphic to the system of simply-typed long normal
forms in lambda-calculus which throughout this paper we will denote by $\lb$.

Lambda terms or variants of them (expressions that may have bound
variables) are a natural means to express proofs (an observation that is called \emph{the}
Curry-Howard isomorphism) in implicational logic. Proof alternatives
(locally, there are only finitely many of them since our logic has no
quantifier that ranges over infinitely many individuals) can be
formally represented by a finite sum of such solution space
expressions, and it is natural to consider those sums up to
equivalence of the \emph{set} of the alternatives. Since whole solution
spaces of (possibly infinite) proof trees are being
modeled, we call these coinductive terms \emph{forests}.

By their coinductive nature, forests are no proper syntactic
objects: they can be defined by all mathematical (meta-theoretic)
means and are thus not ``concrete'', as would be expected from
syntactic elements. This freedom of definition will be demonstrated
and exploited in the canonical definition (Definition~\ref{def:sol})
of forests as solutions to the task of proving a logical sequent (a
formula $A$ in a given context $\Gamma$). In a certain sense, nothing
is gained by this representation: although one can calculate on a
case-by-case basis the forest for a formula of interest and see
that it is described as fixed point of a system of equations
(involving auxiliary forests as solutions for the other
meta-variables that appear in those equations), an arbitrary
forest can only be observed to any finite depth, without ever knowing
whether it is the expansion of a regular cyclic graph structure (the latter being a finite structure).

Therefore, a coinductive representation is more like a semantics, a
mathematical definition; in particular, one cannot extract algorithms
from an analysis based on it. For this reason,
an alternative, \emph{finitary}
representation of solution spaces is desired, and
we develop, for
intuitionistic implication logic, one such representation in the form
of a (``normal'', i.\,e., inductive) typed lambda-calculus. Besides formal sums (to express choice in
the search procedure), this calculus has fixed points, to capture cyclic structure; moreover, fixed-point variables enjoy a relaxed form of binding, since cycle structure has to be captured up to the inference rule of contraction.

Our main result is that the forests that appear as full solution
spaces of logical sequents
can be interpreted as semantics of a typed term in this finitary typed lambda-calculus.
For the Horn fragment (where nesting of implications to
the left is disallowed), this works very smoothly without surprises
(\cite[Theorem 15]{FICS2013}).
The full implicational case, however, needs
some subtleties to capture redundancy that
comes from the introduction of several hypotheses that suppose the
same formula---hypotheses that would be identified by applications of the inference rule of contraction. In the finitary calculus, a relaxed form of binding is adopted for the fixed-point variables over which the
greatest fixed points are formed;  and the interpretation of such finite expressions in terms of forests needs, in the full case, a special operation, defined on forests, that we call
\emph{decontraction} (contraction bottom-up \footnote{This operation was called co-contraction in \cite{FICS2013}.}). 
Without this operation, certain repetitive patterns
in the full solution spaces due to the presence of negative occurrences of
implications could not be matched on the semantical side. With it, we obtain the finitary
representation (Theorem~\ref{thm:FullProp}).

This result lays the foundation for an original approach to proof search. Given a sequent, proof search is run once, not to solve a certain problem (e.\,g., deciding if the sequent is provable), but to generate the finitary representation of the entire solution space. This representation becomes thus available for later use (and reuse), in whatever \emph{a posteriori} analysis we wish to carry out (e.\,g., solve a decision or counting problem); and the analysis consists typically in giving the finitary term representing the solution space to a recursive predicate or function, whose definition is driven by the syntax of the finitary calculus. The potential of this methodology has been proved elsewhere \cite{EspiritoSantoMatthesPintoInhabitation,JESRMLP-FI2019}, in the study of proof search in $\LJT$ and the simply-typed $\lambda$-calculus. But here we will offer new results in the same vein, namely extensive generalizations of the so-called ``monatomic theorem'' \cite{HindleyBasicSimple}.

This paper is a substantially revised and extended version of our first workshop paper \cite{FICS2013}\footnote{Note however that in the present paper we do not treat separately the Horn fragment, as we do in \cite{FICS2013}.} on this topic. Relatively to this work, the main novel aspects of this paper are:
\begin{enumerate}
\item The development of a typing system for the untyped finitary system ${\overline{\lambda}}^\gfpsymb_{\Sigma}$  of \cite{FICS2013} (called $\olfsfix$ in the present paper). The typing system controls the mentioned relaxed form of binding of fixed-point variables that allows the detection of cycles in proof search.
It can be seen as a logic of coinductive proofs, whose soundness and completeness we will prove here.
\item An in-depth analysis of decontraction. This operation is bound to play a central role in reductive proof search, but surprisingly has never been properly studied. We lay down in this paper the basic results of its theory.
\item The revision of the technical details leading to the main theorem of \cite{FICS2013} (Theorem 24), in light of the refinements allowed by the novel typing system, leading to the revised form as Theorem~\ref{thm:FullProp} below.\label{item:maintheorem}
\item An illustration of the potential that our methodology has in the study of proof search in $\LJT$ and the simply-typed $\lambda$-calculus, exemplified with a new extensive generalization of the monatomic theorem mentioned before.
\end{enumerate}
This paper grew out of our technical report \cite{JESRMLP-arXiv2016} to which two subsequent journal publications \cite{EspiritoSantoMatthesPintoInhabitation,JESRMLP-FI2019} refer. It is also meant as a replacement for that technical report, so that future readers of these journal articles would rather consult the present paper. However, from the list above, only item~\ref{item:maintheorem} is needed for this purpose, the other developments deepen the understanding of the concepts and thus are a genuine contribution in this paper (not considering that technical report as a publication). Still, also the main theorem of the workshop paper \cite{FICS2013} has not yet been published in archival quality and thus appears here in such quality for the first time.

The paper is organized as follows. Section~\ref{sec:background}
recalls the system $\LJT/\lb$
and elaborates on proof search in this
system. Section~\ref{sec:coinductive}
develops the coinductive
representation of solution spaces for
$\LJT/\lb$. Section~\ref{sec:decontraction} studies the operation of decontraction. Section~\ref{sec:finitary-calculus} develops the finitary calculus and the
finitary representation of full solution spaces.
Section~\ref{sec:analysis} is dedicated to applications to proof search in $\LJT$ and inhabitation problems in $\lb$.
Section~\ref{sec:final} concludes, also discussing related and future work.


\section{Background}\label{sec:background}

We start by introducing our presentation of the base system $\lb$, of simply-typed long normal
forms in lambda-calculus, which, as mentioned before (and explained later), is in Curry-Howard correspondence with  cut-free $\LJT$ \cite{HerbelinCSL94}.

\subsection{Simply-typed $\lambda$-calculus, reduced to normal forms}

Letters $p,q,r$ are used to range over a base set of propositional
variables (which we also call \emph{atoms}).  Letters $A,B,C$ are used
to range over the set of formulas (= types) built from propositional
variables using the implication connective (that we write $A\impl B$)
that is parenthesized to the right. Throughout the paper, we will use the fact that
any implicational formula can be uniquely decomposed as $A_1\impl A_2\impl
\cdots\impl A_k\impl p$ with $k\geq0$, written in vectorial
notation as $\vec{A}\impl p$. For example, if the vector $\vec{A}$ is
empty the notation means simply $p$, and if $\vec{A}=A_1,A_2$, the
notation means $A_1\impl(A_2\impl p)$.

A term of $\lb$ (also referred to as a proof term) is either a typed lambda-abstraction or a variable
applied to a possibly empty list of terms. For succinctness, instead
of writing lists as a second syntactic category, we will use the
informal notation $\tuple{t_1,\ldots,t_k}$ (meaning $\tuple{}$ if
$k=0$), abbreviated $\tuple{t_i}_i$ if there is no ambiguity on
the range of indices. So,
$\lb$-terms are given by the
following grammar:
$$
\begin{array}{lcrcl}
\textrm{(terms)} &  & t,u & ::= & \,\lambda x^A.t \mid x\,\tuple{t_1,\ldots,t_k}\\
\end{array}
$$
where a countably infinite set of variables ranged over by letters
$w$, $x$, $y$, $z$ is assumed. Note that in lambda-abstractions we
adopt a \emph{domain-full} presentation (a.\,k.\,a.~Church-style
syntax), annotating the bound variable with a formula.
As is common-place with lambda-calculi, we will identify terms up to $\alpha$-equivalence, i.\,e., names of bound variables may be consistently changed, and this is not considered as changing the term.
The term
constructor $x\,\tuple{t_1,\ldots,t_k}$ is usually called
\emph{application}. When $n=0$ we simply write the variable
$x$.
The terms are obviously in one-to-one correspondence with $\beta$-normal ``ordinary'' lambda-terms, the only difference being the explicit tupling of argument terms to variables in the $\ol$ syntax. 

We will view contexts $\Gamma$ as finite sets of declarations $x:A$,
where no variable $x$ occurs twice. The context $\Gamma,x:A$ is
obtained from $\Gamma$ by adding the declaration $x:A$, and will only
be written
if $x$ is not
declared in $\Gamma$.
Context union is written as concatenation $\Gamma,\Delta$ for contexts $\Gamma$ and $\Delta$ if $\Gamma\cap\Delta=\emptyset$.
The letters $\Gamma$, $\Delta$, $\Theta$ are used to range over contexts, and the notation $\dom(\Gamma)$ stands for the set of variables declared in $\Gamma$. We will write $\Gamma(x)$ for the type associated with $x$ for $x\in\dom(\Gamma)$, hence viewing $\Gamma$ as a function on $\dom(\Gamma)$.
Context inclusion $\Gamma\subseteq\Delta$ is just set inclusion.

As usual, in this presentation of $\ol$ there is only one form of sequent, namely
$\Gamma\vdash t:A$. We call a sequent \emph{atomic} when
$A$ is an atom. (Note however that this contrasts to $\LJT/\lambdabar$ \cite{HerbelinCSL94} where two forms of sequents are used, as lists of terms are treated formally.)
The rules of $\ol$ for deriving sequents are in
Figure~\ref{fig:lambda-bar}. $\LVecIntro$ presupposes that the indices
for the $t_i$ range over $1,\ldots,k$ and that $\vec
B=B_1,\ldots,B_k$, for some $k\geq0$. Such obvious constraints for finite vectors will not be spelt out in
the rest of the paper.

In the particular case of $k=0$, in which $(x:p)\in\Gamma$ is the only hypothesis of $\LVecIntro$, we type variables (with atoms).
\begin{figure}[tb]\caption{Typing rules of $\ol$}\label{fig:lambda-bar}
$$
\begin{array}{c}
\infer[\RIntro]{\seqt\Gamma{\lambda x^A.t}{A\impl B}}{\seqt{\Gamma,x:A}
tB}\quad\quad
\infer[\LVecIntro]
 {\seqt{\Gamma}{x\tuple{t_i}_i}{p}}
 {(x:\vec B\impl p)\in\Gamma\quad\forall i,\,\seqt\Gamma{t_i}{B_i}}
\end{array}
$$
\end{figure}
In fact, viewed in terms of the
system $\LJT/\lambdabar$,
 $\LVecIntro$ is a derived rule, combining logical steps of \emph{contraction}, \emph{left implication}, and \emph{axiom}, the latter being \emph{atomic}, formed with atom $p$.

Note that the conclusion of the $\LVecIntro$ rule is an atomic sequent. This is not the case in
$\LJT/\lambdabar$ \cite{HerbelinCSL94}, where list sequents can have a non-atomic formula on the right-hand side. In the variant of cut-free $\LJT/\lambdabar$ we adopted, the only rule available for deriving an implication is $\RIntro$. A consequence of this restriction is that the space of proofs is reduced, allowing only \emph{uniform proofs} \cite{MillerNadathurPfenningScedrovAPAL91}, and all the logical steps of $\LJT$ underlying $\LVecIntro$ are only required with atomic right-hand side (both in the conclusion and in the rightmost premise). Still, our atomic restriction in $\LVecIntro$ will not cause loss of completeness of the system for intuitionistic implication.  This restriction is typically adopted in systems tailored for proof search, as for example systems of \emph{focused proofs}. In fact, our presentation of $\LJT/\lambdabar$
corresponds to a \emph{focused backward chaining} system  where all atoms are \emph{asynchronous} (see e.\,g. \cite{LiangMillerTCS09}). A consequence  of the atomic restriction of $\LVecIntro$ (specifically, of an atomic axiom) in $\lambdabar$ is that it does not type all $\beta$-normal forms, but only those in \emph{$\eta$-long $\beta$-normal form} (see, e.\,g., \S 8A7 of the book \cite{HindleyBasicSimple}, where these terms are called simply \emph{long $\beta$-nf's}, and \S 8A8 of \emph{op.~cit.} for an argument of why any $\beta$-normal form can be \emph{$\eta$-expanded} to a \emph{long $\beta$-nf}).

\subsection{Reductive proof search for $\ol$}
\label{subsec:reductive-proof-search}

We consider proof search problems given by a context $\Gamma$ and an implicational
formula $A$. We express them as \emph{logical sequents} $\seq{\Gamma}A$, corresponding to 
sequents of $\ol$ without proof terms. $\seq{\Gamma}A$ is nothing but the pair consisting of
$\Gamma$ and $A$, but which is viewed as a problem description: to
search for proofs of formula $A$ in context $\Gamma$. We use the letter $\sigma$ to communicate logical sequents but allow ourselves to speak of \emph{sequent} $\sigma$ in the interest of a lighter language.

Even though the system $\ol$ is a focused sequent calculus,
reductive proof search on $\ol$ has well identified points where
choices are needed \cite{DyckhoffPintoLMS99}. This is readily seen
in such a simple setting as ours, where only implication is
considered. Observing the rules in Figure~\ref{fig:lambda-bar}, one
concludes that implications have to be decomposed by $\RIntro$ until
an atom is obtained; here, in order to apply $\LVecIntro$, a choice
has to be made as to which assumption $x$ is to be picked from the
context, generating a control branching of the process (if there is
no $x$ to choose, we mark the choice point with failure); at each
choice, several search sub-problems are triggered, one for each
$B_i$, generating a different branching of the process, more of a
conjunctive nature.\footnote{Of course, this is all too reminiscent
of or- and and-branching in logic programming. But we are not
confined to the Horn fragment.} In all, a \emph{search forest} is
generated, which is \emph{pruned} to a tree, once a choice is made
at each choice point. Such trees we call \emph{solutions} (of the
proof-search problem posed by the given sequent). Sequents with
solutions are called \emph{solvable}. Since the search forest is a
structure where all solutions are superimposed, we also call it
\emph{solution space}.

Finite solutions are exactly the proofs in $\ol$ (hence the provable
sequents are solvable); but solutions need not be finite. For
instance, given the sequent $\sigma=(\seq{f:p\impl p,x:p}p)$, we can
apply forever the $\LVecIntro$ rule with variable $f$ if we wish, producing an
infinite solution. But $\sigma$ also has finite solutions, hence is
provable. On the other hand, the solvable sequent $\seq{f:p\impl
p}p$ has a unique infinite solution, hence is not provable.

\begin{example}\label{ex:types} The illustrating examples of this paper are with the following types.
  \begin{itemize}
  \item $\Boolet:=p\impl p\impl p$, an encoding of the Boolean values as $\lambda x^p.\lambda y^p.x$ and $\lambda x^p.\lambda y^p.y$. This example illustrates that we obtain different solutions when using the differently labeled (with $x$ and with $y$) hypotheses for $p$. We do not apply the so-called total discharge convention and stay plainly in the spirit of lambda-calculus.
  \item $\Inftyt:=(p\impl p)\impl p$, which is obviously uninhabited in lambda-calculus (as would be the type $p$ alone), but, as mentioned before, has a unique infinite solution (see Example~\ref{ex:finf}).
  \item $\Churcht:=(p\impl p)\impl p\impl p$, the type of Church numerals $\lambda f^{p\impl p}.\lambda x^p.f^n\tuple x$, $n\geq 0$. As mentioned above, there is also the solution with an infinite repetition of $f$'s.
  \item $\Peircet:=((p\impl q)\impl p)\impl p$ with different atoms $p$ and $q$ (the Peirce formula, in particular when reading $q$ as falsity), which is a classical tautology but not one of minimal logic and therefore uninhabited in lambda-calculus.
  \item $\DNPeircet:=(\Peircet\impl q)\impl q$, which is provable in minimal logic and therefore inhabited in lambda-calculus (already studied in \cite{FICS2013}).
  \item $\Threet:=((p\impl p)\impl p)\impl p$, the simplest type of rank 3 (the nesting depth) which has inhabitants of the form $\lambda x.x\tuple{\lambda y_1.x\tuple{\lambda y_2.x\tuple{\cdots\tuple{\lambda y_n.y_i}\cdots}}}$, $n\geq 1$ and $1\leq i\leq n$. (The types $(p\impl p)\impl p$ of $x$ and $p$ of all $y_k$ have been omitted for presentation purposes.) Notice that $\Threet$ is $\Peircet$ with identification of the two atoms. It may be seen as a simplification of the $\DNPeircet$ example.
  \end{itemize}
\end{example}

Some of our examples are also covered in Section~1.3.8 of \cite{lambdacalculuswithtypes}. Notice that they write $\Boolet$ as $1_2$ (their
example (i)), $\Churcht$ as $1\to0\to0$ (their example (iv)) and $\Threet$ as $3$ (their example (vii)) in that book. $\Peircet$ is their example (iii).

The type $\Threet\impl p\impl p$ is example (viii) in Section~1.3.8 of the cited book, and is called the ``monster''.  Since $\Threet$ is $\Peircet$ with identification of the two atoms $p$, $q$, the monster type is similarly resembling $\DNPeircet$, but of rank 4 (while the latter has rank 5). For us, both types are equally challenging, insofar as both require an infinite supply of bound variables for enumerating their (normal) inhabitants, which is why we did not include the monster type in our sample of examples.


\section{Coinductive representation of proof search}\label{sec:coinductive}
In this section we develop a coinductive representation of solutions and of solution spaces. This representation combines two ideas: the coinductive reading of the syntax of proofs, and the adoption of formal sums (in the case of solution spaces). Formal sums allow the definition of the operation of decontraction, which will play a crucial role in the relationship to the finitary representation of solution spaces to be developed in the next section.

\subsection{Representation of solutions: the $\cool$-system}

We introduce now $\cool$, a coinductive extension of $\ol$.
Its expressions are formed
without any consideration of well-typedness and will be the raw
syntax that underlies possibly non-wellfounded proofs, i.\,e.,
solutions.

 The raw syntax of these expressions is presented
as follows
$$ N ::=_\cosign \lambda x^A.N\,|\,  x \tuple{N_1,\ldots,N_k}\enspace,$$
yielding the terms of system $\cool$ (read coinductively, as indicated by the index $\cosign$)---still with finite tuples $\tuple{N_i}_i$, which is why we will call these expressions rather \emph{coterms}.

We consider a coinductive definition of syntax with binding as a base
concept, but the reader might appreciate a concrete set-theoretic
construction.  The terms of $\lb$ are then construed as finite trees,
with the grammar elements $\lb x^A$ and
$x\tuple{\cdot_1,\ldots,\cdot_k}$ on the nodes (where the latter
includes with the case $k=0$ also the leaves of the tree). And these
trees are identified modulo $\alpha$-equivalence. The coinductive
reading is then based on finite and infinite trees, again with these
node decorations. If we disregard variable binding for the moment,
this construction can be seen as metric completion of the finite terms
(see, e.\,g., \cite[Sect.~12.2]{terese}). To take into account
variable binding with its necessity to identify $\alpha$-equivalent
terms, we can just follow the description of infinitary
lambda-calculus in \cite[Sect.~12.4]{terese}. Even though the coterms
may be infinite, all the positions in them are of finite length, and a
definition of $\alpha$-equivalence by recursion on the lengths of
positions can be given. This allows to define a metric on
$\alpha$-equivalence classes and then to identify the $\alpha$-equivalence classes of coterms as metric
completion of the $\alpha$-equivalence classes of finite terms. Besides incorporating the
identification of coterms that only differ in the naming of their
bound variables, we consider as equal terms that \emph{finitely
  decompose} in the same way, which is to say that their successive
deconstruction (not taking into account consistent differences in
names of bound variables) according to the grammar must proceed the
same way, and this to arbitrary depth. In the described set-theoretic
construction, this just means that they are $\alpha$-equivalent finite
or infinite trees, which is an extensional concept (that does not
depend on how that infinite tree has been generated by an effective
program).
Thus, the natural notion of equality that we are
using is bisimilarity modulo $\alpha$-equivalence. Following
mathematical practice, this is still written as plain equality (in
type theory, it would have to be distinguished from definitional
equality / convertibility and from propositional equality / Leibniz
equality and would be a coinductive binary relation).

Since the raw syntax is interpreted coinductively, also the typing
rules have to be interpreted coinductively, which is symbolized by the
double horizontal line in Figure~\ref{fig:co-lambda-bar}, a notation
that we learnt from \cite{NUB}. (Of course,
the formulas/types stay inductive.). This defines when $\seqt\Gamma
NA$ holds for a \emph{finite} context $\Gamma$,
a coterm $N$ and a type $A$, and the only difference to the rules in
Figure~\ref{fig:lambda-bar} is their coinductive reading and their
reference to coinductively defined terms. When $\seqt\Gamma
NA$ holds, we say $N$ is a solution of $\sigma$, when $\sigma=\seq\Gamma A$.
\begin{figure}[tb]\caption{Typing rules of $\cool$}\label{fig:co-lambda-bar}
$$
\begin{array}{c}
\infer=[\RIntro_\cosign]{\seqt\Gamma{\lambda x^A.N}{A\impl B}}{\seqt{\Gamma,x:A}
NB}\quad\quad
\infer=[\LVecIntro_\cosign]
 {\seqt{\Gamma}{x\tuple{N_i}_i}{p}}
 {(x:\vec B\impl p)\in\Gamma\quad\forall i,\,\seqt\Gamma{N_i}{B_i}}
\end{array}
$$
\end{figure}
The set-theoretic counterpart of coinductive typing derivations consists
of finite and infinite trees that are suitably labelled with the data
of applications of the typing rules. No extra identification of $\alpha$-equivalent derivations
is needed, and we anyway do not consider identity of proofs for our purposes.

In the rest of the paper, we will gloss over such set-theoretic interpretations of
coinductive concepts. 
\begin{example}
\label{ex:finf}
Consider $\finf:=\lambda f^{p\impl
  p}.N$ with the coterm $N$ the infinitely repeated application of $f$, i.\,e., on the top level, $N$ has an application node with variable $f$ and just one argument, and the latter is the same as $N$ (and it does not even make sense to ask what comes after this infinite succession of applications of $f$).
In other words, $N$ is the unique coterm that is solution of the equation
$N \bisimp f\tuple N$. Using coinduction on the typing relation, we can easily show $\seqt{}\finf{\Inftyt}$, and hence find a (co)inhabitant of a formula that does not correspond to a theorem in most logics.
\end{example}

Another view of the typing system of Figure~\ref{fig:co-lambda-bar} is
a coinductive definition of which logical sequents $\seq\Gamma A$ are
solvable: they are those for which a coterm $N$ exists such that
$\seqt{\Gamma}NA$ is coinductively derivable by these two
rules---where the syntax of $N$ traces the rule applications, hence
$N$ is just the evidence for the existence of a coinductive derivation
in a system similar to Figure~\ref{fig:co-lambda-bar}, but without
proof terms in the sequents. However, there is no requirement of having
a program generating the tree corresponding to the coterm $N$ that is merely meant to exist. Constructive witnesses for derivability---however for sets of solutions instead of individual solutions---will be studied in the finitary system of Section~\ref{sec:finitary-calculus}.

\smallskip
As expected, the restriction of the typing relation to the finite
$\ol$-terms coincides with the typing relation of the $\ol$ system:

\begin{lemma}
\label{lem:equiv-typability-lambda-bar-terms}
For any $t\in\ol$,  $\seqt\Gamma t A$ in $\ol$ iff $\seqt\Gamma t A$ in $\cool$.
\end{lemma}
\begin{proof}
By induction on $t$, and using inversion of typing in $\ol$.
\end{proof}

After having recalled the coinductive reading of syntax with variable binding and what typing means for it, we now move to original material.

\subsection{Representation of solution spaces: the $\coolfs$ system}\label{sec:coindrepr}

We now come to the coinductive representation of whole search spaces in $\ol$.

The set of
coinductive cut-free $\ol$-terms with finite numbers of elimination
alternatives is denoted by $\coolfs$ and is given by the following
grammar:
$$
\begin{array}{lcrcl}
\textrm{(terms)} &  & N & ::=_\cosign & \lambda x^A.N\,|\, \ns{E_1}{E_n}\\
\textrm{(elim. alternatives)} &  & E & ::=_\cosign & x \tuple{N_1,\ldots,N_k}\\
\end{array}
$$
where both $n,k\geq0$ are arbitrary. The terms of $\coolfs$ are also called \emph{forests}. If we do not want to specify the syntactic category (terms or elimination alternatives), we consider them just as expressions and generically name them $T$, to reflect their nature as terms in a wide sense.

Note that summands cannot be lambda-abstractions.\footnote{The division into two syntactic categories also forbids the generation of an infinite sum (for which $n=2$ would suffice had the categories for $N$ and $E$ been amalgamated).} We will often use $\sum_iE_i$
instead of $\ns{E_1}{E_n}$---in generic situations or
if the dependency of $E_i$ on $i$
is clear, as well as the number of elements.
If $n=0$, we write $\oo$
for $\ns{E_1}{E_n}$.
If $n=1$, we write $E_1$ for $\ns{E_1}{E_n}$
(in particular
this injects the category of elimination alternatives into
the category of (co)terms) and do as if $+$ was a binary operation on
(co)terms. However, this will always have a unique reading in terms of
our raw syntax of $\coolfs$. In particular, this reading makes $+$
associative and $\oo$ its neutral element.

The coinductive typing rules of $\coolfs$ are the ones of $\cool$, together with the rule given in
Figure~\ref{fig:co-lambda-bar-sum}, where the sequents for coterms
and elimination alternatives are not distinguished notationally.

\begin{figure}[tb]\caption{Extra typing rule of $\coolfs$ w.\,r.\,t.~$\cool$}\label{fig:co-lambda-bar-sum}
$$
\begin{array}{c}
\infer=[\Alts]{\seqt{\Gamma}{\sum_iE_i}p}{\forall i,\,\seqt\Gamma{E_i}p}
\end{array}
$$
\end{figure}
Notice that $\seqt{\Gamma}{\oo}p$ for all $\Gamma$ and $p$.
This phenomenon makes an alternative view analogously to the one described for system $\cool$ after Example~\ref{ex:finf} rather uninteresting. Atoms
should not be seen as a kind of coinductive consequence of the rules governing implication. Later in this section, we will introduce the notion of membership
in forests, and $\oo$ obviously will then not have any members according to that definition. However, this kind of emptiness is undecidable in general, hence
the forests in the derivations do play an important role for any interpretation of derivations. 

Since, like the coterms, forests are not built in finitary ways from finitary
syntax (although the number of elimination alternatives is always finite, as is the number of elements of the tuples), their most natural notion of equality is again bisimilarity modulo $\alpha$-equivalence.
However, in forests, we even want to neglect the precise order
of the summands and their (finite) multiplicity. We thus consider the sums of
elimination alternatives as if they were sets of alternatives, i.\,e.,
we further assume that $+$ is symmetric and idempotent.  This means,
in particular, that this identification is used recursively when
considering bisimilarity (anyway recursively modulo
$\alpha$-equivalence). This approach is convenient for a mathematical treatment
but would be less so for a formalization on a computer: It has been
shown by Picard and the second author \cite{PicardMatthesCMCS12} that
bisimulation up to permutations in unbounded lists of children can be
managed in a coinductive type even with the interactive proof
assistant Coq, but it did not seem feasible to abstract away from the
number of occurrences of an alternative (which is the meaning of
idempotence of $+$ in presence of symmetry), where multiplicity
depends on the very same notion of equivalence that is undecidable in
general.

As for $\cool$, we just use mathematical equality for this notion of
bisimilarity on expressions of $\coolfs$, and so the sums of
elimination alternatives can plainly be treated as if they were finite
sets of elimination alternatives (given by finitely many elimination
alternatives of which several might be identified through
bisimilarity).

We are now heading for a concise description of the full (as explained later) solution spaces for logical sequents by means of our extended coinductive syntax.
\begin{definition}[Full solution spaces]
\label{def:sol}
  The function $\solletter$, which takes a sequent $\sigma=(\seq{\Gamma}A)$
  and returns a forest, is given corecursively as follows:
  In the case of an implication,
$$\sol{\Gamma}{A\supset B} := \lambda x^A.\sol{\Gamma,x:A}{B}\enspace.$$
In the case of an atom $p$, for the definition of $\sol{\Gamma}{p}$, let $y_i:A_i$ be the $i$-th declaration in some enumeration of $\Gamma$ with $A_i$ of the form $\vec{B_i}\impl p$. Let $\vec{B_i} = B_{i,1},\ldots,B_{i,k_i}$. Define $N_{i,j}:=\sol{\Gamma}{B_{i,j}}$. Then, $E_i:=y_i\fl j{N_{i,j}}$, and finally,
$$\sol{\Gamma}{p} := \sum_iE_i\enspace.$$
This is more sloppily written as
$$\sol{\Gamma}{p} :=  \fs{{(y:\vec{B}\supset p)\in\Gamma}} {y\fl{j}{\sol{\Gamma}{B_j}}}\enspace.$$
In this manner, we can even write the whole definition in one line:
\begin{equation}\label{eq:one-line-sol-space}
\sol{\Gamma}{\vec A \impl p} :=  \lambda \vec x:\vec A.\fs{{(y:\vec{B}\supset p)\in\Delta}} {y\fl{j}{\sol{\Delta}{B_j}}}
\end{equation}
with $\Delta:=\Gamma,\vec x:\vec A$. The usual convention on bound variables ensures that ($x$'s are fresh enough so that) $\Delta$ is a context.
\end{definition}
This definition has to be read with Proposition~\ref{prop:adequacy-general-case} in mind that will be stated and proven later in this section and that guarantees that all and only the solutions of a logical sequent $\sigma$ are in a precise sense contained in $\solfunction\sigma$.

A crucial element (for the succinctness of this definition and the rather structure-oriented further analysis) is that $\RIntro$ is the only way to prove an implication, hence that the leading lambda-abstractions are inevitable. Then, the extended (finite) context $\Delta$ is traversed
to pick variables $y$ with formulas of the form $\vec B\impl p$, thus with the right atom $p$ in the conclusion. And this spawns tuples of search spaces, for all the $B_j$, again w.\,r.\,t.~the extended context $\Delta$.
Notice that this is a well-formed definition: for every sequent $\sigma$, $\solfunction\sigma$ is a forest, regardless of the result of proof search for the given sequent $\sigma$, and this forest has the type prescribed by $\sigma$:
\begin{lemma}[Type soundness of $\solletter$]\label{lem:type-of-sol-spaces}
  Given $\Gamma$ and $A$, the typing  $\seqt\Gamma{\sol\Gamma A}A$ holds in $\coolfs$.
\end{lemma}
In particular, all free variables of $\sol\Gamma A$ are declared in $\Gamma$.

Let us illustrate  the  function $\cal S$ at work with some examples.

\begin{example}
\label{ex:boole}
One sees immediately that $\sol{}\Boolet=\lambda x^p.\lambda y^p.x+y$.
\end{example}

\begin{example}
\label{ex:infty}
Observe that $\sol{}{\Inftyt}\bisimp\finf$ (applying our notational conventions, and reflecting the fact that there is a unique alternative at each sum). In other words, $\finf$ solves the same equation as is prescribed for $\sol{}{\Inftyt}$, and so it is \emph{the} solution (modulo $\bisimp$).
\end{example}

\begin{example}
\label{ex:Church}
Consider the sequent \mbox{$\seq{}{\Churcht}$}.
We have:
$$
\Churchforest:=\sol{}{\Churcht}\\
=\lambda f^{p\impl p}.\lambda x^p. \sol{f:p\impl p,x:p}p
$$
Now, observe that $\sol{f:p\impl p,x:p}p\bisimp\bs{f\tuple{\sol{f:p\impl p,x:p}p}}x$ is asked for.
We
identify $\sol{f:p\impl p,x:p}p$ as the (unique) solution as a forest for $N$ of the equation $N\bisimp\bs{f\tuple N}x$.
Using $\fix$ as means to communicate solutions of fixed-point equations on the \emph{meta-level}\footnote{This notation does not imply any form of designating or even programming the fixed point by a suitable language; we allow all mathematical means to justify the existence of the fixed point as a forest, and for this we have the underlying set-theoretic view as (equivalence classes of) potentially infinite trees at our disposal.},
we have
$$
\sol{}{\Churcht}
\bisimp\lambda f^{p\impl p}.\lambda x^p. \fix\, N. \bs{f\tuple N}x
$$

By \emph{unfolding} of the fixed point and by making a \emph{choice} at each of the elimination alternatives, we can \emph{collect} from this coterm as the finitary solutions of the sequent all the Church numerals ($\lambda f^{p\impl p}.\lambda x^p.f^n\tuple{x}$ with $n\in\mathbb{N}_0$), together with the infinitary solution $\lambda f^{p\impl p}.\lambda x^p.\fix\, N. f\tuple N$
(corresponding to always making the
$f$-choice at the elimination alternatives).
\end{example}

\begin{example}
\label{ex:Horn}
We consider now an example without nested implications (in the Horn fragment).
Let $\Gamma=x:p\impl
q\impl p,y:q\impl p\impl q,z:p$, with $p\neq q$. Note that the full
solution spaces of $p$ and $q$ relative to this sequent are mutually
dependent and they give rise to the following system of equations:
$$
\begin{array}{rcl}
N_p&\bisimp&\bs{x\tuple{N_p,N_q}}z\\
N_q&\bisimp&y\tuple{N_q,N_p}\\
\end{array}
$$
and so we have
$$
\begin{array}{rcl}
\sol\Gamma p&\bisimp& \fix\, N_p.\bs{x\tuple{N_p,\fix\, N_q.y\tuple{N_q,N_p}}}z\\
\sol\Gamma q&\bisimp& \fix\, N_q.y\tuple{N_q,\fix\, N_p. \bs{x\tuple{N_p,N_q}}z}\\
\end{array}
$$
Whereas for $p$ we can collect one finite solution ($z$), for $q$ we
can only collect infinite solutions.
\end{example}

\begin{example}
\label{ex:dn-Peirce}
Let us consider $\DNPeircet$ of Example~\ref{ex:types}. When $q$ is viewed as absurdity, $\Peircet$ is
Peirce's law, and thus $\DNPeircet$ can be viewed as double negation of
Peirce's law. We have the calculation in Figure~\ref{fig:calc-sol-dnpeirce} (where in sequents we omit
formulas on the left-hand side).
\begin{figure}[tb]\caption{Steps towards calculating $\sol{}{\DNPeircet}$}\label{fig:calc-sol-dnpeirce}
\[
\begin{array}{rcl}
N_0&=&\sol{}{\DNPeircet}=\lambda x^{\Peircet\impl q}. N_1\\
N_1&=&\sol{x}{q}= x\tuple{N_2}\\
N_2&=&\Sol{x}{\Peircet}=\lambda y^{(p\impl q)\impl p}. N_3\\
N_3&=&\sol{x,y}{p}= y\tuple{N_4}\\
N_4&=&\sol{x,y}{p\impl q}=\lambda z^{p}. N_5\\
N_5&=&\sol{x,y,z}{q}= x\tuple{N_6}\\
N_6&=&\Sol{x,y,z}{\Peircet}=\lambda y_1^{(p\impl q)\impl p}. N_7\\
N_7&=&\sol{x,y,z,y_1}{p}= \bs{\bs{y\tuple{N_8}}z}{y_1\tuple{N_8}}\\
N_8&=&\sol{x,y,z,y_1}{p\impl q}= \lambda z_1^{p}. N_9\\
N_9&=&\sol{x,y,z,y_1,z_1}{q}
\end{array}
\]
\end{figure}
Now, in $N_9$ observe that $y,y_1$ both have type $(p\impl q)\impl p$
and $z,z_1$ both have type $p$, and we are back at $N_5$ but with the
duplicates $y_1$ of $y$ and $z_1$ of $z$. Later, we will call
 this duplication phenomenon
 \emph{decontraction}, and we will give a finitary description of $N_0$ and, more generally, of all $\solfunction\sigma$ (again, see Theorem~\ref{thm:FullProp}). Of course, by taking the middle alternative in $N_7$, we obtain a finite proof, showing that $\DNPeircet$ is provable in $\ol$.
\end{example}

\begin{example}
  \label{ex:Three}
  For completeness, we describe the beginning of the calculations for
  $\Threet$ (for $\Peircet$ see Example~\ref{ex:Peirce}).
  $\sol{}\Threet=\lambda x^{(p\impl p)\impl p}.x\tuple{\lambda
    y^p.N}$, abbreviating $N$ for $\sol{x:(p\impl p)\impl
    p,y:p}p$. Then, $N=x\tuple{\lambda z^p.N'}+y$, with
  $N'=\sol{x:(p\impl p)\impl p,y:p,z:p}p$. We could further unravel
  the definition and provide a description of $\sol{}\Threet$ up to
  any finite depth, but we prefer a more symbolic solution in
  Section~\ref{sec:finitary-calculus} which exploits decontraction in
  the same way as for the preceding example.
\end{example}

We give a membership semantics for expressions of $\coolfs$ in terms of sets of terms in $\cool$.
More precisely, the \emph{membership relations} $\colr MN$ and $\colra
ME$ are contained in $\cool\times\coolfs$ and $\cool\times\coolfsE$
respectively (where $\coolfsE$ stands for the set of elimination
alternatives of $\coolfs$) and are given coinductively by the rules in
Figure~\ref{fig:collect}. In particular there is no $M$ such that $\colra M\oo$.
In this sense, $\oo$ is an empty solution space, but there are many others,
such as $\lb x^A.\oo$ and $x\tuple{\oo}$. These three forests are pairwise distinct,
i.\,e., not bisimilar. We do thus not identify forests having the same members
(and examples can be given that even have the same types in the same contexts).

We allow ourselves a small interlude: A natural question is if one can avoid $\oo$ as subexpression of a forest, so that
having an empty solution space does not ``come as a surprise'' but is visible from the outset: having $\oo$ at the root.
As far as full solution spaces $\solfunction\sigma$ are concerned, there is a refined definition involving the concepts
developed in Section~\ref{sec:finitary-calculus} of the present paper that can achieve forests of that special form,
and this leads to a ``König's lemma for simple types'' \cite[Theorem 4.25]{JESRMLP-FI2019} saying that those obtained forests
are infinite iff $\sigma$ has an infinite solution. In other words, those forests cannot accumulate an infinite amount of nodes that ``in the end''
turn out not to contribute anything to a solution. This being said, it does not seem feasible to develop a theory of solution spaces without having
$\oo$ as a building block. End of the interlude.

\begin{figure}[tb]\caption{Membership relations}\label{fig:collect}
$$
\begin{array}{c}
\infer=[]{\colr{\lambda x^A.M}{\lambda x^A.N}}{\colr{M}{N}}\quad\quad
\infer=[]{\colra{x\tuple{M_i}_i}{x\tuple{N_i}_i}}{\forall i,\,\colr{M_i}{N_i}}\quad\quad
\infer=[\forsomej]{\colr{M}{\sum_iE_i}}{\colra{M}{E_j}}\\\\
\end{array}
$$
\end{figure}

\smallskip
Coterms have the types of the forests they are members of.
\begin{lemma}[Soundness w.r.t.~membership semantics]\label{lem:typing-of-members}\quad
\begin{enumerate}
\item For $N\in\cool$, $T\in\coolfs$, if $\seqt\Gamma {T} A$ in $\coolfs$ and $\colr N{T}$ then $\seqt\Gamma N A$ in $\cool$.
\item For $t\in\ol$, $T\in\coolfs$, if $\seqt\Gamma {T} A$ in $\coolfs$ and $\colr t{T}$ then $\seqt\Gamma t A$ in $\ol$.
\end{enumerate}
\end{lemma}
\begin{proof}
We just prove the first statement, the second statement follows immediately from the first by virtue of Lemma~\ref{lem:equiv-typability-lambda-bar-terms}. 

It suffices to show for $N\in\cool$, $N'\in\coolfs$, if $\seqt\Gamma {N'} A$ in $\coolfs$ and $\colr N{N'}$ then $\seqt\Gamma N A$ in $\cool$ (replacing expression $T$ by term $N'$), since from this follows easily the result for elimination alternatives (replacing $T$ by $E\in\coolfs$).
Let
$$
\begin{array}{l}
R:=\{(\Gamma,N,A)\mid \exists N'\in\coolfs\cdot\colr N{N'}\wedge\seqt\Gamma {N'} A\}
\end{array}$$
By coinduction, to prove that this relation is contained in the typing relation of $\cool$, it suffices to show that it is \emph{closed backward} relatively to the rules defining that typing relation---which means, roughly speaking, that for each element of $R$ there is a typing rule which produces such element from premisses in $R$. This is the most fundamental principle of coinduction for coinductively defined predicates. It exploits that the coinductively defined predicate is maximal among the post-fixedpoints of the set operator underlying the coinductive definition. 
In our present application of the principle, we need to show that for any $(\Gamma,N,A)\in R$, one of the following holds:
\begin{enumerate}
\item $A=A_0\impl A_1$, $N=\lambda x^{A_0}.N_1$, and $(\Gamma\!,\!x:A_0\,,\,N_1\,,\,A_1)\in R$;
\item $A=p$, and there is $y:\vec{B}\supset p\in\Gamma$ so that $N=y\fl i{N_i}$, and, for all $i$, $(\Gamma, {N_i}, {B_i})\in R$.
\end{enumerate}

Let $(\Gamma,N,A)\in R$. Then $\colr N{N'}$ and $\seqt\Gamma {N'} A$, for some $N'\in\coolfs$. The proof proceeds by case analysis on $A$.

Case $A=A_0\impl A_1$. By definition of the typing relation, we must have $N'=\lambda x^{A_0}.N_1'$ and $\seqt{\Gamma,x:A_0}{N_1'}{A_1}$, for some $N_1'$; and by definition of $\colbase$, we must have $N=\lambda x^{A_0}.N_1$, and $\colr {N_1}{N_1'}$, for some $N_1$; therefore, $(\Gamma\!,\!x:A_0\,,\,N_1\,,\,A_1)\in R$, by definition of $R$.

Case $A=p$. By definition of the typing relation, we have $N'=\fs{j} E_j$ and $\seqt\Gamma {E_j} p$, for all $j$. Then, by definition of $\colbase$,
we must have, $\colr{N}{E_j}$, for some $j$. Let $E_j=y\fl i{N_i'}$. Again by definition of $\colbase$, $N=y\fl i{N_i}$, with $\colr{N_i}{N_i'}$ for all $i$. Since $\seqt\Gamma {y\fl i{N_i'}} p$, we must have, again by definition of the typing relation, $y:\vec{B}\supset p\in\Gamma$ and $\seqt{\Gamma}{N_i'}{B_i}$ for all $i$. Hence, for all $i$, $(\Gamma\,,\,N_i\,,\,B_i)\in R$, by definition of $R$.
\end{proof}

Now, we prove that in fact, for any search problem $\sigma=\seq \Gamma A$, the members of $\solfunction\sigma$ are exactly the solutions of $\sigma$.

\begin{proposition}[Soundness and completeness of full solution spaces]\label{prop:adequacy-general-case}~
\begin{enumerate}
\item
For $N\in\cool$, $\colr N{\sol\Gamma A}$ iff $\seqt\Gamma N A$ in $\cool$.
\item
For $t\in\ol$, $\colr t{\sol\Gamma A}$ iff $\seqt\Gamma t A$ in $\ol$.
\end{enumerate}
\end{proposition}
\begin{proof}

We prove the first statement in detail as a further example of coinductive reasoning, the second statement follows immediately from the first by virtue of Lemma~\ref{lem:equiv-typability-lambda-bar-terms}.

``If'' (stating completeness, i.\,e., that the $\solletter$ function indeed gathers all solutions). Consider the relations
$$
\begin{array}{l}
R_1:=\{(N,{\sol\Gamma A})\mid \seqt\Gamma N A\}\\
R_2:=\{(x\fl i{N_i},x\fl i{\sol\Gamma{B_i}})\mid (x:B_1,\cdots,B_k\impl p)\in\Gamma\wedge\seqt\Gamma {x\tuple{N_1,\ldots,N_k}} p\}
\end{array}
$$
It suffices to show that $R_1\subseteq\colbase$, but this cannot be
proven alone since $\colbase$
is defined simultaneously for coterms and elimination alternatives.
We also prove $R_2\subseteq\colebase$, and to prove both by
coinduction on the membership relations, it suffices to show that the
relations $R_1$, $R_2$ are closed backward relatively to the rules defining the membership predicate, that is:
\begin{enumerate}
\item for any $(M,N)\in R_1$, one of the following holds:
\begin{enumerate}
\item $(M,N)=({\lambda x^A.M'},{\lambda x^A.N'})$, and $(M',N')\in R_1$;\label{prop:adequacy-general-case-case1a}
\item $N=\fs i{E_i}$, and for some $i$, $(M,E_i)\in R_2$;\label{prop:adequacy-general-case-case1b}
\end{enumerate}
\item for any $(M,E)\in R_2$, $M={x\fl i{M_i}}$, and $E={x\fl i{N_i}}$, and for all $i$, $({{M_i}},{{N_i}})\in R_1$
\end{enumerate}

1. Take an arbitrary element of $R_1$, i.\,e., take $(M,{\sol\Gamma A})$ s.\,t.~$\seqt\Gamma M A$. One of the following happens:
\begin{enumerate}
\item[i)] $A=A_0\impl A_1$, $M=\lambda x^{A_0}.M'$, and $\seqt{\Gamma,x:A_0}{M'}{A_1}$;
\item[ii)] $A=p$, and there is $y:\vec{B}\supset p\in\Gamma$ so that $M=y\fl i{M_i'}$, and, for all $i$, $\seqt\Gamma {M_i'} {B_i}$.
\end{enumerate}

Case i). Note that $\sol\Gamma A=\lambda x^{A_0}.\sol{\Gamma,x:A_0}{A_1}$. So, in order to prove (\ref{prop:adequacy-general-case-case1a}), we need to show $(M',\sol{\Gamma,x:A_0}{A_1})\in R_1$, which follows from $\seqt{\Gamma,x:A_0}{M'}{A_1}$.

Case ii). Note that $\sol\Gamma A=\fs{{z:\vec{C}\supset p\in\Gamma}} {z\fl{j}{\sol{\Gamma}{C_j}}}$. So, since $y:\vec{B}\supset p\in\Gamma$, for the proof of (\ref{prop:adequacy-general-case-case1b}), it suffices to show  $(M,{y\fl{i}{\sol{\Gamma}{B_i}}})\in R_2$, which holds because $y:\vec{B}\supset p\in\Gamma$ and $\seqt\Gamma {y\fl i{M_i'}} p$
(the latter being a consequence of $y:\vec{B}\supset p\in\Gamma$, and $\seqt\Gamma {M_i'} {B_i}$, for all $i$).

2. Take an arbitrary element of $R_2$. So, it must be of the form $(x\fl{i}{N_i},x\fl{i}{\sol\Gamma{B_i}})$ s.t. $(x:\vec{B}\supset p)\in\Gamma$ and $\seqt\Gamma {x\fl{i}{N_i}} p$. From the latter follows  $\seqt\Gamma {N_i} {B_i}$, for all $i$. So, by definition of $R_1$, $({N_i},\sol\Gamma{B_i})\in R_1$, for all $i$.

``Only if'' (stating soundness, i.\,e., that the $\solletter$ function only collects solutions). Follows from Lemmas \ref{lem:type-of-sol-spaces} and \ref{lem:typing-of-members}.
\end{proof}

\begin{example}
\label{ex:Peirce}
Let us  consider the case of Peirce's law that is not valid intuitionistically. We have (for $p\neq q$):
$$
\sol{}{\Peircet}
=\lambda x^{(p\impl q)\impl p}. x\tuple{\lambda y^p.\oo}
$$
The fact that we arrived at $\oo$ and found no elimination alternatives on the way \emph{annihilates} the coterm and implies there are no
terms in the full solution space of $\seq{}{\Peircet}$ (hence no
proofs, nor even infinite solutions).
\end{example}

\section{Decontraction}\label{sec:decontraction}

In this section, divided into three subsections, we introduce and study the \emph{decontraction} operation on forests. The main result of this section is Lemma~\ref{lem:cleavage-2}, in the third subsection, because of its role in the proof of Theorem~\ref{thm:FullProp}---the main theorem of the paper. Lemma~\ref{lem:cleavage-2} shows that decontraction is the right operation to apply to a full solution space $T=\sol{\Gamma}C$ to express the effect on the full solution space of growing the context $\Gamma$ to an \emph{inessential extension} $\Gamma'$---this growth is made precise below and denoted by $\Gamma\leq\Gamma'$. Before, in the second subsection, the more general situation, where $T$ is any expression in $\coolfs$ (not necessarily a full solution space) is analyzed in Lemma~\ref{lem:deco-pres-types}, a result that shows in what sense decontraction witnesses the inversion of the inference rule of contraction. Finally, inversion of contraction is related to (and follows from) a kind of inversion of substitution, whose most general form is contained in Lemma~\ref{lem:undo-subst-2}, to be found already in the first subsection.

The decontraction operation on forests, denoted  $[\Gamma'/\Gamma]N$, is defined only when $\Gamma\leq\Gamma'$. Roughly speaking, the decontraction effect at the level of forests is to add new elimination alternatives, made possible by the presence of more variables in $\Gamma'$. This effect is best seen in the last clause of Definition~\ref{def:de-cont-forests} (in Figure~\ref{fig:def-decontr}) that applies the decontraction operation to a single elimination alternative.
\begin{definition}\label{def:leq}
\begin{enumerate}
\item $|\Gamma|=\{A\mid\textrm{there is $x$ s.\,t.}\,(x:A)\in\Gamma\}$.
\item $\Gamma\leq\Gamma'$ if $\Gamma\subseteq\Gamma'$ and
$|\Gamma|=|\Gamma'|$.
\end{enumerate}
\end{definition}
Notice that $|\Gamma|$ has only one element for each type occurring in the declarations of $\Gamma$. It thus abstracts away from multiple hypotheses of the same formula.

\begin{definition}[Decontraction for forests]\label{def:de-cont-forests}
Let $\Gamma\leq\Gamma'$. For $T$ an expression of  $\coolfs$, we define $[\Gamma'/\Gamma]T$ by
corecursion as described in Figure~\ref{fig:def-decontr}.
\begin{figure}[tb]\caption{The decontraction operation on forests}\label{fig:def-decontr}
\[
\begin{array}{lcll}
{[}\Gamma'/\Gamma](\lb x^A.N)&=&\lb x^A.[\Gamma'/\Gamma]N\\
{[}\Gamma'/\Gamma]\s i{E_i}&=&\s i{[\Gamma'/\Gamma]E_i}\\
{[}\Gamma'/\Gamma]\big(z\fl i{N_i}\big)&=&z\fl i{[\Gamma'/\Gamma]N_i}&\textrm{if $z\notin dom(\Gamma)$}\\
{[}\Gamma'/\Gamma]\big(z\fl
i{N_i}\big)&=&\kern-1em\s{(w:A)\in\Delta_z}{w}\fl
i{[\Gamma'/\Gamma]N_i}&\textrm{if $z\in dom(\Gamma)$}
\end{array}
\]
\end{figure}
In the last defining clause, $A:=\Gamma(z)$ and $\Delta_z:=\{(z:A)\}\cup(\Gamma'\setminus\Gamma)$.
The usual convention on bound variables applies, which requires in the first clause that the name $x$
is chosen so that it does not appear in $\Gamma'$.
\end{definition}
The effect of the last clause is to replace the summand
$z\fl i{N_i}$ with $z$ of type $\Gamma(z)$ according to $\Gamma$ with
the sum of all $w\fl i{N_i}$ that receive this type according to the
potentially bigger context $\Gamma'$, excluding the other variables of
$\Gamma$ but including the case $w=z$, and to continue the operation
corecursively in the argument terms.\footnote{In the
  workshop version~\cite{FICS2013}, we had a more ``aggressive''
  version of decontraction (called co-contraction in that paper) that did not exclude the other variables
  of $\Gamma$ in the last clause, and for which we further added the binding $x:A$ to $\Gamma$ and $\Gamma'$ in the corecursive call in the lambda-abstraction case.
  On solutions, these differences are
  immaterial, c.\,f. the example after Lemma~\ref{lem:sol-maxdecontr}.}

\begin{lemma}\label{lem:concontr-expands}
  If $\colr MT$ and $\Gamma\leq\Gamma'$ then $\colr M{[\Gamma'/\Gamma]T}$.
\end{lemma}
\begin{proof}
  A coinductive proof can confirm the obvious intuition of the effect of decontraction: either a summand is maintained, with corecursive application of decontraction to the subterms, or it is replaced by a sum with even extra summands.
\end{proof}

\begin{lemma}\label{lem:cocontr-identity}
$[\Gamma/\Gamma]T\bisimp T$.
\end{lemma}
\begin{proof}
  Obvious coinduction for all expressions.
\end{proof}

We formally extend the decontraction data from contexts to sequents $\sigma$. (This overloading of the operation will only be used in the next section.)
\begin{definition}[Decontraction for sequents]\label{def:de-cont-sequents}
Let $\sigma=(\seq{\Gamma}A)$ and $\sigma'=(\seq{\Gamma'}A')$.
\begin{enumerate}
\item $\sigma\leq\sigma'$  if $\Gamma\leq\Gamma'$ and $A=A'$;
\item if $\sigma\leq\sigma'$, then $[\sigma'/\sigma]T:=[\Gamma'/\Gamma]T$.
\end{enumerate}
\end{definition}

\subsection{Decontraction and substitution}

Decontraction is a form of undoing substitution, in the following sense ($N\in\cool$):
\begin{equation}\label{eq:undo-subst}
\colr N{[\Gamma,x:A,y:A/\Gamma,x:A][x/y]N}
\end{equation}
In fact, we prove a stronger result. Let $[x/x_1,\cdots,x_n]N$ denote $[x/x_1]\cdots[x/x_n]N$. We will even allow ourselves to abbreviate $x_1,\cdots,x_n$ by $\vec x$, when variable $n$ is in the context of discourse.
\begin{lemma}[Undoing substitution -- a general principle]\label{lem:undo-subst}
For $N\in\cool$, $T\in\coolfs$,
$$\colr{[x_1/x_1,\cdots,x_n]N}{T}\Rightarrow\colr N{[\Gamma,x_1:A,\ldots,x_n:A/\Gamma,x_1:A]T}\enspace.$$
\end{lemma}
\begin{proof}
Obviously, it suffices to show the statement with a term $N'$ in place of the expression $T$. This will follow from $R_1$ below being included in the membership relation with terms as second argument.
Let $\Delta:=\Gamma,x_1:A$ and $\Delta':=\Gamma,x_1:A,\ldots,x_n:A$. Let
$$
\begin{array}{rcl}
R_1&:=&\{(N,[\Delta'/\Delta]N')\mid \colr{[x_1/\vec x]N}{N'}\}\\
R_2&:=&\{(z\fl i{N_i},z\fl i{[\Delta'/\Delta]N_i'})\mid \forall i,\,\colr{N_i}{[\Delta'/\Delta]N_i'}\in R_1\}
\end{array}
$$
We argue by coinduction on membership. The proof obligations named (1)(a), (1)(b), and (2) in the proof of Proposition \ref{prop:adequacy-general-case} are renamed here Ia, Ib, and II, respectively.

Let $(N,[\Delta'/\Delta]N')\in R_1$, hence
\begin{equation}\label{eq:proof-of-lem-undo-subst-1}
\colr{[x_1/\vec x]N}{N'}\enspace .
\end{equation}
We have to show that Ia or Ib holds. We proceed by case analysis of $N$.

Case $N=\lb z.N_0$. Then $\colr{\lb z.[x_1/\vec x]N_0}{N'}$, hence, by definition of membership, we must have $N'=\lb z.N_0'$ and
\begin{equation}\label{eq:proof-of-lem-undo-subst-2}
\colr{[x_1/\vec x]N_0}{N_0'}\enspace,
\end{equation}
hence $[\Delta'/\Delta]N'=\lb z.[\Delta'/\Delta]N_0'$. From (\ref{eq:proof-of-lem-undo-subst-2}) and definition of $R_1$ we get $(N_0,[\Delta'/\Delta]N_0')\in R_1$, so Ia holds.

Otherwise, that is, if $N$ is not a lambda-abstraction, then the same is true of $[x_1/\vec x]N$, hence (\ref{eq:proof-of-lem-undo-subst-1}) implies that $N'=\s j{E_j'}$, with
\begin{equation}\label{eq:proof-of-lem-undo-subst-3}
\colr{[x_1/\vec x]N}{E_j'}
\end{equation}
for some $j$, hence
\begin{equation}\label{eq:proof-of-lem-undo-subst-4}
[\Delta'/\Delta]N'=\s j{[\Delta'/\Delta]E_j'}\enspace.
\end{equation}
To fulfil Ib, we need $(N,E)\in R_2$, for some summand $E'$ of (\ref{eq:proof-of-lem-undo-subst-4}). From (\ref{eq:proof-of-lem-undo-subst-3}) and the definition of membership we must have $N=z\fl i{N_i}$, for some $z$, hence
\begin{equation}\label{eq:proof-of-lem-undo-subst-5}
[x_1/\vec x]N=w\fl i{[x_1/\vec x]N_i}\enspace,
\end{equation}
with $w$ a variable determined by $z$ and $\vec x$ as follows: if $z\in\{x_1,\ldots,x_n\}$, then $w=x_1$, else $w=z$. Facts (\ref{eq:proof-of-lem-undo-subst-3}) and (\ref{eq:proof-of-lem-undo-subst-5}) give $E_j'=w\fl i{N_i'}$ and, for all $i$,
\begin{equation}\label{eq:proof-of-lem-undo-subst-6}
\colr{[x_1/\vec x]N_i}{N_i'}\enspace,
\end{equation}
hence
\begin{equation}\label{eq:proof-of-lem-undo-subst-7}
(N_i,[\Delta'/\Delta]N_i')\in R_1\enspace.
\end{equation}

Now we will see that $z\fl i{[\Delta'/\Delta]N_i'}$ is a summand of $[\Delta'/\Delta]E_j'$, sometimes the unique one. There are two cases:

First case: $z\in\{x_1,\ldots,x_n\}$. Then $[\Delta'/\Delta]E_j'=\sum_{k=1}^n x_k\fl i{[\Delta'/\Delta]N_i'}$, since $w=x_1$.

Second case: otherwise, $w=z$. Now, by definition of decontraction, $z\fl i{[\Delta'/\Delta]N_i'}$ is always a summand of $[\Delta'/\Delta](z\fl i{N_i'})$, and the latter is $[\Delta'/\Delta]E_j'$ since $w=z$.

Therefore, $z\fl i{[\Delta'/\Delta]N_i'}$ is a summand of sum (\ref{eq:proof-of-lem-undo-subst-4}). Moreover, $(N,z\fl i{[\Delta'/\Delta]N_i'})\in R_2$ by definition of $R_2$ and (\ref{eq:proof-of-lem-undo-subst-7}). So Ib holds.

Now let $(z\fl i{N_i},z\fl i{[\Delta'/\Delta]N_i'})\in R_2$. Proof obligation II is fulfilled, as $(N_i,[\Delta'/\Delta]N_i')\in R_1$ holds for all $i$, by definition of $R_2$.
\end{proof}
Fact (\ref{eq:undo-subst}) follows from the previous lemma by taking $n=2$, $x_1=x$, $x_2=y$ and $T=[x_1/x_1,x_2]N$.

The converse of the implication in Lemma \ref{lem:undo-subst} fails if other declarations with type $A$ exist in $\Gamma$.
\begin{example}
Let $\Gamma:=\{z:A\}$, $\Delta:=\Gamma,x:A$, $\Delta':=\Gamma,x:A,y:A$, $N:=y$ and $T:=z$. Then $N$ is a member of ${[\Delta'/\Delta]T}$, since $[\Delta'/\Delta]T=z+y$, but $[x/y]N=x$ and $x$ is not a member of $T$.
\end{example}

The result of a decontraction $[\Gamma,x_1:A,\cdots,x_n:A/\Gamma,x_1:A]T$, where $\Gamma$ has no declarations with type $A$, does not depend on $\Gamma$ nor $A$, so it deserves a lighter notation as $[x_1+\cdots +x_n/x_1]T$. We will even allow ourselves to abbreviate $x_1+\cdots +x_n$ by $\sum{\vec x}$, when variable $n$ is in the context of discourse.
This particular case of the operation satisfies the equations in Figure~\ref{fig:eqns-special-decontr}.
\begin{figure}[tb]\caption{Corecursive equations governing $[\sum{\vec x}/x_1]$}\label{fig:eqns-special-decontr}
\[
\begin{array}{lcll}
{[}\sum{\vec x}/x_1](\lb x^A.N)&=&\lb x^A.[\sum{\vec x}/x_1]N\\
{[}\sum{\vec x}/x_1]\s i{E_i}&=&\s i{[\sum{\vec x}/x_1]E_i}\\
{[}\sum{\vec x}/x_1]\big(z\fl i{N_i}\big)&=&z\fl i{[\sum{\vec x}/x_1]N_i}&\textrm{if $z\neq x_1$}\\
{[}\sum{\vec x}/x_1]\big(x_1\fl
i{N_i}\big)&=&\sum_{j=1}^n x_j\fl
i{[\sum{\vec x}/x_1]N_i}&
\end{array}
\]
\end{figure}
For this particular case, we get a pleasing formula:
\begin{lemma}[Undoing substitution -- a tighter result for a special case]\label{lem:undo-subst-2}
For $N\in\cool$, $T\in\coolfs$,
$$\colr{[x_1/x_1,\cdots,x_n]N}{T}\Leftrightarrow\colr N{[x_1+\cdots +x_n/x_1]T}\enspace,$$
provided $x_i\notin FV(T)$, $i=2,\ldots,n$.
\end{lemma}
\begin{proof}
``Only if''. Particular case of Lemma \ref{lem:undo-subst}.

``If''. Let $\phi(T)$ denote the proviso on $T$. Let
$$
\begin{array}{rcl}
R_1&:=&\{([x_1/\vec x]N,N')\mid \phi(N')\wedge\colr{N}{[\sum{\vec x}/x_1]N'}\}\\
R_2&:=&\{(z\fl i{[x_1/\vec x]N_i},z\fl i{N_i'})\mid \forall i,\,([x_1/\vec x]N_i,N_i')\in R_1\}
\end{array}
$$
We argue by coinduction on membership and thus obtain the ``if'' part with $T$ replaced by $N'$, from which the general case immediately follows. The proof obligations named (1)(a), (1)(b), and (2) in the proof of Proposition \ref{prop:adequacy-general-case} are renamed here Ia, Ib, and II, respectively.

Let $([x_1/\vec x]N,N')\in R_1$, hence $\phi(N')$ and
\begin{equation}\label{eq:proof-2nd-lem-undo-subst-1}
\colr{N}{[\sum{\vec x}/x_1]N'}\enspace.
\end{equation}
The proof proceeds by case analysis of $N$.

Case $N=\lb z.N_0$, so $[x_1/\vec x]N=\lb z.[x_1/\vec x]N_0$. By (\ref{eq:proof-2nd-lem-undo-subst-1}) and definitions of membership and of $[\sum{\vec x}/x_1]N'$, $N'=\lb z.N_0'$, hence $\phi(N_0')$ (because $z$ is not one of $x_2,\cdots,x_n$), $[\sum{\vec x}/x_1]N'=\lb z.[\sum{\vec x}/x_1]N_0'$ and
\begin{equation}\label{eq:proof-2nd-lem-undo-subst-2}
\colr{N_0}{[\sum{\vec x}/x_1]N_0'}\enspace.
\end{equation}
So $([x_1/\vec x]N_0,N_0')\in R_1$, by definition of $R_1$, (\ref{eq:proof-2nd-lem-undo-subst-2}) and $\phi(N_0')$, which completes proof obligation Ia.

Case $N=z\fl i{N_i}$. Then $[x_1/\vec x]N=y\fl i{[x_1/\vec x]N_i}$, with $y=x_1$ when $z\in\{x_1,\ldots,x_n\}$, and $y=z$ otherwise. From (\ref{eq:proof-2nd-lem-undo-subst-1}) and definitions of membership and  of $[\sum{\vec x}/x_1]N'$, one gets $N'=\s j{E_j'}$, hence $\phi(E_j')$ for all $j$, and $[\sum{\vec x}/x_1]N'=\s j{[\sum{\vec x}/x_1]E_j'}$. In order to fulfil proof obligation Ib, we need $([x_1/\vec x]N,E')\in R_2$, for some summand $E'$ of $N'$. From (\ref{eq:proof-2nd-lem-undo-subst-1}) again, we get, for some $j$,
\begin{equation}\label{eq:proof-2nd-lem-undo-subst-3}
\colr{z\fl i{N_i}}{[\sum{\vec x}/x_1]E_j'}\enspace.
\end{equation}
Let $E'_j=w\fl i{N_i'}$, hence $\phi(N_i')$ for all $i$. We now have two cases:

First case: $w=x_1$. Then $[\sum{\vec x}/x_1]E_j'=\sum_{k=1}^nx_k\fl i{[\sum{\vec x}/x_1]N_i'}$. From (\ref{eq:proof-2nd-lem-undo-subst-3}) we get, for some $k$,
\begin{equation}\label{eq:proof-2nd-lem-undo-subst-4}
\colr{z\fl i{N_i}}{x_k\fl i{[\sum{\vec x}/x_1]N_i'}}
\end{equation}
hence, for all $i$,
\begin{equation}\label{eq:proof-2nd-lem-undo-subst-5}
\colr{N_i}{[\sum{\vec x}/x_1]N_i'}\enspace.
\end{equation}
From (\ref{eq:proof-2nd-lem-undo-subst-4}), $z=x_k$, hence $y=x_1$. We prove $([x_1/\vec x]N,E'_j)\in R_2$,  that is $(x_1\fl i{[x_1/\vec x]N_i},x_1\fl i{N_i'})\in R_2$. By definition of $R_2$, we need $([x_1/\vec x]N_i,N_i')\in R_1$, for all $i$. This follows from (\ref{eq:proof-2nd-lem-undo-subst-5}), $\phi(N_i')$ and the definition of $R_1$.

Second case: $w\neq x_1$. Then $[\sum{\vec x}/x_1]E_j'=w\fl i{[\sum{\vec x}/x_1]N_i'}$. From (\ref{eq:proof-2nd-lem-undo-subst-3}), $z=w$; from $\phi(E'_j)$ and $w\neq x_1$, $z\notin\{x_1,\ldots,x_n\}$. Still from (\ref{eq:proof-2nd-lem-undo-subst-3}), we get again (\ref{eq:proof-2nd-lem-undo-subst-5})
and now $([x_1/\vec x]N,E'_j)=(z\fl i{[x_1/\vec x]N_i},z\fl i{N_i'})\in R_2$ follows as before.

Let $(z\fl i{[x_1/\vec x]N_i},z\fl i{N_i'})\in R_2$, hence proof obligation II holds by definition of $R_2$.
\end{proof}

The proviso about variables $x_2,\cdots,x_n$ in the previous lemma is necessary for the ``if'' implication. Otherwise, one has the following counter-example: $n:=2$, $N:=x_2$, and $T=x_2$. $N$ is a member of $[x_1+x_2/x_1]T=x_2$ but $x_1=[x_1/x_1,x_2]N$ is not a member of $T$.

\subsection{Decontraction and contraction}

Decontraction is related to the inference rule of contraction. By \emph{contraction} we mean the rule in the following lemma.
\begin{lemma}[Contraction]\label{lem:invert-contraction}
In system $\ol$ the following rule is admissible and invertible:
$$
\infer{\seqt{\Gamma,x:A}{[x/y]t}{B}}{\seqt{\Gamma,x:A,y:A}{t}{B}}\enspace.
$$
That is: for all $t\in\ol$, $\seqt{\Gamma,x:A,y:A}{t}{B}$ iff $\seqt{\Gamma,x:A}{[x/y]t}{B}$.
\end{lemma}
\begin{proof}
Routine induction on $t$, using inversion of $\RIntro$ and $\LVecIntro$.
\end{proof}

If $\Gamma\leq\Gamma'$, then, from a proof of $\seq{\Gamma'}B$, we get a proof of $\seq{\Gamma}B$ by a number of contractions. The following result justifies the terminology ``decontraction''.
\begin{lemma}[Decontraction and types]
\label{lem:deco-pres-types} Let $T$ be an expression of $\coolfs$ and $\Gamma'\cup\Delta$ be a context.
If $\Gamma\cup\Delta\vdash T:B$ and $\Gamma\leq\Gamma'$ then $\Gamma'\cup\Delta\vdash[\Gamma'/\Gamma]T:B$.
\end{lemma}
\begin{proof}
  (Notice that we exceptionally consider not necessarily disjoint unions of contexts. This is immaterial for the proof but will be needed in Lemma~\ref{lem:inter-pres-types}.) Immediate by coinduction.\footnote{With this lemma in place, invertibility in Lemma \ref{lem:invert-contraction} follows from general reasons. Take $N=t$ in fact (\ref{eq:undo-subst}) and then apply this lemma and Lemma \ref{lem:typing-of-members}.}
\end{proof}
In particular, if $\Gamma\vdash u:B$ in $\ol$ and $\Gamma\leq\Gamma'$, then indeed $\Gamma'\vdash[\Gamma'/\Gamma]u:B$ --- but $[\Gamma'/\Gamma]u$ is not guaranteed to be a proof (\emph{i.\,e.}, a term in $\ol$).
\begin{example}\label{ex:invert-contraction}
Let $\Gamma:=\{f:p\impl p\impl q,x:p\}$, $\Gamma':=\{f:p\impl p\impl q,x:p,y:p\}$, and $u:=f\tuple{x,x}$, hence $\Gamma\leq\Gamma'$ and $\seqt{\Gamma}uq$.
Then, $[\Gamma'/\Gamma]u=f\tuple{x+y,x+y}$, and the given particular case of the previous lemma entails $\seqt{\Gamma'}{f\tuple{x+y,x+y}}q$. The term $f\tuple{x+y,x+y}$  is no $\ol$-term, but rather has several members. Due to Lemma \ref{lem:undo-subst-2}, these are exactly the (four, in this case) $t\in\ol$ such that $[x/y]t=u$. Thanks to Lemma \ref{lem:invert-contraction}, it follows that each member $t$ of $f\tuple{x+y,x+y}$ satisfies $\seqt{\Gamma'}{t}q$.
\end{example}
On the other hand, if $T$ in Lemma \ref{lem:deco-pres-types} is the full solution space $\sol{\Gamma}B$ (rather than a mere member $u$ of it), then $[\Gamma'/\Gamma]T$ is indeed the full solution space $\sol{\Gamma'}B$ --- but we have to wait until Lemma \ref{lem:cleavage-2} to see the proof.
\begin{example}
Continuing Example \ref{ex:invert-contraction}, since $\sol{\Gamma}{q}=u$, one has $[\Gamma'/\Gamma]\sol{\Gamma}{q}=f\tuple{x+y,x+y}$. Lemma \ref{lem:cleavage-2} will guarantee that $f\tuple{x+y,x+y}$ (a term obtained from $u$ by decontraction) is the full solution space $\sol{\Gamma'}{q}$. Thanks to Proposition \ref{prop:adequacy-general-case}, one sees again that each member of $t$ of $f\tuple{x+y,x+y}$ satisfies $\seqt{\Gamma'}{t}q$.
\end{example}

\subsection{Decontraction and full solution spaces}

The intuitive idea of the next notion is to capture saturation of sums, so to speak.
\begin{definition}[Maximal decontraction]\label{def:maxdecontr}
Let $T\in\coolfs$ and $\Gamma$ be a context.
\begin{enumerate}
\item Consider an occurrence of $x$ in $T$. Consider the traversed
  lambda-abstractions from the root of $T$ to the given
  occurrence of $x$, and let $y_1^{A_1},\ldots,y_n^{A_n}$ be the respective variables. We call $\Gamma,y_1:A_1\ldots,y_n:A_n$ the \emph{local extension} of $\Gamma$ for the given occurrence of $x$.
\item $T$ in $\coolfs$ is maximally decontracted w.\,r.\,t.~$\Gamma$ if:
\begin{enumerate}
\item all free variables of $T$ are declared in $\Gamma$; and
\item every occurrence of a variable $x$ in $T$ is as
  head of a summand $x\fl i{N_i}$ in a sum in which also $y\fl i{N_i}$
  is a summand (modulo bisimilarity), for every variable $y$ that gets the same type as $x$ in the
  local extension of $\Gamma$ for the occurrence of $x$.
\end{enumerate}
\end{enumerate}
\end{definition}

\begin{lemma}[Full solution spaces are maximally decontracted]\label{lem:sol-maxdecontr}
  Given sequent $\seq\Gamma C$, the full solution space $\sol{\Gamma}C$ is
  maximally decontracted w.\,r.\,t.~$\Gamma$.
\end{lemma}
\begin{proof}
  By coinduction. For the variable occurrences that are on display in
  the one-line formula (\ref{eq:one-line-sol-space}) for $\sol\Gamma{\vec A\impl p}$---that is, for each of the $y$'s that are head variables of the displayed summands---the local
  context is $\Delta=\Gamma,\vec x:\vec A$, and if $y_1$ and $y_2$
  have the same type in $\Delta$ with target atom $p$, both variables
  appear as head variables with the same lists of argument terms. For variable occurrences hidden in the $j$-th argument of some $y$, we use two facts: (i) the $j$-th argument is maximally decontracted w.\,r.\,t. $\Delta$ by coinductive hypothesis; (ii) $\Delta$ collects the variables $\lambda$-abstracted on the path from the root of the term to the root of $j$-th argument.
\end{proof}
\begin{example}
Let $\Gamma:=\{z:p\}$, $\Delta:=\Gamma,x:p$, $N:=\lb x^p.z\tuple{}$ and $N':=\lb x^p.z\tuple{}+x\tuple{}$. The term $N$ is not maximally decontracted w.\,r.\,t. $\Gamma$. Intuitively, the sum $z\tuple{}$ is not saturated, as it does not record all the alternative proofs of $\seq{\Delta}p$. Hence $N$ cannot be the full solution space $\sol{\Gamma}{p\impl p}$ --- the latter is $N'$, hence $N'$ is maximally decontracted w.\,r.\,t. $\Gamma$, by the previous lemma. The output of decontraction $[\Gamma/\Gamma]N$ (being $N$) is not maximally decontracted\footnote{This is in contrast with the definition of co-contraction in \cite{FICS2013}, which outputs maximally decontracted terms, e.\,g., $[\Gamma/\Gamma]N=N'$ in this case.}. We will be interested mostly in applying decontraction to already maximally decontracted terms, e.\,g., full solution spaces.
\end{example}

\begin{lemma}\label{lem:de-cont-extension-1}
  If $|\Gamma'\setminus\Gamma|$ and $|\Delta|$ are disjoint, $\Gamma',\Delta$ is a context
  and $\Gamma\leq\Gamma'$ then
  $[\Gamma',\Delta/\Gamma,\Delta]T\bisimp[\Gamma'/\Gamma]T$.
\end{lemma}
\begin{proof}
  Easy coinduction.
\end{proof}

The disjointness condition of
the previous lemma is rather severe. It can be replaced by maximal decontraction of the given term.

\begin{lemma}\label{lem:de-cont-extension-2}
  If $\Gamma',\Delta$ is a context, $\Gamma\leq\Gamma'$ and $T$ is maximally decontracted w.\,r.\,t.~$\Gamma,\Delta$, then
  $[\Gamma',\Delta/\Gamma,\Delta]T\bisim[\Gamma'/\Gamma]T$.
\end{lemma}
\begin{proof}
  By coinduction. The proof then boils down to showing for any subterm
  $z\fl i{N_i}$ of $T$, if a $w\neq z$ is found according to the last
  clause of the definition of decontraction with
  $[\Gamma',\Delta/\Gamma,\Delta]$, then one can also find $w$
  according to the last clause of the definition of decontraction
  with $[\Gamma'/\Gamma]$. Assume such a $w$. Since it comes from the
  last clause, we have $z\in\dom(\Gamma,\Delta)$ (hence, by the usual
  convention on the naming of bound variables, $z$ is even a free
  occurrence in $T$), and
  $(w:(\Gamma,\Delta)(z))\in\Gamma'\setminus\Gamma$.  If
  $z\in\dom(\Gamma)$, then we are obviously done.
  Otherwise, $z\in\dom(\Delta)$, and so
  $(w:\Delta(z))\in\Gamma'\setminus\Gamma$.  Since
  $|\Gamma'|=|\Gamma|$, there is $(x:\Delta(z))\in\Gamma$. Since $T$
  is maximally decontracted w.\,r.\,t.~$\Gamma,\Delta$, the subterm
  $z\fl i{N_i}$ is one summand in a sum which also has the summand
  $x\fl i{N_i}$, and for the latter summand, the last clause of the
  definition of decontraction with $[\Gamma'/\Gamma]$ can be used
  with $(w:\Gamma(x))\in\Gamma'\setminus\Gamma$.
\end{proof}

\begin{corollary}\label{cor:de-cont-extension-2}
If $\Gamma',\Delta$ is a context, $\Gamma\leq\Gamma'$, then
  $[\Gamma',\Delta/\Gamma,\Delta]\sol{\Gamma,\Delta}C\bisim[\Gamma'/\Gamma]\sol{\Gamma,\Delta}C$.
\end{corollary}
\begin{proof}
  Combine the preceding lemma with
  Lemma~\ref{lem:sol-maxdecontr}.\footnote{The notion of being
    maximally decontracted is not essential for this paper. Only this
    corollary will be used in the sequel, and it could also be proven directly, in the
    style of the proof of the following lemma. For this to work
    smoothly, the statement should be generalized to: If
    $\Gamma',\Delta,\Theta$ is a context, $\Gamma\leq\Gamma'$, then
    $[\Gamma',\Delta/\Gamma,\Delta]\sol{\Gamma,\Delta,\Theta}C\bisim[\Gamma'/\Gamma]\sol{\Gamma,\Delta,\Theta}C$.}
\end{proof}

The following main result of this section says that the full solution space w.\,r.\,t.~an
inessential extension of a context is obtained by applying the decontraction operation to the full solution space corresponding to the original context.
\begin{lemma}[Decontraction and full solution spaces]
\label{lem:cleavage-2}
If $\Gamma\leq\Gamma'$ then the following holds:
$\sol{\Gamma'}C\bisim[\Gamma'/\Gamma](\sol{\Gamma}C)$.
\end{lemma}
\begin{proof} Let
$R:=\{(\sol{\Gamma'}C,[\Gamma'/\Gamma](\sol{\Gamma}C))\mid\Gamma\leq\Gamma',C\textrm{ arbitrary}\}$.
\noindent We prove that $R$ is closed backward relative to the
notion of bisimilarity taking sums of alternatives as if they were sets. From this, we conclude $R\subseteq\bisim$.
\begin{equation}\label{eq:lem-cleavage2-fst}
\sol{\Gamma'}C=\lb z_1^{A_1}\cdots z_n^{A_n}.\s{(z:\vec B\impl
p)\in\Delta'}{z\fl j{\sol{\Delta'}{B_j}}}
\end{equation}
\noindent and
\begin{equation}\label{eq:lem-cleavage2-snd}
[\Gamma'/\Gamma](\sol{\Gamma}C)=\lb z_1^{A_1}\cdots
z_n^{A_n}.
\s{(y:\vec B\impl p)\in\Delta}{\s{\setbox0=\hbox{\scriptsize{\phantom{$\vec B$}}}\copy0\kern-\wd0(w:\Delta(y))\in\Delta'_y}{w\fl j{[\Gamma'/\Gamma]\sol{\Delta}{B_j}}}}
\end{equation}
\noindent where $\Delta:=\Gamma,z_1:A_1,\ldots,z_n:A_n$,
$\Delta':=\Gamma',z_1:A_1,\ldots,z_n:A_n$, for $y\in\dom(\Gamma)$, $\Delta'_y:=\{(y:\Delta(y))\}\cup(\Gamma'\setminus\Gamma)$,
and for $y=z_i$, $\Delta'_y=\{(y:\Delta(y))\}$.

From $\Gamma\leq\Gamma'$ we get $\Delta\leq\Delta'$,
hence
$$
(\sol{\Delta'}{B_j},[\Delta'/\Delta]\sol{\Delta}{B_j})\in R\enspace,
$$
which fits with the summands in~(\ref{eq:lem-cleavage2-snd}) since, by Corollary~\ref{cor:de-cont-extension-2}, $[\Delta'/\Delta]\sol{\Delta}{B_j}\bisim[\Gamma'/\Gamma]\sol{\Delta}{B_j}$.
\noindent To conclude the proof, it suffices to show that (i) each
head-variable $z$ that is a ``capability'' of the summation in
(\ref{eq:lem-cleavage2-fst}) is matched by a head-variable $w$ that
is a ``capability'' of the summation in
(\ref{eq:lem-cleavage2-snd}); and (ii) vice-versa.

(i) Let $z\in dom(\Delta')$. We have to exhibit $y\in dom(\Delta)$
such that $(z:\Delta(y))\in\Delta'_y$. First case: $z\in
dom(\Delta)$. Then, $(z:\Delta(z))\in\Delta'_z$. So
we may take $y=z$. Second and last case:
$z\in\dom(\Gamma')\setminus\dom(\Gamma)$. By definition of $\Gamma\leq\Gamma'$, there is
$y\in\dom(\Gamma)$ such that $(z:\Gamma(y))\in\Gamma'$. Since $\Gamma(y)=\Delta(y)$ and $z\notin\dom(\Delta)$, we get
$(z:\Delta(y))\in\Delta'_y$.

(ii) We have to show that, for all $y\in dom(\Delta)$, and all
$(w:\Delta(y))\in\Delta'_y$, $(w:\Delta(y))\in\Delta'$. But this is
immediate. \end{proof}

Notice that we cannot expect that the summands appear in the same
order in (\ref{eq:lem-cleavage2-fst}) and
(\ref{eq:lem-cleavage2-snd}). Therefore, we are obliged to use symmetry of $+$.
It is even
convenient to disregard multiplicity, as seen in the following example.

\begin{example}\label{ex:idempotence-needed}
  Let $\Gamma:=x:p$, $\Gamma':=\Gamma,y:p$, $\Delta:=z:p$, $\Theta:=\Gamma,\Delta$,
  $\Theta':=\Gamma',\Delta$ and $C:=p$. Then $\sol{\Theta}C=x+z$ and
  $\sol{\Theta'}{C}=x+y+z$. This yields
  $[\Theta'/\Theta]\sol{\Theta}C=(x+y)+(z+y)$ and
  $[\Gamma'/\Gamma]\sol{\Theta}C=(x+y)+z$, where parentheses are only
  put to indicate how decontraction has been calculated. Taken together, these calculations
  contradict the strengthening of Lemma~\ref{lem:cleavage-2} without
  idempotence of $+$, when the parameters $\Gamma$, $\Gamma'$, of the lemma are taken as $\Theta$, $\Theta'$,
  and they also contradict the analogous strenghtening of
  Corollary~\ref{cor:de-cont-extension-2} when the parameters $\Gamma$,
  $\Gamma'$, $\Delta$, $C$ of the corollary are as given here.

  The summand-wise and therefore rather elegant definition of decontraction is the root cause for
  this blow-up of the decontracted terms. However, mathematically, there is no blow-up since we
  identify $(x+y)+(z+y)$ with $x+y+z$, as they represent the same set of elimination alternatives.
\end{example}

In the light of Lemma~\ref{lem:sol-maxdecontr}, Lemma~\ref{lem:cleavage-2} shows
that $\sol{\Gamma}C$, which is maximally decontracted
w.\,r.\,t. $\Gamma$, only needs the application of the decontraction
operation $[\Gamma'/\Gamma]$ for $\Gamma\leq\Gamma'$ to obtain a term
that is maximally decontracted w.\,r.\,t. $\Gamma'$.

\begin{example}[Example~\ref{ex:dn-Peirce} continued]\label{ex:dn-Peirce-cont1}
Thanks to Lemma~\ref{lem:cleavage-2}, $N_9$ is obtained by
decontraction from $N_5$:
$$N_9\bisim[x:\cdot,y:(p\impl q)\impl p,z:p,y_1:(p\impl q)\impl p,z_1:p\,/\,x:\cdot,y:(p\impl q)\impl p,z:p]N_5\enspace,$$
where the type of $x$ has been omitted. Hence, $N_6$, $N_7$, $N_8$ and $N_9$ can be eliminated, and $N_5$ can be expressed as the (meta-level) fixed point:
$$N_5\bisim\fix\, N.x\tuple{\lambda y_1^{(p\impl q)\impl p}.y\tuple{\lambda z_1^p.[x,y,z,y_1,z_1/x,y,z]N}+z+y_1\tuple{\lambda z_1^p.[x,y,z,y_1,z_1/x,y,z]N}}\enspace,$$
now missing out all types in the decontraction operation(s).
Finally, we obtain the closed forest
$$\sol{}{\DNPeircet}\bisim\lambda x^{\Peircet\impl q}.x\tuple{\lambda y^{(p\impl q)\impl p}.y\tuple{\lambda z^{p}. N_5}}$$
This representation also makes evident that, by exploiting the different decontracted copies of $y$, there are infinitely many $M\in\cool\setminus\ol$ such that $\colr M{\sol{}{\DNPeircet}}$, in other words, $\seq{}\DNPeircet$ has infinitely many infinite solutions.
\end{example}
\begin{example}[Example~\ref{ex:Three} continued]\label{ex:Three-cont1}
Likewise, Lemma~\ref{lem:cleavage-2} shows that, with the notation of Example~\ref{ex:Three} and omitting the types in the decontraction operation, $N'=[x,y,z/x,y]N$, hence
$$\sol{}\Threet=\lambda x^{(p\impl p)\impl p}.x\tuple{\lambda y^p.\nu N.x\tuple{\lambda z^p.[x,y,z/x,y]N}+y}$$
Visibly, the only infinite solution is obtained by choosing always the
left alternative, creating infinitely many vacuous bindings, thus it
can be described as $\lambda x^{(p\impl p)\impl p}.N_0$ with
$N_0=x\tuple{\lambda \_^p.N_0}$ (where $\_$ is the name of choice for
a variable that has no bound occurrences).
\end{example}

We have now seen succinct presentations of the full solution spaces of all
of the examples in Example~\ref{ex:types}. Although described with few
mathematical symbols, they are still on the informal level of
infinitary terms with meta-level fixed points, but this will be
remedied by a finitary system in the next section.


\section{A typed finitary system for solution spaces}\label{sec:finitary-calculus}

In this section we develop a finitary lambda-calculus to represent solution spaces of proof search problems in $\ol$. The main points in the design of the calculus are:
\begin{enumerate}
\item $\ol$ is extended with fixed-point variables and formal greatest fixed points, as well as formal sums;
\item Fixed-point variables stand for spaces of solutions;
\item Fixed-point variables are typed by logical sequents;
\item A relaxed form of binding of fixed-point variables has to be allowed, and controlled through the typing system.
\end{enumerate}

The calculus is called finitary because its terms are generated inductively; and its terms are called finitary forests due to the presence of formal sums. There is a semantics of the finitary forests, by way of an interpretation into forests. The relaxed form of binding is matched, on the semantical side, by the special operation of decontraction. This is developed in the first subsection, with the typing system only coming in the second subsection. To each sequent, one can associate a finitary forest (third subsection) whose interpretation is the forest that represents the full solution space of the sequent: this is our foundational theorem (fourth subsection), showing the completeness of the semantics w.\,r.\,t.~those forests that represent solution spaces. The fifth and final subsection presents a variation of the semantics, that will be needed in the applications described in Section~\ref{sec:analysis}.

\subsection{The untyped system $\olfsfix$}\label{sec:untypedfinitary}

The set of inductive cut-free lambda-terms with finite numbers of
elimination alternatives, and a fixed-point operator is denoted by
$\olfsfix$ and is given by the following grammar (read inductively):

$$
\begin{array}{lcrcl}
\textrm{(terms)} &  & N & ::= & \lambda x^A.N\mid \gfp\,{X^\rho}.\ns{E_1}{E_n}\mid X^\rho\\
\textrm{(elim. alternatives)} &  & E & ::= & x \tuple{N_1,\ldots,N_k}\\
\end{array}
$$
where $X$ is assumed to range over a countably infinite set of
\emph{fixed-point variables} (also letters $Y$, $Z$ will range over them), and where, as for $\coolfs$,
both $n,k\geq0$ are arbitrary. We extend our practice established for $\coolfs$ of writing the sums $\ns{E_1}{E_n}$ in the form $\sum_iE_i$ for $n\geq0$. Also the tuples continue to be communicated as $\fl i{N_i}$. As for $\coolfs$, we will identify expressions modulo symmetry and idempotence of $+$, thus treating sums of elimination alternatives as if they were the set of those elimination alternatives. Again, we will write $T$ for expressions of $\olfsfix$, i.\,e., for terms and elimination alternatives.

In the term formation rules, letter $\rho$ appears. It is supposed to stand for ``restricted'' logical sequents in that we require them to be \emph{atomic}, i.\,e., of the form $\seq\Gamma p$ with atomic conclusion. Henceforth, this restriction is indicated when using the letter $\rho$, possibly with decorations.
Let $\FPV(T)$ denote
the set of free occurrences of typed fixed-point variables in
$T$, defined with the expected cases as follows: $\FPV(X^\rho):=\{X^\rho\}$, $\FPV(\lambda x^A.N):=\FPV(N)$, $\FPV(x\tuple{N_1,\ldots,N_k}):=\FPV(N_1)\cup\ldots\cup\FPV(N_k)$.
However, in $\gfp\,{X^\rho}.\sum_iE_i$ the fixed-point construction $\gfp$ binds
\emph{all} free occurrences of $X^{\rho'}$ in the elimination alternatives $E_i$, not
just $X^\rho$, as long as $\rho\leq\rho'$. To be precise, the definition is as follows:
$$\FPV(\gfp\,{X^\rho}.\sum_iE_i):=\bigl(\FPV(E_1)\cup\ldots\cup\FPV(E_n)\bigr)\setminus
    \{X^{\rho'}\mid\mbox{$\rho'$ atomic logical sequent and $\rho\leq\rho'$}\}$$
In fact, the sequent $\rho$ serves a different
purpose than being the precise type of bound fixed-point variables $X$, see below on
well-bound expressions that require at least that only $X^{\rho'}$ with $\rho\leq\rho'$ are free in the body of the $\gfp$-abstraction with binding variable $X^\rho$.

In the sequel, when we refer to \emph{finitary forests} we have
in mind the terms of $\olfsfix$. The fixed-point operator is called
$\gfp$ (``greatest fixed point'') to indicate that its semantics is
(now) defined in terms of infinitary syntax, but there, fixed points
are unique. Hence, the reader may just read this as ``the fixed
point''.

We next present a general-purpose interpretation of expressions of
$\olfsfix$ in terms of the coinductive syntax of $\coolfs$ (using the
$\fix$ operation on the meta-level). We stress its general purpose by
putting $g$ as upper index to the semantics brackets. This is for
contrast with the special-purpose interpretation we introduced under
the name ``simplified semantics'' in our subsequent
work~\cite{EspiritoSantoMatthesPintoInhabitation} and that will be presented at the end of this Section~\ref{sec:finitary-calculus}.
The general-purpose interpretation of finitary forests is based on the same ideas as our original
interpretation \cite{FICS2013} but is more precise on the conditions that
guarantee its well-definedness. (Nonetheless, in the cited paper, no problem arises with
the less precise definitions since only representations of full solution
spaces were interpreted, see below.)

We call an expression $T$ \emph{trivially regular} if $\FPV(T)$ has no duplicates: A
set $S$ of typed fixed-point variables is said to \emph{have no
  duplicates} if the following holds: if $X^{\rho_1},X^{\rho_2}\in
S$, then $\rho_1=\rho_2$. In other words: $X$ does not appear with two different types in $S$.
We \emph{do not} confine our
investigation to trivially regular expressions, see
\ref{sec:app-regular} for an example where we require more
flexibility.

\begin{definition}[Regularity in  $\olfsfix$]
  Let $T\in\olfsfix$. $T$ is regular if for all fixed-point variable
  names $X$, the following holds: if $X^\rho\in\FPV(T)$ for some
  sequent $\rho$, then there is a sequent $\rho_0$ such that, for
  all $X^{\rho'}\in\FPV(T)$, $\rho_0\leq\rho'$.
\end{definition}
Obviously, every trivially regular $T$ is regular (using
$\rho_0:=\rho$ and reflexivity of $\leq$ since $\rho'=\rho$).
Trivially, every closed $T$, i.\,e., with $\FPV(T)=\emptyset$, is
trivially regular.

As is to be expected, interpretation of expressions of $\olfsfix$ is done with the help of \emph{environments}, a notion which will be made more precise than in \cite{FICS2013}.
Since interpretations of $T$ only depend on the values of the environment on
$\FPV(T)$, we rather assume that environments are partial functions
with a finite domain. Hence, an environment $\xi$ is henceforth a
partial function from typed fixed-point variables $X^\rho$ to
(co)terms of $\coolfs$ with finite domain $\dom(\xi)$ that has no
duplicates (in the sense made precise above).

The interpretation function will also be made partial: $\interp T\xi$ will only be defined when environment $\xi$ is \emph{admissible} for $T$:
\begin{definition}[Admissible environment]
An environment $\xi$ is admissible for expression $T$ of $\olfsfix$ if for every $X^{\rho'}\in\FPV(T)$, there is an $X^\rho\in\dom(\xi)$ such that $\rho\leq\rho'$.
\end{definition}
Notice that the required sequent $\rho$ in the above definition is
unique since $\xi$ is supposed to be an environment. This observation
even implies the following characterization of regularity:
\begin{lemma}\label{lem:regular-versus-admissible}
$T\in\olfsfix$ is regular iff there is an environment $\xi$ that is admissible for $T$.
\end{lemma}
\begin{proof}
  Obvious.
\end{proof}

We have to add a further restriction before defining the
interpretation function:
\begin{definition}[Well-bound expression]
We call an expression $T$ of $\olfsfix$
\emph{well-bound} iff
for any of its subterms $\gfp\,{X^\rho}.\sum_iE_i$ and any free occurrence of $X^{\rho'}$ in any $E_i$,  $\rho\leq\rho'$.
\end{definition}
According to our definition of $\FPV$, an expression that is not well-bound has a subterm $N:=\gfp\,{X^\rho}.\sum_iE_i$ such that
$\FPV(N)$ contains some $X^{\rho'}$ that ``escapes'' the binding because $\rho\leq\rho'$ does not hold. Finitary forests we will construct to represent search spaces therefore ought to be well-bound, and this will be strengthened in Lemma~\ref{lem:typablewellbound} to a question of typability.

\begin{definition}[General-purpose interpretation of finitary forests as forests]
For a well-bound expression $T$ of $\olfsfix$, the interpretation $\interp T\xi$ for an environment $\xi$ that is admissible for
$T$ is given
by structural recursion on $T$ in Figure~\ref{fig:g-interp}.
\begin{figure}\caption{Definition of general-purpose interpretation}\label{fig:g-interp}
\[
\begin{array}{rcll}
\interp{X^{\rho'}}\xi& = & [\rho'/\rho]\xi(X^{\rho})\quad\textrm{for the unique $\rho\leq\rho'$ with $X^\rho\in\dom(\xi)$}\\
\interp{\gfp\,{X^{\rho}}.\s{i}{E_i}}\xi&= & \fix\, N.\s i{\interp {E_i}{\xi\cup[X^{\rho}\mapsto N]}}\\
\interp{\lambda x^A.N}{\xi}& = & \lambda x^A.\interp N\xi\\
\interp{x \fl i{N_i}}\xi&= & x \tuple{{\interp{N_i}\xi}}_i\\
\end{array}
\]
\end{figure}
Notice that the case of $\gfp$
uses the extended
environment $\xi\cup[X^{\rho}\mapsto N]$ that is admissible for
$E_i$ thanks to our assumption of well-boundness. (Moreover,
by renaming $X$, we may suppose that there is no $X^{\rho'}$ in
$\dom(\xi)$.) The meta-level fixed point over $N$ is well-formed
since every elimination alternative starts with a head/application
variable, and all occurrences of $N$ in the summands are thus
\emph{guarded} by constructors for elimination alternatives, and
therefore the fixed-point definition is \emph{productive} (in the
sense of producing more and more data of the fixed point through
iterated unfolding) and uniquely determines a forest, unlike an
expression of the form $\fix\, N.N$ that does not designate a
forest and would only come from the syntactically illegal term $\gfp
X^\rho.X^{\rho}$.
\end{definition}
We better not use the shorthand $\interp{\cdot}{\xi}$ with the placeholder for the expression from $\olfsfix$ to
be interpreted since the question of admissibility of
$\xi$ depends on the actual argument $T$.

The interpretation $\interp T\xi$ only depends on the values of $\xi$
for arguments $X^\rho$ for which there is a sequent $\rho'$ such
that $X^{\rho'}\in\FPV(T)$. In more precise words, the
interpretations $\interp T\xi$ and $\interp T{\xi'}$ coincide whenever
$\xi$ and $\xi'$
agree (already w.\,r.\,t.\ definedness) on all typed fixed-point variables
$X^\rho$ for which there is a sequent $\rho'$ such that
$X^{\rho'}\in\FPV(T)$.

If $T$ is closed, i.\,e., $\FPV(T)=\emptyset$, then the empty function is an admissible environment for $T$, and the environment index in the interpretation is left out, hence the interpretation is abbreviated to $\interpwe T$. Anyway, the interpretation of a closed $T$ does not depend on the environment.

If no $X^{\rho'}$ occurs free in $\sum_iE_i$ for any sequent
$\rho'$, we allow ourselves to abbreviate the finitary forest
$\gfp\,{X^{\rho}}.\sum_iE_i$ as $\sum_iE_i$. Thanks to our
observation above on the dependence of $\interp T\xi$ on $\xi$, we
have
$\interp{\s{i}{E_i}}\xi\bisimp\s{i}{\interp{E_i}\xi}$.

\subsection{Typing system for $\olfsfix$}\label{sec:finitary-typing}

We now have finitary means to represent solution spaces, in other words, the finitary forests of $\olfsfix$ can now serve as
witnesses for the existence of suitable forests (terms of $\coolfs$) that arise as their general-purpose semantics. According to the idea of the Curry-Howard correspondence,
these witnesses appear as the proof objects that are being typed in a typing system that is to be understood as a constructive deductive system for solution spaces. Due to the semantics into forests of $\coolfs$, this deductive system, when seen as a logic, is a ``logic of coinductive proofs''---or rather a ``logic of coinductive proof spaces'', due to the use of sums.

The main desiderata for the typing system we are about to introduce
are soundness and completeness in the following sense: 
Soundness means that if a proof object $T$ is typed in the typing system by a certain sequent, then the semantics of $T$ is a forest in $\coolfs$ whose members are proofs (more generally, solutions) of that sequent. Completeness means that \emph{every} sequent types a proof object $T$ such that its semantics is a forest in $\coolfs$ whose members are exactly \emph{all} proofs (more generally, solutions) of that sequent.

The deductive system we are introducing is in the form of a typing system for $\olfsfix$, and it is given through
inference rules for deriving sequents of the general form $\judge\Xi\Gamma T B$, shown in Figure~\ref{fig:typingolfsfix}.
Here are some explanations of the figure. The first
context $\Xi$ has the form  $\vect{X:\rho}$, so fixed-point variables are typed by atomic sequents. The sequents accumulated in $\Xi$ are the ``coinductive hypotheses'' of the typing derivation, when the latter is seen as a proof in the ``logic of coinductive proofs''. The first typing rule in Figure~\ref{fig:typingolfsfix} implies that fixed-point variables enjoy a relaxed form of binding.

The context $\Xi$ is such that no fixed-point variable name $X$ occurs twice (there is no
condition concerning duplication of sequents). So, $\Xi$ can be (and
will be) seen as a partial function, and $\Xi$, when regarded as a set
of typed fixed-point variables, has no duplicates. If $\Xi$ is empty,
then we write $\Gamma\vdash T:B$ instead of $\judge\Xi\Gamma T B$.

\begin{figure}\caption{Typing system for $\olfsfix$}\label{fig:typingolfsfix}
$$
\begin{array}{c}
\infer{\judge\Xi\Gamma {X^{\rho'}} p}{(X:\rho)\in\Xi\qquad\rho\leq\rho'=(\seq {\Theta'}p)\qquad\Theta'\subseteq\Gamma}\\
\\
\infer{\judge\Xi\Gamma{\gfp\,{X^\rho}.\sum_iE_i}p}{\mbox{for all $i$,}\,\,\,\judge{\Xi,X:\rho}\Gamma{E_i}p\qquad\rho=(\seq\Theta p)\qquad \Theta\subseteq\Gamma}\\
\\
\infer{\judge\Xi\Gamma{\lb x^A.N}{A\supset B}}{\judge{\Xi}{\Gamma,x:A}N B}\qquad
\infer{\judge\Xi\Gamma{x\tuple{N_i}_i}p}{(x:\vec B\impl
p)\in\Gamma\qquad\mbox{for all $i$,}\,\,\,\judge{\Xi}\Gamma{N_i}B_i}
\end{array}
$$
\end{figure}

\begin{lemma}[Weakening]
  If $\judge\Xi\Gamma TB$, $\Xi\subseteq\Xi'$ and $\Gamma\subseteq\Gamma'$ then $\judge{\Xi'}{\Gamma'} TB$.
\end{lemma}
\begin{proof}
  Obvious since, for the $\Xi$ argument, there is only look-up, and
  for the $\Gamma$ argument, weakening is directly built into the
  rules concerning fixed-point variables and goes through inductively for
  the others.
\end{proof}

\begin{lemma}
If $\judge\Xi\Gamma TB$ then the free term variables of $T$ are in $\dom(\Gamma)$.
\end{lemma}
Notice that the free term variables of $X^{\seq\Gamma p}$ are $\dom(\Gamma)$ and that $\dom(\Gamma)$ enters the free term variables of $\gfp X^{\seq\Gamma p}.\sum_iE_i$.
\begin{proof}
  Induction on $T$.
\end{proof}

\begin{lemma}\label{lem:typable-Xi-admissible}
  If $\judge\Xi\Gamma TB$ and $X^{\rho'}\in\FPV(T)$ then there is a sequent $\rho$ such that $(X:\rho)\in\Xi$ and $\rho\leq\rho'$.
\end{lemma}
\begin{proof}
  Induction on $T$.
\end{proof}

\begin{corollary}\label{cor:typable-Xi-admissible}
If  $\judge\Xi\Gamma TB$, and $\xi$ is a
partial function from typed fixed-point variables $X^\rho$ to
(co)terms of $\coolfs$ with domain $\Xi$, then $\xi$ is an environment, and it is admissible for $T$.
\end{corollary}

As a consequence of the last lemma, we obtain by induction on $T$:
\begin{lemma}[Typable terms are well-bound]\label{lem:typablewellbound}
 If $\judge\Xi\Gamma TB$ then $T$ is well-bound.
\end{lemma}
\begin{proof}
  Induction on $T$.
\end{proof}

\begin{definition}[Well-typed environment]
An environment $\xi$ is well-typed w.\,r.\,t.~context $\Gamma$ if for all $X^{\seq\Theta q}\in\dom(\xi)$,  $\Theta\subseteq\Gamma$ and $\Gamma\vdash\xi(X^{\seq\Theta q}):q$ (in $\coolfs$).
\end{definition}

\begin{lemma}[Soundness of the typing system for $\olfsfix$ w.r.t.~the in\-ter\-pret\-ation $\interp \cdot\cdot$]\label{lem:inter-pres-types}
Let  $\judge\Xi\Gamma TB$ in $\olfsfix$ and $\xi$ be a well-typed environment w.\,r.\,t.~$\Gamma$ with $\dom(\xi)=\Xi$. Then $\Gamma\vdash\interp T\xi: B$ in $\coolfs$.
In particular (for empty $\Xi$), if $\Gamma\vdash T:B$ in $\olfsfix$,
then $\Gamma\vdash\interpwe T:B$ in $\coolfs$.
\end{lemma}
A proof sketch is as follows: Induction on $T$, using Lemma~\ref{lem:deco-pres-types} in the base case of a fixed-point variable and using an embedded coinduction in the case of a greatest fixed point. To see how this works intuitively, we set $T:=\gfp\,{X^\rho}.\sum_iE_i$ and assume $\judge\Xi\Gamma Tp$, which comes in particular from $\judge{\Xi'}\Gamma{E_i}p$ for all $i$, with $\Xi':=\Xi,X:\rho$. By definition, $\interp T\xi=\sum_i\interp{E_i}{\xi'}$, with $\xi':=\xi\cup[X^\rho\mapsto\interp T\xi]$. Our goal is to show that $\Gamma\vdash\interp T\xi:p$. Since the typing relation is coinductive, we spawn a coinduction for this goal. It suffices to show for every $i$ that $\Gamma\vdash\interp{E_i}{\xi'}:p$. For this, we apply the induction hypothesis on $\judge{\Xi'}\Gamma{E_i}p$ and the environment $\xi'$ that is well-typed w.\,r.\,t.~$\Gamma$ and has the right domain. Well-typedness demands in particular that $\Gamma\vdash\xi'(X^\rho):p$, but this is $\Gamma\vdash\interp T\xi:p$ that we assumed coinductively. And this reasoning is not circular since the appeal to the coinductive hypothesis in the so constructed argument for the typing judgement is guarded through the application of typing rules. This application of guarded coinduction is still peculiar since the goal to be proved enters the provisos of the lemma. However, a rather straightforward argument depending on ``observation depth'' can be given, which we will develop now.

\begin{definition}[Typability in $\coolfs$ with observation depth]
  We define for $n\geq0$ the $n$-typabil\-ity relation
  $\Gamma\vdash_nT:A$ with the same data $\Gamma$, $T$ and $A$ as for
  typability in $\coolfs$, by recursion on $n$. The recursive
  definition is presented in the style of inductive derivation rules
  in Figure~\ref{fig:ntyping}.
\end{definition}
The first three rules take up the coinductive rules for typing in
$\coolfs$ but replace the coinductive reading by the descent from
index $n+1$ to $n$ when read backwards. The last rule expresses that
if we are interested in observations up to depth $0$ only, every
forest is accepted. Hence index $0$ does not stand for an observation
at the root but for no observation at all.
\begin{figure}[tb]\caption{Typing rules with observation depth of $\coolfs$}\label{fig:ntyping}
\[
\begin{array}{c}
\infer{\seqn{n+1}\Gamma{\lambda x^A.N}{A\impl B}}{\seqn n{\Gamma,x:A}
NB}\qquad
\infer{\seqn{n+1}{\Gamma}{x\tuple{N_i}_i}{p}}
  {(x:\vec B\impl p)\in\Gamma\quad\forall i,\,\seqn n\Gamma{N_i}{B_i}}
  \qquad\infer{\seqn{n+1}{\Gamma}{\sum_iE_i}p}{\forall i,\,\seqn n\Gamma{E_i}p}
  \qquad\infer{\seqn 0\Gamma TA}{}
\end{array}
\]
\end{figure}

\begin{lemma}[Antitonicity of $n$-typability]
\label{lem:ntyping-antitone}
Given a context $\Gamma$, a formula $A$ and a forest $T$. Then for all
$n\geq0$, if $\seqn{n+1}\Gamma TA$ then $\seqn n \Gamma TA$.
\end{lemma}
\begin{proof}
  Obvious induction on $n$.
\end{proof}
We refine Lemma~\ref{lem:deco-pres-types} to $n$-typability.
\begin{lemma}[Closedness under decontraction of $n$-typability]
\label{lem:ntyping-decontraction}
Let $T$ be an expression of $\coolfs$ and $\Gamma'\cup\Delta$ be a
context.  If $\seqn n{\Gamma\cup\Delta}TB$ and $\Gamma\leq\Gamma'$
then the following holds:
\begin{enumerate}
\item If $T$ is a term $N$, then
  $\seqn n{\Gamma'\cup\Delta}{[\Gamma'/\Gamma]N}B$;
\item If $T$ is an elimination alternative $E$, then
  $\seqn n{\Gamma'\cup\Delta}{E'}B$ holds for every summand $E'$ of
  $[\Gamma'/\Gamma]E$.
\end{enumerate}
\end{lemma}
\begin{proof}
  Simultaneous induction on $n$ (the formulation is general enough to get the induction through).
\end{proof}
We will use this lemma for $\Gamma'\subseteq\Delta$ and terms $N$,
hence in the following form: If $\seqn n{\Delta}NB$ and
$\Gamma\leq\Gamma'\subseteq\Delta$, then
$\seqn n{\Delta}{[\Gamma'/\Gamma]N}B$.

Since $n$-typability simply counts derivation depths of typability and
the rules only have finitely many premisses (even if their number is
unbounded), we regain typability if $n$-typability holds for all $n$.
\begin{lemma}[Inductive characterization of typability in $\coolfs$]\label{lem:ntypingok}
 Given a context $\Gamma$, a formula $A$ and a forest $T$. Then $\seqt\Gamma TA$ iff  $\seqn n\Gamma TA$ for all $n\geq0$.
\end{lemma}
\begin{proof}
  From left to right: induction on $n$. From right to left: the usual
  coinductive argument, exploiting antitonicity to put together
  indices obtained for the finitely many premisses.
\end{proof}
\begin{lemma}[Ramification of Lemma~\ref{lem:inter-pres-types}]
  Let $\judge\Xi\Gamma TB$ in $\olfsfix$ and $\xi$ be an environment
  with $\dom(\xi)=\Xi$ such that for all
  $X^{\seq\Theta q}\in\dom(\xi)$, $\Theta\subseteq\Gamma$. Then, the
  following implication holds for all $n\geq0$: If for all
  $X^{\seq\Theta q}\in\dom(\xi)$,
  $\seqn n\Gamma{\xi(X^{\seq\Theta q})}q$, then
  $\seqn n\Gamma{\interp T\xi}B$.
\end{lemma}
\begin{proof}
  Let us denote by $\bigass n\xi$ the assumption of the implication to prove,
  for all $n\geq0$. The proof is by induction on $T$
  (equivalently, on the derivation of $\judge\Xi\Gamma TB$).

  Case $T=X^{\rho'}$. Then $B=p$, $(X:\rho)\in\Xi$,
  $\rho=(\seq\Theta p)\leq(\seq{\Theta'}p)=\rho'$ with
  $\Theta'\subseteq\Gamma$. Recall that $\rho\leq\rho'$ is equivalent to $\Theta\leq\Theta'$.
  Let $n\geq0$ and assume $\bigass n\xi$, in
  particular $\seqn n\Gamma{\xi(X^{\rho})}p$. By
  Lemma~\ref{lem:ntyping-decontraction}, since $\xi(X^{\rho})$ is a
  term of $\coolfs$, we get
  $\seqn n\Gamma{[\Theta'/\Theta](\xi(X^{\rho}))}p$, but
  $[\Theta'/\Theta](\xi(X^{\rho}))=[\rho'/\rho](\xi(X^{\rho}))=\interp{T}\xi$.

  Case $T=\gfp\,{X^\rho}.\sum_iE_i$, with $\rho=(\seq{\theta}p)$. Then
  $B=p$, $\Theta\subseteq\Gamma$ and for all $i$,
  $\judge{\Xi'}\Gamma {E_i}p$, with $\Xi':=\Xi,X:\rho$. We prove the
  announced implication by induction on $n$. The case $n=0$ is trivial
  by the definition of $0$-typability. We thus assume as side
  induction hypothesis that $\bigass n\xi$ implies
  $\seqn n\Gamma{\interp T\xi}p$. We assume $\bigass {n+1}\xi$ and
  have to show $\seqn {n+1}\Gamma{\interp T\xi}p$. As in the proof
  sketch for Lemma~\ref{lem:inter-pres-types},
  $\interp T\xi=\sum_i\interp{E_i}{\xi'}$, with $\xi':=\xi\cup[X^\rho\mapsto\interp T\xi]$.
  Let $i$ be one of the indices. We have to show that $\seqn n\Gamma{\interp {E_i}{\xi'}}p$. We apply the main induction hypothesis
  for $E_i$, with the derivation $\judge{\Xi'}\Gamma {E_i}p$ and environment $\xi'$ which satisfies the global condition for the
  environment in the lemma since for $X^{\seq\Theta p}\in\dom(\xi')\setminus\dom(\xi)$, we also have $\Theta\subseteq\Gamma$.
  It thus remains to show $\bigass{n}{\xi'}$. We already assumed
  $\bigass {n+1}\xi$, hence by Lemma~\ref{lem:ntyping-antitone}, we also have $\bigass {n}\xi$, and the side induction hypothesis ensures
  $\seqn n\Gamma{\interp T\xi}p$, but the latter is $\seqn n\Gamma{\xi'(X^\rho)}p$. This was the missing verification to pass from $\bigass {n}\xi$
  to $\bigass{n}{\xi'}$.

  The other cases require no technical intricacies and use
  Lemma~\ref{lem:ntyping-antitone} and the admissible rule of context
  weakening for $n$-typability (in the case of lambda-abstraction).
\end{proof}
By virtue of both directions of Lemma~\ref{lem:ntypingok}, Lemma~\ref{lem:inter-pres-types} follows from its ramification.

\medskip
By composing the soundness properties of the general-purpose interpretation and the membership semantics of Section \ref{sec:coinductive} we obtain the following result, which says that, if $T$ is typable in the typing system for $\olfsfix$, then $\interpwe T$ only has ``correct'' members (finite or infinite).

\begin{theorem}[Soundness of the typing system for $\olfsfix$]\label{thm:soundness}
If $\Gamma\vdash T:A$ in $\olfsfix$, then:
\begin{enumerate}
\item For $N\in\cool$, if $\colr N{\interpwe T}$ then $\seqt\Gamma N A$ in $\cool$.
\item For $t\in\ol$, if $\colr t{\interpwe T}$ then $\seqt\Gamma t A$ in $\ol$.
\end{enumerate}
\end{theorem}
\begin{proof}
By Lemmas \ref{lem:typing-of-members} and \ref{lem:inter-pres-types}.
\end{proof}

\subsection{Finitary representation of full solution spaces}
\label{sec:finitary-representation}

Full solution spaces for $\ol$ can be shown to be finitary, with the help
of the \emph{finitary representation mapping} $\finrep{\sigma}{\Xi}$,
which we introduce now.

\begin{definition}[Finitary representation of full solution spaces]
\label{def:finrep} Let $\Xi:=\vect{X:\rho}$ be a vector
of $m\geq 0$ declarations $(X_i:\rho_i)$ with $\rho_i=\seq{\Theta_i}{q_i}$ where no
fixed-point variable name occurs twice. The definition of $\finrep{\seq\Gamma{\vec
A\impl p}}{\Xi}$ is as follows:

If, for some $1\leq i\leq m$, $p=q_i$, $\Theta_i\subseteq\Gamma$
and $|\Theta_i|=|\Gamma|\cup\{A_1,\ldots,A_n\}$,
then
$$
\finrep{\seq\Gamma{\vec A\impl p}}{\Xi}:=\lb z_1^{A_1}\cdots
z_n^{A_n}.X_i^{\rho}\enspace,$$
where $i$ is taken to be the biggest such index.\footnote{In the original definition \cite[Definition 22 of function $N$]{FICS2013}, the need for this disambiguation was neglected, with an insufficient extra condition that no sequent occurs twice among the $\rho_i$.}
Otherwise,
$$\finrep{\seq\Gamma{\vec A\impl p}}{\Xi}:=
\lb z_1^{A_1}\cdots
z_n^{A_n}.\gfp\,{Y^{\rho}}.{\s{(y:\vec {B}\impl p)\in\Delta}{y \fl
j{\finrep{\seq{\Delta}{B_j}}{\Xi,Y:\rho}}}}
$$
where, in both cases, $\Delta:=\Gamma,z_1:A_1,\ldots,z_n:A_n$ with a context $z_1:A_1,\ldots,z_n:A_n$ of ``fresh'' variables (not occurring in $\Gamma$ or any $\Theta_i$),
and $\rho:=\seq{\Delta}p$.
In the latter case, $Y$ is tacitly supposed not to occur in $\Xi$ (otherwise, the extended list of declarations would not be well-formed).
\end{definition}
Notice that, in the first case, the leading lambda-abstractions
bind variables in the type superscript $\rho$ of $X_i$, and that the
condition $\Theta_i\subseteq\Gamma$---and not
$\Theta_i\subseteq\Delta$---underlines that the fresh variables $z_1,\ldots,z_n$  cannot
be consulted although their types enter well into the next condition
$|\Theta_i|=|\Gamma|\cup\{A_1,\ldots,A_n\}$, which is equivalent to
$|\Theta_i|=|\Delta|$ (of which only $|\Theta_i|\supseteq|\Delta|$
needs to be checked). The first case represents the situation when the
full solution space is already captured by a purported solution $X_i$ for
the sequent $\seq{\Theta_i}p$ with the proper target atom, with all
hypotheses in $\Theta_i$ available in $\Gamma$ and, finally, no more
formulas available for proof search in the extended current context
$\Delta$ than in $\Theta_i$. Hence, the purported solution $X_i$ only
needs to be expanded by decontraction in order to cover the full solution
space for $\rho$ (as will be confirmed by
Theorem~\ref{thm:FullProp}). That $\finrepsymb$ indeed is a total function will be proven below in Lemma~\ref{lem:termination}.

In the sequel, we will omit the second argument $\Xi$ to $\finrepsymb$ in case $\Xi$ is the empty vector of declarations ($m=0$ in the definition).

Note that, whenever one of the sides of the following equation is defined according to the first or second case, then so is the other, and the equation holds (of course, it is important
to use variables $z_i$ that are fresh w.\,r.\,t.~$\Xi$):
$$\finrep{\seq\Gamma{\vec A\impl p}}{\Xi}\bisimp\lb z_1^{A_1}\cdots z_n^{A_n}.\finrep{\seq{\Gamma,z_1:A_1,\ldots,z_n:A_n}{p}}{\Xi}$$

\begin{example}[Examples~\ref{ex:dn-Peirce} and \ref{ex:dn-Peirce-cont1} continued]\label{ex:dn-Peirce-cont2}
  We calculate the finitary forest representing the full solution space for
  the twice negated Peirce formula $A:=\DNPeircet$, writing $A_0$ for $\Peircet$. The successive steps are seen in Figure~\ref{fig:calc-f-dnpeirce} where we continue with
  the omission of formulas in the left-hand sides of sequents. For
  brevity, we do not repeat the sequents associated with the fixed-point
  variables. The names of intermediary terms are chosen for easy
  comparison with Example~\ref{ex:dn-Peirce}.
\begin{figure}\caption{Steps towards calculating $\finrepempty{\seq{} \DNPeircet}$}\label{fig:calc-f-dnpeirce}
\[
\begin{array}{rcl}
\finrepempty{\seq{} A}&=&\lambda x^{A_0\impl q}. N'_1\\
N'_1&=&\gfp\, X_1^{\seq xq}.x\tuple{\finrep{\seq{x}{A_0}}{X_1}}\\
\finrep{\seq{x}{A_0}}{X_1}&=&\lambda y^{(p\impl q)\impl p}. N'_3\\
N'_3&=&\gfp\, X_2^{\seq {x,y}p}.y\tuple{\finrep{\seq{x,y}{p\impl q}}{X_1,X_2}}\\
\finrep{\seq{x,y}{p\impl q}}{X_1,X_2}&=&\lambda z^{p}. N'_5\\
N'_5&=&\gfp\, X_3^{\seq {x,y,z}q}. x\tuple{\finrep{\seq{x,y,z}{A_0}}{X_1,X_2,X_3}}\\
\finrep{\seq{x,y,z}{A_0}}{X_1,X_2,X_3}&=&\lambda y_1^{(p\impl q)\impl p}. N'_7\\
N'_7&=&\gfp\, X_4^{\seq {x,y,z,y_1}p}.\\
&&y\tuple{\finrep{\seq{x,y,z,y_1}{p\impl q}}{X_1,X_2,X_3,X_4}}+z+\\
&&y_1\tuple{\finrep{\seq{x,y,z,y_1}{p\impl q}}{X_1,X_2,X_3,X_4}}\\
\finrep{\seq{x,y,z,y_1}{p\impl q}}{X_1,X_2,X_3,X_4}&=&\lambda z_1^{p}. N'_9\\
N'_9&=&X_3^{\seq{x,y,z,y_1,z_1}{q}}\\
\end{array}
\]
\end{figure}
The fixed-point variables $X_1$, $X_2$ and $X_4$ thus have no occurrences in $\finrepempty{\seq{} A}$, and, as announced before, we will omit them in our resulting finitary forest
$$\finrepempty{\seq{} \DNPeircet}=\lambda x^{\Peircet\impl q}.x\tuple{\lambda y^{(p\impl q)\impl p}.y\tuple{\lambda z^{p}. N'_5}}$$\\[-4ex]
with
$$N'_5=\gfp\, X_3^{\seq {x,y,z}q}.x\tuple{\lambda y_1^{(p\impl q)\impl p}.y\tuple{\lambda z_1^p.X_3^{\seq{x,y,z,y_1,z_1}{q}}}+z+y_1\tuple{\lambda z_1^p.X_3^{\seq{x,y,z,y_1,z_1}{q}}}}\enspace,$$
still omitting the formulas in the left-hand sides of the sequents.
\end{example}
\begin{example}\label{ex:finitary} For the other examples, we have the following representations.
  \begin{itemize}
  \item $\finrepempty\Boolet=\lambda x^p.\lambda y^p.x+y$.
  \item $\finrepempty\Inftyt=\lambda f^{p\impl p}.\gfp\,{X^{\seq{ f:p\impl p}p}}. {f\tuple {X^{\seq{ f:p\impl p}p}}}$.
 \item $\finrepempty\Churcht=\lambda f^{p\impl p}.\lambda x^{p}.\gfp\,{X^{\rho}}. {f\tuple {X^{\rho}}+x}$ with
$\rho:=\seq{ f:p\impl p,x:p}p$.
  \item $\finrepempty\Peircet=\lambda x^{(p\impl q)\impl p}. x\tuple{\lambda y^p.\oo}$ (using $\oo$ for the empty sum under the omitted $\gfpsymb$).
  \item $\finrepempty\Threet=\lambda x^{(p\impl p)\impl p}.x\tuple{\lambda y^p.\gfp Y^{\rho_1}.x\tuple{\lambda z^p.Y^{\rho_2}}+y}$ with $\rho_1:=\seq{ x:(p\impl p)\impl p,y:p}p$, $\rho_2:=\seq{ x:(p\impl p)\impl p,y:p,z:p}p$, hence $\rho_1\leq\rho_2$.
  \end{itemize}
Notice that for $\Inftyt$, $\Churcht$ and $\Threet$, the presentation of the full solution spaces had already been brought close to this format thanks to cycle analysis that guided the unfolding process, and Theorem~\ref{thm:FullProp} below ensures that this works for any sequent.
\end{example}

\medskip
Strictly speaking, Definition~\ref{def:finrep} is not justified since
the recursive calls do not follow an obvious pattern that guarantees termination. The following
lemma spells out the measure that is recursively decreasing in the
definition of $\finrepsymb$.

To this end, we introduce some definitions. Given a finite set $\cal A$ of formulas
$${\cal
A}^{sub}:=\{B\mid\textrm{there exists $A\in{\cal A}$ such that $B$
is subformula of $A$}\}\enspace.$$
We say $\cal A$ is \emph{subformula-closed} if ${\cal A}^{sub}={\cal A}$.
A \emph{stripped sequent} is a pair $({\cal B},A)$, where $\cal
B$ is a finite set of formulas. A \emph{stripped restricted sequent} additionally has that $A$ is an atom.
If $\sigma=\seq{\Gamma}A$, then
its \emph{stripping} $|\sigma|$ denotes the stripped sequent $(|\Gamma|,A)$. We say
$({\cal B},A)$ \emph{is over} $\cal A$ if ${\cal B}\cup \{A\}\subseteq{\cal A}$.
There are $size({\cal A}):=a\cdot2^k$ stripped restricted
sequents over $\cal A$, if $a$ (resp.~$k$) is the number of atoms
(resp.~formulas) in $\cal A$.

\begin{lemma}[Termination of $\finrepsymb$]\label{lem:termination}
For all sequents $\sigma$ and vectors $\Xi$ as in Definition~\ref{def:finrep}, the finitary forest $\finrep{\sigma}{\Xi}$ is well-defined.
\end{lemma}
\begin{proof}
As in the definition, we consider a sequent $\sigma$ of the form $\seq\Gamma C$ with $C=\vec A\impl p$.
Let us call \emph{recursive call} a ``reduction''
\begin{equation}\label{eq:recursive-call}
\finrep{\seq\Gamma{\vec A\impl p}}{\vect{X:\seq{\Theta}q}}\leadsto
\finrep{\seq{\Delta}{B_j}}{\vect {X:\seq{\Theta}q},Y:\rho}
\end{equation}
\noindent where the if-guard in Definition~\ref{def:finrep} fails; $\Delta$ and
$\rho$ are defined as in the same definition; and, for some $y$,
$(y:\vec {B}\impl p)\in\Delta$. We want to prove that every sequence
of recursive calls from $\finrep{\seq\Gamma C}\Xi$ is finite.

Observe that the context of the first argument to $\finrepsymb$ is
monotonically increasing during any sequence of recursive calls: in the
reduction, one passes from $\Gamma$ to its extension $\Delta$ by fresh
variables. Since only fresh variables are added to $\Gamma$, this
means that whenever for some $i$, $\Theta_i$ is not a subset of
$\Gamma$ in the original call to $\finrepsymb$, this will hold of all
the further contexts occurring in the recursive calls. In other words,
the if-guard in Definition~\ref{def:finrep} fails forever, hence $X_i$ will
not enter the result of the computation. Therefore, without loss of
generality, we may assume that for all $i$, $\Theta_i\subseteq\Gamma$.
Trivially, this condition is then inherited to the recursive calls: for $1\leq i\leq m$,
$\Theta_i\subseteq\Gamma\subseteq\Delta$, and $\Theta_{m+1}=\Delta$ which is the context
in the new first argument of $\finrepsymb$.

In the original definition \cite[Definition 22 of function
$N$]{FICS2013}, it was required that no sequent occurs twice among the
$(\seq{\Theta_i}{q_i})=\rho_i$. Of course, if $\rho_j=\rho_i$ with $j<i$, then the
first case of the definition of $\finrepsymb$ will not take into
account $X_j:\rho_j$ (since the biggest $i$ with the required
properties is chosen), hence it will never be taken into
account. Therefore, without loss of generality, we may assume that all
$\rho_i$ are different.

But we can do better: Since we may already assume that for all $i$,
$\Theta_i\subseteq\Gamma$, we infer from $|\rho_j|=|\rho_i|$ with
$j<i$ that the first case of the definition of $\finrepsymb$ will not
take into account $X_j:\rho_j$ (for the same reason as
before). Therefore, without loss of generality, we may assume that all
the stripped (restricted) sequents $|\rho_i|$ are different, in
other words, $\size(\Xi)=m$, where $m\geq 0$ is the length of vector
$\Xi$ and $\size(\Xi)$ is the number of elements of $|\Xi|$ and
$|\Xi|:=\{|\rho|:\rho\in\Xi\}$. Also this condition is inherited to the recursive calls:
Since $\Theta_i\subseteq\Gamma$ for all $i$, if $|\rho|=|\Theta_i|$ for some $i\leq m$, the first
clause of Definition~\ref{def:finrep} would have applied, but we assumed to be in the recursive case.
As a consequence of this extra assumption, the first clause of the definition will never be with
two possible indices $i$ out of which the biggest would have to be chosen.

Let ${\cal A}:=(|\Gamma|\cup\{C,q_1,\ldots,q_m\})^{sub}$. By our
assumptions, the strippings of $\sigma$ and all $\rho_i$ are over ${\cal A}$.
In particular, $m\leq\size({\cal A})$.

We will now show that for subformula closed ${\cal A}$, if the strippings of $\sigma$
and all $\rho_i$ are over ${\cal A}$, then this also holds for the
arguments $\finrepsymb$ is called with in the recursive call:
$|\Delta|=|\Gamma|\cup\{A_1,\ldots,A_n\}\subseteq{\cal A}$
since $\vec A\impl p\in\cal A$ and $\cal A$ is subformula-closed. For the same reason $p\in\cal A$.
$B_j$ is a subformula of $\vec
B\impl p$ and $\vec B\impl p\in|\Delta|$ because $(y:\vec B\impl
p)\in\Delta$, for some $y$.

Since in subsequent recursive calls, the strippings of the arguments
are all over ${\cal A}$, we continue to have $m'\leq\size({\cal A})$
for all subsequent lengths $m'$ of the second argument of
$\finrepsymb$. Of course, this is a fixed bound on the recursion depth which is therefore finite.
Put differently, termination is guaranteed since the measure $\size({\cal A})-m\geq0$ strictly
decreases.\end{proof}

We have justified the definition of $\finrepempty\sigma$ for
all sequents $\sigma$.

Notice that yet more detailed invariants could be established above
(under the same restrictions we were allowed to ask for without loss
of generality): $\Theta_1\subseteq\ldots\subseteq\Theta_m$ would also
be preserved under reduction, as well as that the last $\Theta_m$ is
$\Gamma$, unless $m=0$. Yet another invariant is that all $q_i$ are in
$|\Gamma|^{sub}$. All of them can be trivially initiated with empty
$\Xi$ and thus are observed in $\finrepempty\sigma$.

Also notice that, while the growing size of $\Xi$ is our argument for
termination, an implementation for calculating $\finrepempty\sigma$
would rather not store all of $\Xi$ in its recursive calls: as soon as a reduction occurs
where $|\Delta|$ is a strict superset of $|\Gamma|$, it is clear that
the if-case of Definition~\ref{def:finrep} can never apply for some element
in $\Xi$ in the recursive calculation of
$\finrep{\seq{\Delta}{B_j}}{\Xi,Y:\rho}$, so (the old) $\Xi$ does not need to be
stored in those further recursive calls.

An important objective of the typing system in Section~\ref{sec:finitary-typing} is attained by the following result:
\begin{lemma}[Finitary representation is well-typed]\label{lem:finrep-typed}
 $$\judge\Xi\Gamma{\finrep{\seq\Gamma C}\Xi}C\enspace.$$
In particular, $\Gamma\vdash\finrepempty{\seq\Gamma C}:C$.
\end{lemma}
\begin{proof}
  By structural recursion on the obtained finitary forest $\finrep{\seq\Gamma C}\Xi$. Notice that the context weakening built into the $\gfpsymb$ rule in Figure~\ref{fig:typingolfsfix} is not needed for this result (i.\,e., $\Theta$ and $\Gamma$ of that rule can always agree).
\end{proof}
\begin{corollary}[Finitary representation is well-bound]\label{cor:finrep-wellbound}
  $\finrep{\sigma}{\Xi}$ is well-bound, and $\finrepempty\sigma$ is moreover closed.
\end{corollary}
\begin{proof}
  Use Lemma~\ref{lem:typablewellbound} for the first part. Notice that
  this is needed to argue that free fixed-point variables of
  $\finrep{\sigma}{\Xi}$ have necessarily names that occur in
  $\Xi$. But we can just apply Lemma~\ref{lem:typable-Xi-admissible}
  for empty $\Xi$ to obtain the second part.
\end{proof}

\subsection{Equivalence of representations and completeness of the typing system for $\olfsfix$}\label{sec:equivalence}

Now, we establish 
the result on the equivalence of the coinductive and
inductive representations of the full solution spaces.
For this, we need that forests are identified not only up to bisimilarity,
because of the rather rough way decontraction operates that
takes identification up to symmetry and idempotence of the sum
operation for the elimination alternatives for granted.
The proof below is a revision of the proof of \cite[Theorem 24]{FICS2013} in the light of the new notion of environments and their admissibility w.\,r.\,t.~a term, but with the help from the typing system for finitary forests.
\begin{theorem}[Equivalence]
\label{thm:FullProp}
For any sequent $\sigma$,
$\interpwe{\finrepempty{\sigma}}\bisim\solfunction\sigma$.
\end{theorem}
\begin{proof}

For a vector $\Xi=\vect{X:\rho}$ satisfying the requirements in Definition~\ref{def:finrep},
the mapping $\xi_\Xi$ obtained by setting $\xi_\Xi(X_i^{\rho_i}):=\solfunction{\rho_i}$ is an environment.
By Corollary~\ref{cor:finrep-wellbound}, $\finrep{\sigma}{\Xi}$ is well-bound. Moreover, using Corollary~\ref{cor:typable-Xi-admissible}, we have that $\xi_\Xi$ is admissible for $\finrep{\sigma}{\Xi}$. Therefore, $\interp{\finrep{\sigma}{\Xi}}{\xi_\Xi}$ is well-defined. We will show that $\interp{\finrep{\sigma}{\Xi}}{\xi_\Xi}\bisim\solfunction\sigma$ -- whose right-hand side is independent from $\Xi$, and thus also its left-hand side.

The theorem follows by taking for $\Xi$ the empty vector, since by
convention the empty environment is omitted from the notation for the
general-purpose interpretation. (Anyway, by
Corollary~\ref{cor:finrep-wellbound}, $\finrepempty\sigma$ is closed and
thus its general-purpose interpretation does not depend on an
environment.)

The proof is by structural induction on the term  $\finrep{\sigma}{\Xi}$.
Let $\sigma=\seq\Gamma{\vec A\impl p}$ and $\Delta:=\Gamma,z_1:A_1,\ldots,z_n:A_n$, as in Definition~\ref{def:finrep}.
We will again assume that $\rho_i$ is given as $\seq{\Theta_i}{q_i}$.

Case $p=q_i$ and $\Theta_i\subseteq\Gamma$ and
$|\Theta_i|=|\Delta|$, for some $1\leq i\leq m$,
which implies $\rho_i\leq(\seq{\Delta}{p})$ (*).
The proof of this case is completed in Figure~\ref{fig:proofcaseinequivalencethm}.
\begin{figure}\caption{Part (iii) of first main case in proof of Theorem~\ref{thm:FullProp}}\label{fig:proofcaseinequivalencethm}
\[
\begin{array}{rcll}
\lhs&=&\lb z_1^{A_1}\cdots
z_n^{A_n}.\interp{X_i^{\seq{\Delta}{p}}}{\xi_\Xi}&\textrm{(by definition)}\\
&=&\lb z_1^{A_1}\cdots
z_n^{A_n}.[(\seq{\Delta}{p})/\rho_i]\xi_\Xi(X_i^{\rho_i})&\textrm{(by definition and (*) above)}\\
&=&\lb z_1^{A_1}\cdots
z_n^{A_n}.[(\seq{\Delta}{p})/\rho_i]\solfunction{\rho_i}&\textrm{(by definition of $\xi_\Xi$)}\\
&\bisim&\lb z_1^{A_1}\cdots
z_n^{A_n}.\sol{\Delta}{p}&\textrm{(by Lemma \ref{lem:cleavage-2} and (*))}\\
&=&\rhs&\textrm{(by definition)}
\end{array}
\]
\end{figure}

The inductive case is essentially an
extension of the inductive case in \cite[Theorem 15]{FICS2013} for the Horn fragment. In this other case, we calculate as follows.

$\lhs=\lb z_1^{A_1}\cdots
z_n^{A_n}.N^\infty$, where $N^\infty$ is the unique solution of the following equation
\begin{eqnarray}
\label{eq:N-infty-full-case} N^\infty&\bisimp&\s {(y:\vect {B}\impl
p)\in\Delta}{y\fl j {\interp
{\finrep{\seq{\Delta}{B_j}}{\Xi,Y:\rho}}{\xi_\Xi\cup[Y^{\rho}\mapsto
N^\infty]}}}
\end{eqnarray}
\noindent where $\rho:=\seq\Delta p$.
Now observe that, by
inductive hypothesis (applied to the subexpressions $\finrep{\seq{\Delta}{B_j}}{\Xi,Y:\rho}$ of $\finrep{\sigma}{\Xi}$), the following equations (\ref{eq:sol-full-case}) and
(\ref{eq:sol-two-full-case}) are equivalent.
\begin{eqnarray}
\label{eq:sol-full-case}
\solfunction{\rho}&\bisim& \s {(y:\vect {B}\impl p)\in\Delta}{y\fl j {\interp {\finrep{\seq{\Delta}{B_j}}{\Xi,Y:\rho}}{\xi_{(\Xi,Y:{\rho})}}}}\\
\label{eq:sol-two-full-case}
\solfunction{\rho}&\bisim& \s {(y:\vect {B}\impl
p)\in\Delta}{y\fl j {\sol\Delta {B_j}}}
\end{eqnarray}
By definition of
$\solfunction{\rho}$, (\ref{eq:sol-two-full-case})
holds;
since $\xi_{(\Xi,Y:{\rho})}=\xi_\Xi\cup[Y^{\rho}\mapsto
\solfunction{\rho}]$ and because of (\ref{eq:sol-full-case}),
$\solfunction{\rho}$ is the solution $N^\infty$ of
(\ref{eq:N-infty-full-case}).
Therefore $\lhs\bisim\lb z_1^{A_1}\cdots
z_n^{A_n}.
\solfunction{\rho}$, and the latter is $\rhs$ by definition of
$\sol\Gamma{\vec A\impl p}$.
\end{proof}

\begin{corollary}\label{cor:finitary-regular}
  $\finrep{\sigma}{\Xi}$ is regular.
\end{corollary}
\begin{proof}
  By Lemma~\ref{lem:regular-versus-admissible}, $\finrep{\sigma}{\Xi}$ is regular since $\xi_\Xi$ in the proof above is
  admissible for it.
\end{proof}
See~\ref{sec:app-regular} for an even stronger result than regularity.

\begin{corollary}\label{cor:equivalence}
  For every $M\in\cool$, $\colr M{\interpwe{\finrepempty{\sigma}}}$ iff $\colr M{\solfunction\sigma}$.
\end{corollary}
\begin{proof}
  Immediate consequence of Theorem~\ref{thm:FullProp}. Obviously, membership is not affected by bisimilarity (modulo $\alpha$-equivalence and our identifications for the sum operation).
\end{proof}

The equivalence theorem is the last building block for the announced completeness result on the typing system of $\olfsfix$, which says that, in that typing system, any sequent has an inhabitant $T$ such that the members of $\interpwe T$ are exactly the ``correct'' ones (finite or infinite).

\begin{theorem}[Completeness of the typing system of $\olfsfix$] \label{thm:finitarycomplete}
For every logical sequent $\sigma=\seq\Gamma A$, there is a closed finitary forest $T$ such that $\Gamma\vdash T:A$ and:
\begin{enumerate}
\item For $N\in\cool$, $\colr N{\interpwe T}$ iff $\seqt\Gamma N A$ in $\cool$.
\item For $t\in\ol$, $\colr t{\interpwe T}$ iff $\seqt\Gamma t A$ in $\ol$.
\end{enumerate}
\end{theorem}
\begin{proof}
Take $T:=\finrepempty\sigma$. Its recursive definition does terminate (Lemma~\ref{lem:termination}), $T$ receives the right type (Lemma~\ref{lem:finrep-typed}), $T$ is closed (Corollary~\ref{cor:finrep-wellbound}), $\interpwe T$ and $\solfunction\sigma$ have the same members (Corollary \ref{cor:equivalence})
and these are the ``correct'' ones (Proposition~\ref{prop:adequacy-general-case}).
\end{proof}
Such
completeness cannot be expected at the level of individual
solutions. Take, for instance, $\Gamma=x_0:p\impl
p,\ldots,x_9:p\impl p$. Then $\sol\Gamma p$ is the forest
$N$ such that $N\bisimp x_0<N>+\cdots+x_9<N>$, one of whose members is,
say, the decimal expansion of $\pi$.

Although full solution spaces may have irrational members, they have
``rationality'' as a collection, since essentially---not taking
into account contraction phenomena---they are generated by
repeating infinitely a choice from a fixed menu. It is this
``rationality'' that can be expressed by finitary forests.

\subsection{Special-purpose semantics}

With the equivalence theorem above, the general-purpose
semantics in form of interpretation $\interp{T}{\xi}$ for finitary forests $T$ and suitable environments $\xi$ has
demonstrated its usefulness. However, when it comes to verifying properties of logical sequents through our approach, the full solution spaces given by $\solfunction\sigma$ play an important role.

In fact, when inspecting the proof of Theorem~\ref{thm:FullProp} we
observe that the considered environments all have \emph{form
  $\xi_\Xi$}, always mapping fixed-point variables $X^\rho$ to full
solution spaces $\solfunction{\rho}$---this even true for the extended
environment $\xi_\Xi\cup[Y^{\rho}\mapsto N^\infty]$, as comes out of
the proof that $N^\infty=\solfunction{\rho}$ by using
equations~(\ref{eq:N-infty-full-case}), (\ref{eq:sol-full-case}) and
the definition of $\solletter$.

This motivates the more radical
step of not only mapping all free fixed-point variables to ``their''
full solution space, but any occurrence, free or bound. This gives rise to
the special-purpose semantics that was mentioned in
Section~\ref{sec:untypedfinitary}. To recall from above, we introduced it under
the name ``simplified semantics'' in~\cite[Definition~15]{EspiritoSantoMatthesPintoInhabitation}.

\begin{definition}[Special-purpose interpretation of finitary forests as forests]
For an expression $T$ of $\olfsfix$, the special-purpose interpretation $\interps T$ is an expression of $\coolfs$ given
by structural recursion on $T$:
$$
\begin{array}{rclcrcl}
\interps{X^{\rho}}& = &\solfunction{\rho}&\qquad&\interps{\lambda x^A.N}& = & \lambda x^A.\interps N\\
\interps{\gfp\,{X^{\rho}}.\s{i}{E_i}}&= & \s i{\interps {E_i}}&\qquad&\interps{x \fl i{N_i}}&= & x \tuple{{\interps{N_i}}}_i\\
\end{array}
$$
\end{definition}
Note that the base case profits from the sequent annotation at
fixed-point variables, and the interpretation of the $\gfp$-constructor
has nothing to do with a greatest fixed point. Of course,
this may be ``wrong'' according to our understanding of a (greatest)
fixed point. So, we have to single out those expressions in $\olfsfix$ for which this interpretation
serves its special purpose.
\begin{definition}[Proper expressions]\label{def:proper}
  An expression $T\in\olfsfix$ is \emph{proper} if for any of its subterms
  $T'$ of the form $\gfp\,{X^{\rho}}.\s{i}{E_i}$, it holds that
  $\interps{T'}=\solfunction{\rho}$.
\end{definition}

For proper expressions, the special-purpose semantics agrees with the
general-purpose semantics we studied before -- for the special case of
environments we used in the proof of the equivalence theorem. Of
course, this can only make sense for expressions which have that
previous semantics, in other words for well-bound and regular
expressions.

\begin{lemma}[Lemma~22 in \cite{EspiritoSantoMatthesPintoInhabitation}]\label{lem:simplequal}
  Let $T$ be well-bound and $\xi$ be an admissible environment for $T$
  such that for all $X^\rho\in\dom(\xi)$:
  $\xi(X^\rho)=\solfunction\rho$. If $T$ is proper, then
  $\interp T\xi=\interps T$.
\end{lemma}
\begin{corollary}\label{cor:simplequal}
  For well-bound, closed and proper $T$, $\interpwe T=\interps T$.
\end{corollary}
The corollary is sufficient for our purposes since
$\finrepempty\rho$ is not only well-bound and closed, but also
proper, which is the more difficult part of the following result.

\begin{lemma}[Equivalence for special-purpose semantics -- Thm.~19 in~\cite{EspiritoSantoMatthesPintoInhabitation}]\label{lem:FullProp-simpl}
  Let $\sigma$ be a sequent and $\Xi$ as in Definition~\ref{def:finrep}.
  \begin{enumerate}
  \item $\finrep\sigma\Xi$ is proper.\label{lem:FullProp-simpl.1}
  \item $\interps{\finrep\sigma\Xi}=\solfunction\sigma$.\label{lem:FullProp-simpl.2}
  \end{enumerate}
\end{lemma}
In particular, $\interps{\finrep\sigma\Xi}$ is independent of $\Xi$,
and this conforms with the initial motivation for the special-purpose
semantics, as described above, that leaves no room for different interpretations
of ``purported solutions'' $X_i$ (cf.~our discussion right after Definition~\ref{def:finrep}).


\section{Application: analysis of proof search}\label{sec:analysis}

Given a (proof-)search problem, determined by a given logical sequent, one
is usually interested in its \emph{resolution}\footnote{This word here is of course not to be taken in its well-known, technical sense.} (the finding of the
solution), what is searched for is a finite solution (a proof), and the
unique \emph{analysis} done of the problem is the one that results
from the success or failure of the search---the given sequent is or
is not provable. In addition, since one wants a finite solution, a
layer of algorithmic \emph{control} (failure and loop detection,
followed by backtracking \cite{Howe97}) has to be added to the
purely logical structure of the search. Finally, this mix of
bottom-up proof-search and control is a generic recipe for
\emph{decision procedures} for the logic at hand.\footnote{See for
instance the textbook proof of decidability of propositional
intuitionistic logic in \cite{SorensenUrzyczyn2006}. Already
Gentzen's proof of decidability for the same logic \cite{Gentzen69}
is based on algorithmic control of proof search; however, in his
case, \emph{deductive} proof search is employed.}

How does this picture change, given the representations of proof search developed before? First, we may separate all the above concerns relating to
proof-search problems: we may postpone control considerations, by giving prominence to solutions rather than proofs; and we may separate analysis
from resolution: resolution is just one possible analysis one can make of the representation of the whole collection of solutions that we have at our disposal. Second, we obtain decision procedures just by doing analysis of representations of solution spaces, that is, without
``running'' the search again: the search is run only once, to generate the finitary representation of the (full) solution space. Third, the decision algorithms are syntax-directed, recursive procedures, driven by the syntax of the finitary calculus, avoiding the mentioned mix of bottom-up proof search and \emph{ad hoc} algorithmic control.

In this section we give an indication of how the approach to proof search described and justified in the previous sections, and with the characteristics identified above, can be applied, in the context of implicational logic and the simply-typed lambda-calculus, to give new answers to well-known problems about proof search, like decision and counting problems (Section~\ref{subsec:new-solutions}), to pose and solve new problems (Section~\ref{subsec:new-questions}), and to generalize known theorems (Section~\ref{subsec:new-results}). The material in Section~\ref{subsec:new-results} is new, while the material sketched in the other two was detailed elsewhere~\cite{EspiritoSantoMatthesPintoInhabitation,JESRMLP-FI2019}.

\subsection{New solutions for old problems}\label{subsec:new-solutions}

Our finitary representation of full solution spaces $\finrepempty{\sigma}$ allows new syntax-directed solutions for inhabitation and counting problems in simply-typed $\lambda$-calculus, as shown in detail in \cite{EspiritoSantoMatthesPintoInhabitation}. Here we briefly illustrate these new solutions.

Given a sequent $\sigma=(\seq\Gamma A)$, let $\inhabs\sigma$ denote the set of ($\eta$-long $\beta$-normal) \emph{inhabitants} of $A$ relative to context $\Gamma$ in $\ol$, i.\,e., $\inhabs\sigma:=\{t\in\ol\mid \seqt \Gamma tA\;\textrm{in}\;\ol\}$. For $T\in\coolfs$, let $\finext T$ denote the \emph{finite extension} of $T$, i.\,e., $\finext T=\{t\in\ol\mid\mem tT\}$. Observe that, due to Proposition~\ref{prop:adequacy-general-case}.2 and Theorem~\ref{thm:FullProp},
$$\inhabs\sigma = \finext{\solfunction{\sigma}}=\finext{\interpwe{\finrepempty{\sigma}}}.
$$

The \emph{inhabitation problem} in simply-typed $\lambda$-calculus can be formulated as the problem ``given sequent $\sigma$, is the set $\inhabs\sigma$ nonempty?'' (as is well-known, the answer to this question does not depend on whether all $\lambda$-terms are considered or only the $\beta$-normal ones or even the $\eta$-long $\beta$-normal terms). Our solution to this problem starts by defining two predicates $\exfinsymb$ and $\allinfsymb$ on expressions in $\coolfs$ (Figure~5 of~\cite{EspiritoSantoMatthesPintoInhabitation}), which are complementary ($\exfin T$ iff $\allinf T$ does not hold~\cite[Lemma 20]{EspiritoSantoMatthesPintoInhabitation}), and capture emptiness of the set of inhabitants ($\allinf T$ iff
$\finext{T}$ is empty~\cite[Lemma 21]{EspiritoSantoMatthesPintoInhabitation}). Next, we define companion predicates $\nbinfpsymb P$
and $\binfpsymb P$ on expressions in $\olfsfix$ that are parameterized by a predicate $P$ on sequents satisfying the proviso: $P\subseteq\exfinsymb\circ\solletter$ and $P$ decidable. The syntax-directed  definitions of the two predicates are recalled in Figure~\ref{fig:nbinf}.  Again these predicates are complementary ($\nbinfp P{T}$ iff $\binfp P{T}$ does not hold \cite[Lemma 22]{EspiritoSantoMatthesPintoInhabitation}), and the syntax-directedness of their definitions allows to immediately conclude that they are decidable. Then, the following holds:

\begin{figure}[tb]\caption{$\nbinfpsymb P$
		and $\binfpsymb P$ predicates, for $P$ satisfying the \emph{proviso}: $P\subseteq\exfinsymb\circ\solletter$ and $P$ decidable.}\label{fig:nbinf}
	$$
	\begin{array}{c}
	\infer[]{\nbinfp P{X^\sigma}}{P(\sigma)}\quad\quad
	\infer[]{\nbinfp P{\lambda x^A.N}}{\nbinfp P N}\quad\quad
	\infer[\forsomej]{\nbinfp P{\gfp X^\sigma.\sum_iE_i}}{\nbinfp P{E_j}}\quad\quad
	\infer[]{\nbinfp P{x\tuple{N_i}_i}}{\forall i,\,\nbinfp P{N_i}}\\[2ex]
	\infer[]{\binfp P{X^\sigma}}{\neg P(\sigma)}\quad\quad
	\infer[]{\binfp P{\lambda x^A.N}}{\binfp P N}\quad\quad
	\infer[]{\binfp P{\gfp X^\sigma.\sum_iE_i}}{\forall i,\,\binfp P{E_i}}\quad\quad
	\infer[\forsomej]{\binfp P{x\tuple{N_i}_i}}{\binfp P{N_j}}
	\end{array}
	$$
\end{figure}

\begin{lemma}[Deciding the existence of inhabitants in $\ol$ -- Theorem~24 of \cite{EspiritoSantoMatthesPintoInhabitation}]
	\label{lem:decide-fin-inhab}\quad
	\begin{enumerate}
		\item \label{lem:decide-fin-inhab.1}
		For any $T\in\olfsfix$  well-bound, proper and closed,  $\nbinfp P T$ iff  $\exfin{\interps T}$.
		\item $\nbinfp \emptyset{\finrepempty\sigma}$ iff $\exfin{\solfunction\sigma}$ iff $\inhabs\sigma$ is non-empty.\label{lem:decide-fin-inhab.2}
		\item The problem, ``given $\sigma$, is $\inhabs\sigma$ non-empty'' is decided by deciding $\nbinfp \emptyset{\finrepempty\sigma}$.
		\label{lem:decide-fin-inhab.3}
	\end{enumerate}
\end{lemma}
Summing up, the inhabitation problem of simply-typed lambda-calculus can be decided by first computing ${\finrepempty\sigma}$, and then traversing its structure to decide $\nbinfp \emptyset{\finrepempty\sigma}$.
The result allows definitions of \emph{sharper} versions of the predicates $\nbinfsymb$ and $\binfsymb$ that are still decidable: $\nbinfcansymb:=\nbinfpsymb {P_*^\nbinfsymb}$ and
$\binfcansymb:=\binfpsymb{P_*^\nbinfsymb}$ for $P_*^\nbinfsymb:=\nbinfpsymb\emptyset\circ\finrepsymb$ (which meets the proviso of Figure~\ref{fig:nbinf} thanks to Lemma~\ref{lem:decide-fin-inhab}.\ref{lem:decide-fin-inhab.2} and decidability of $\nbinfp \emptyset{\finrepempty\sigma}$).
The main result on these predicates is Lemma~27 of~\cite{EspiritoSantoMatthesPintoInhabitation} that, without any condition on $T$, we have $\nbinfcan T$ iff  $\exfin{\interps T}$.

An easy consequence (that also uses Lemma~\ref{lem:decide-fin-inhab}.\ref{lem:decide-fin-inhab.2}) for $\nbinfcansymb$ we need in this paper, is that, if $N=\gfp X^\rho.\sum_iE_i$ with $N$ proper (but not necessarily closed), then $\nbinfcan N$ is equivalent to $\nbinfp{\emptyset}{\finrepempty{\rho}}$. 

A second consequence needed for this paper makes use of Lemma~\ref{lem:FullProp-simpl}.\ref{lem:FullProp-simpl.2} (more precisely, the remark immediately after the lemma): for all sequents $\sigma$ and declarations $\Xi$ and $\Xi'$, $\nbinfcan{\finrep\sigma\Xi}$ iff $\nbinfcan{\finrep\sigma{\Xi'}}$, which can in particular be used for empty $\Xi'$.

Following the same steps, but making use of the already obtained decidable predicates $\nbinfcansymb$ and $\binfcansymb$, a syntax-directed solution can be construed also for the not so well-known problem ``given a sequent $\sigma$, is $\inhabs\sigma$ finite'' (studied for example in \cite{benyellesthesis,HindleyBasicSimple}).
So, we define complementary predicates $\inffinsymb$ and $\finfinsymb$ on expressions in $\coolfs$ such that $\finfin T$ iff $\finext T$ \cite[Figure 7, Lemmas 28 and 29]{EspiritoSantoMatthesPintoInhabitation}. Then we define the companion, complementary predicates $\FFpsymb P$ and $\NFFpsymb P$ on expressions in $\olfsfix$, parameterized by a predicate $P$ on sequents satisfying the proviso: $P\subseteq\finfinsymb\circ\solletter$ and $P$ decidable. Again, to appreciate the syntax-directedness of these definitions we recall them in Figure~\ref{fig:FF}. 

\begin{figure}[tb]
	\caption{$\FFpsymb P$ and $\NFFpsymb P$ predicates, for $P$ satisfying the \emph{proviso}: $P\subseteq\finfinsymb\circ\solletter$ and $P$ decidable.}
	\label{fig:FF}
	$$
	\begin{array}{c}
	\infer[]{\FFp P{ X^\sigma}}{P(\sigma)}\quad\quad
	\infer[]{\FFp P{\lambda x^A.N}}{\FFp P N}\quad\quad
	\infer[]{\FFp P{\gfp X^\sigma.\sum_iE_i}}{\forall i,\,\FFp P{E_i}}\quad\quad
	\infer[]{\FFp P{x\tuple{N_i}_i}}{\forall i,\,\FFp P{N_i}}\qquad
	\infer[]{\FFp P{x\tuple{N_i}_i}}{\binfcan{N_j}}\\[2ex]
	\infer[]{\NFFp P{ X^\sigma}}{\neg P(\sigma)}\quad\quad
	\infer[]{\NFFp P{\lambda x^A.N}}{\NFFp P N}\quad\quad	
	\infer[]{\NFFp P{\gfp X^\sigma.\sum_iE_i}}{\NFFp P{E_j}}\quad\quad
	\infer[]{\NFFp P{x\tuple{N_i}_i}}{
		\NFFp P{N_j}\quad\forall i,
		\, \nbinfcan{N_i}}
	\end{array}
	$$
\end{figure}

\begin{lemma}[Deciding type finiteness in $\ol$ -- Theorem~33 of \cite{EspiritoSantoMatthesPintoInhabitation}]\label{lem:decide-finhab-finite}\quad
	\begin{enumerate}
		\item \label{lem:decide-finhab-finite.1}
		For any $T\in\olfsfix$  well-bound, proper and closed,  $\FFp P T$ iff  $\finfin{\interps T}$.
		\item $\FFp \emptyset{\finrepempty\sigma}$ iff $\finfin{\solfunction\sigma}$ iff $\inhabs\sigma$ is finite.\label{lem:decide-finhab-finite.2}
			\item The problem, ``given $\sigma$, is $\inhabs\sigma$ finite'' is decided by deciding $\FFp \emptyset{\finrepempty\sigma}$.
	\end{enumerate}
      \end{lemma}
One can then also define \cite[Definition~35]{EspiritoSantoMatthesPintoInhabitation} \emph{sharper} versions of the predicates $\FFsymb$ and $\NFFsymb$ that are still decidable: $\FFcansymb:=\FFpsymb {P_*^\FFsymb}$ and
$\NFFcansymb:=\NFFpsymb{P_*^\FFsymb}$ for $P_*^\FFsymb:=\FFpsymb\emptyset\circ\finrepsymb$ (which meets the proviso of Figure~\ref{fig:FF} thanks to Lemma~\ref{lem:decide-finhab-finite}.\ref{lem:decide-finhab-finite.2}  and decidability of $\FFp \emptyset{\finrepempty\sigma}$). A generalization of this construction for other notions of finiteness is found in \cite[Definition~4.17]{JESRMLP-FI2019}. However, we will not even make use of $\FFcansymb$ in the remainder of this paper.  

In \cite{EspiritoSantoMatthesPintoInhabitation}, we show that the decision of finiteness of simple types can be supplemented with a syntax-directed procedure to count the number of inhabitants (when there are finitely many of them). This is done through a \emph{counting function} $\cnt T$. In its \emph{finitary version} (defined only for a subset of $\olfsfix$ -- the so-called \emph{head-variable controlled} expressions -- but big enough to contain all the finitary representations of full solution spaces $\finrepempty{\sigma}$), $\cnt T$ has the following extremely simple definition:
\[
\begin{array}{rcl@{\qquad}rcl}
\cnt{X^\sigma}&:=&0&\cnt{\gfp X^\sigma.\sum_iE_i}&:=&\sum_i\; \cnt{E_i}\\
\cnt{\lambda x^A.N}&:=&\cnt{N}&\cnt{x\tuple{N_i}_i}&:=&\prod_i\; \cnt{N_i}
\end{array}
\] 
Then the following instance of \cite[Theorem 42]{EspiritoSantoMatthesPintoInhabitation} is obtained:
\begin{lemma}[Counting theorem]\label{lem:counting}
If $\inhabs\sigma$ is finite then $\cnt{\finrepempty\sigma}$ is the cardinality of $\inhabs\sigma$.
\end{lemma}

\subsection{New questions asked and answered}\label{subsec:new-questions}

The ``finiteness'' of a simple type $A$ usually means the finiteness of the collection  of its inhabitants (the meaning taken just above).
However, as shown in \cite{JESRMLP-FI2019}, this concept of finiteness is just an instance of a ``generalized'' concept of finiteness  that emerges when a simple type is viewed through its full solution space, and solutions are taken as first-class citizens. This generalization encompasses other rather natural concepts of ``finiteness''  for simple types, such as,  finiteness of any solution of $A$ (i.\,e., the collection of all solutions of $A$ contains only (finite) $\lambda$-terms), or finiteness of the full solution space itself (i.\,e., the forest $\solfunction{\seq{}{A}}$ is a finite expression), and one may ask how these concepts relate, or whether the new concepts are still decidable.

The generalized concept of finiteness is defined through a parametrized predicate $\fingsymb{\Pi}$ on expressions in $\coolfs$, where the parameter $\Pi$ is again a predicate on expressions in $\coolfs$ \cite[Figure 5]{JESRMLP-FI2019}. Exploring this concept, one may conclude that: finiteness of the full solution space implies finiteness of all solutions, which in turn implies (much less obviously) finiteness of the collection of inhabitants \cite[Proposition 3.1]{JESRMLP-FI2019}. Following the methodology explained above to decide $\exfin{\solfunction\sigma}$  and $\finfin{\solfunction\sigma}$, also the generalized finiteness predicate $\fing\Pi{\solfunction\sigma}$ is shown to be decidable (for $\Pi$ subject to some mild conditions)  \cite[Theorem 4.3]{JESRMLP-FI2019}. This, in particular, implies decidability of the two alternative concepts of finiteness of simple types described above.

An ingredient needed to establish decidability of $\fing\Pi{\solfunction\sigma}$ is a separate result establishing decidability of the predicate $\emptysol{\solfunction\sigma}$, which holds when $\sigma$ has no solution (finite or infinite) \cite[Theorem 4.2]{JESRMLP-FI2019}. This  result also has a different application, the definition of the \emph{pruned solution space} of a sequent - the one where branches of the full solution space that are leading to no solution are chopped off. Then, the following version of K\"onig's lemma for simple types holds: a simple type has an infinite solution exactly when the pruned solution space is infinite \cite[Theorem 4.5]{JESRMLP-FI2019}.

\subsection{New results from old ones}\label{subsec:new-results}

It happened to us that, when trying to prove a well-known theorem with our tools, a generalization of the results suggested itself. The theorem is one by Ben-Yelles \cite{benyellesthesis} (see also Hindley's book \cite[Theorem 8D9]{HindleyBasicSimple}) about \emph{monatomic types}, i.\,e., types where only occurrences of a single atom are allowed.

\begin{definition}[Infinity-or-nothing]\label{def:ion}~
	\begin{enumerate}
        \item We say $T\in\olfsfix$ has the \emph{infinity-or-nothing property} (abbreviated as $T$ is i.o.n.) if $\nbinfcan{T}$ implies $\NFFion{T}$, where $\NFFionsymb:=\NFFpsymb{P_\ionindex}$ with
          $P_\ionindex:=\binfpsymb{\emptyset}\circ\finrepsymb$.  (Note that $P_\ionindex$ meets the required proviso of Figure~\ref{fig:FF}: (i) we already observed that $\binfpsymb{\emptyset}\circ\finrepsymb$ is decidable; (ii) $ P_\ionindex(\sigma)$ implies $\finfinsymb(\solletter(\sigma))$ (thanks to Lemmas~\ref{lem:decide-fin-inhab}.\ref{lem:decide-fin-inhab.2} and \ref{lem:decide-finhab-finite}.\ref{lem:decide-finhab-finite.2}, this is equivalent to the obviously true requirement: $\inhabs{\sigma}$ empty implies $\inhabs{\sigma}$ finite).
        \item Sequent $\sigma$ is an i.o.n.\ sequent if $\finrepempty{\sigma}$ is an i.o.n.\ finitary forest.
        \item $A$ is an i.o.n.\ type if $\seq{}A$ is an i.o.n.~sequent.
	\end{enumerate}
      \end{definition}
      As a first simple observation, we have that $N$ i.o.n.\ implies
      $\lambda x^A.N$ i.o.n.\ (the abstraction case of $\nbinfcansymb$
      can be inverted, and there is a matching abstraction case for
      $\NFFionsymb$).
      We remark that every $X^\rho$ is i.o.n., since $\nbinfcan{X^\rho}$
      and $\NFFion{X^\rho}$ both boil down to
      $\nbinfp{\emptyset}{\finrepempty{\rho}}$ (due to the
      complementarity of the two predicates in
      Figure~\ref{fig:nbinf}). Of course, this exploits the uncanonical
      setting with $P_\ionindex$ as parameter to $\NFFsymb$. Had one taken
      $\NFFcansymb=\NFFpsymb{\FFpsymb{\emptyset}\circ\finrepsymb}$
      instead (as introduced after Lemma~\ref{lem:decide-fin-inhab} above), the
      implication would have been equivalent to the wrong implication
      that $\inhabs\rho$ non-empty implies $\inhabs\rho$
      infinite. Also notice that the definition of $T$ i.o.n.\ for
      finitary forests that are not closed (where fixed-point
      variables $X^\rho$ are the extreme case) is rather of a
      technical nature (to be used to get proofs by induction through).
      Since the parameters for the predicates do not
      play a role for well-bound, proper and closed expressions of $\olfsfix$
      (by Lemma~\ref{lem:decide-fin-inhab}.\ref{lem:decide-fin-inhab.1} and
      Lemma~\ref{lem:decide-finhab-finite}.\ref{lem:decide-finhab-finite.1}),
      we have that for those $T$, $T$ is i.o.n.\ iff
      $\nbinfp{\emptyset}{T}$ implies $\NFFp{\emptyset}{T}$.

      The name of the property just introduced is justified by the following result.
\begin{lemma}\label{lem:ion} \quad
	Let $\sigma$ be i.o.n. Then $\inhabs\sigma$ is either empty or infinite, in other words: if $\inhabs\sigma$ is non-empty, then it is infinite. Similarly for an i.o.n.\ type $A$.
\end{lemma}
\begin{proof}
  If $\inhabs\sigma$ is non-empty, then by Lemma~\ref{lem:decide-fin-inhab}.\ref{lem:decide-fin-inhab.2},
  $\nbinfp{\emptyset}{\finrepempty{\sigma}}$. By monotonicity of
  $\nbinfsymb$ in its parameter, we get
  $\nbinfcan{\finrepempty{\sigma}}$. Since $\sigma$ is i.o.n., this
  gives $\NFFion{\finrepempty{\sigma}}$. Since $\NFFsymb$ is antitone
  in its parameter, we get $\NFFp{\emptyset}{\finrepempty{\sigma}}$,
  hence
  Lemma~\ref{lem:decide-finhab-finite}.\ref{lem:decide-finhab-finite.2} and the complementarity of the predicates in Figure~\ref{fig:FF}
  yield infinity of $\inhabs\sigma$.
\end{proof}

We now identify sufficient conditions with syntactic flavor for the i.o.n.\ property. The first one is over finitary forests and concerns occurrences of variables: roughly, in a sum, we need to see an alternative that does not consist of a ``shallow'' variable, i.\,e., a naked variable with empty tuple, and that the tuple components correspond to solution spaces of inhabited sequents (when representing solution spaces) among which one recursively satisfies the same criterion.

\begin{definition}[Deep finitary forests, sequents, and types]~
  \begin{enumerate}
  \item An expression $T$ in $\olfsfix$ is called deep if this can be derived by the following inductive definition:
    \begin{itemize}
    \item A typed fixed-point variable $X^\rho$ is deep.
    \item $\lambda x^A.N$ is deep if $N$ is deep.
    \item $\gfp X^\rho.\sum_i E_i$ is deep if $\nbinfp{\emptyset}{\finrepempty{\rho}}$ implies that there is a deep summand $E_i$.
    \item $x\tuple{N_1,\ldots,N_k}$ is deep if $\nbinfcan{N_j}$ for all $1\leq j\leq k$, and $N_{j}$ is deep for some $1\leq j\leq k$ (hence $k>0$ and the head variable $x$ can be considered as being deeply inside). 
    \end{itemize}
  \item A sequent $\sigma$ is called deep if $\finrepempty{\sigma}$ is a deep finitary forest.
  \item A type $A$ is called deep if $\seq{}A$ is a deep sequent.
  \end{enumerate}

\end{definition}
\begin{theorem}[Deep sequents/types are i.o.n.]\label{thm:deep}
Every deep sequent $\sigma$ is an i.o.n.\ sequent. In particular, every deep type $A$ is an i.o.n.\ type.
\end{theorem}
\begin{proof}
  We have to prove that for every sequent $\sigma$,
  $\finrepempty\sigma$ deep implies $\finrepempty\sigma$ i.o.n. More
  generally, we prove for every sequent $\sigma$ and vector $\Xi$ of
  declarations as in Definition~\ref{def:finrep}: if $\finrep\sigma\Xi$ is
  deep, then it has the i.o.n.\ property. The proof is by induction on
  the structure of the finitary forest $\finrep\sigma\Xi$.

  In case the if-guard in the definition of $\finrepsymb$ holds,
  $\finrep\sigma\Xi$ is a possibly multiply lambda-abstracted
  fixed-point variable $X^\rho$, thus a deep finitary forest. As
  argued after Definition~\ref{def:ion}, $X^\rho$ is i.o.n., and
  lambda-abstractions preserve this property. Hence,
  $\finrep\sigma\Xi$ is i.o.n.

  Otherwise, we use the symbols of Definition~\ref{def:finrep}, but
  abbreviate by $N$ the outer fixed-point expression, headed by $\gfp Y^\rho$, with $\rho=\seq\Delta p$ (where $\Delta=\Gamma,z_1:{A_1}\cdots z_n:{A_n}$), so that
  $\finrep\sigma\Xi=\lambda z_1^{A_1}\cdots z_n^{A_n}.N$. By
  assumption, $\finrep\sigma\Xi$ is deep, hence so is $N$. 
  Therefore: if $\nbinfp{\emptyset}{\finrepempty{\rho}}$, then there is a deep summand 
  $E$ relative to some $(y:\vec B\impl p)\in\Delta$. We
  want to show that $\finrep\sigma\Xi$ is i.o.n. Assume
  $\nbinfcan{\finrep\sigma\Xi}$. Then also $\nbinfcan{N}$. We have to
  show that $\NFFion{\finrep\sigma\Xi}$. Since $\finrep\sigma\Xi$ is proper, so is its
  subexpression $N$. By the ``easy consequence'' mentioned after
  Lemma~\ref{lem:decide-fin-inhab}, we get
  $\nbinfp{\emptyset}{\finrepempty{\rho}}$ from
  $\nbinfcan{N}$. Therefore, there is a deep summand  $E:=y \fl j{\finrep{\seq{\Delta}{B_j}}{\Xi,Y:\rho}}$. To
  show $\NFFion{\finrep\sigma\Xi}$, it suffices to
  show $\NFFion{E}$. Let
  $N_j:=\finrep{\seq{\Delta}{B_j}}{\Xi,Y:\rho}$ for all $j$.  Since
  $E$ is deep, we have $\nbinfcan{N_j}$ for all $j$, and there is
  $j^*$ s.\,t.\ $N_{j^*}$ is deep. $N_{j^*}$ is a sub-expression of
  $\finrep\sigma\Xi$, hence the induction hypothesis applies, by which
  $N_{j^*}$ is i.o.n., hence also $\NFFion{N_{j^*}}$. By definition
  of $\NFFsymb$, we obtain $\NFFion{E}$, as desired.
\end{proof}

We will now identify a class of deep types:
this is our second example of a syntactic restriction that guarantees the i.o.n.\ property.

Let $A=\vec{A}\impl p$. We say $p$ is the \emph{target} atom of $A$ and that the $\vec A$ are the argument types of $A$. Let $\sigma=(\seq\Gamma A)$, with $\Gamma=\{x_1:C_1,\cdots,x_n:C_n\}$. Put $A_{\sigma}:=\vec{C}\impl\vec{A}\impl p$ (the order of the $C_i$'s does not matter). In particular, if $\sigma$ is $\seq{}A$, then $A_{\sigma}=A$.

\begin{definition}[Generalized triple negation]
	\begin{enumerate}
		\item Let us say that a type of the form $A\impl p$ is a \emph{negation at $p$}
		and that a type of the form $(A\impl p)\impl p$ is a \emph{double negation at $p$}.
		\item We introduce the notion of \emph{generalized double negation at $p$}:
		this is any type of the form $\vec B\impl p$ with non-empty $\vec B$
		so that each of the argument types $B_i$ has target atom $p$.
		\item A type $A=\vec A\impl p$ is called a \emph{generalized triple negation} (abbrev: g.t.n.) if one of the argument types $A_i$ is a generalized
		double negation at $p$.
              \item A sequent $\sigma$ is a g.t.n.\ if $A_\sigma$ is a g.t.n.\ (this is indifferent to the order of context formulas used for defining $A_\sigma$).
	\end{enumerate}
\end{definition}

For example $p\impl p$ and $(q\impl p)\impl p\impl p$ are generalized double negations at $p$.
As examples of g.t.n.'s, we mention
$(p\impl p)\impl p$ (only an infinite solution) and
$(p\impl p)\impl p\impl p$ (infinitely many inhabitants corresponding
to the natural numbers).

\begin{lemma}[Sequents/types with generalized triple negation are deep]\label{lem:gtn}
	If $\sigma$ is a g.t.n., then $\sigma$ is deep. Hence, every g.t.n.\ type $A$ is deep.
\end{lemma}
\begin{proof} Assume that $\sigma$ is a g.t.n. We will prove more than
  only that $\sigma$ is deep, i.\,e., $\finrepempty\sigma$ is deep. More generally, we prove for every
  vector $\Xi$ of declarations as in Definition~\ref{def:finrep} that
  $\finrep\sigma\Xi$ is deep.  The proof is by induction on the
  structure of the finitary forest $\finrep\sigma\Xi$.

  In case the if-guard in the definition of $\finrepsymb$ holds,
  $\finrep\sigma\Xi$ is a possibly multiply lambda-abstracted
  fixed-point variable $X^\rho$, thus a deep finitary forest,
  so we do not need the assumption that $\sigma$ is a g.t.n.

  Otherwise, we use the symbols of Definition~\ref{def:finrep}, but
  abbreviate by $N$ the outer fixed-point expression, so that
  $\finrep\sigma\Xi=\lambda z_1^{A_1}\cdots z_n^{A_n}.N$. By
  assumption, $\sigma$ is a g.t.n., and this means $A_\sigma$ is a
  g.t.n., but we can assume that $A_\sigma$ is the same formula as
  $A_\rho$, for $\rho:=\seq\Delta p$, as usual. By definition of
  generalized triple negation, there is a double negation at $p$ among
  the formulas of $\Delta$. Let $(y:\vec B\impl p)\in\Delta$ be the
  corresponding association with non-empty $\vec B$ and so that each
  of the argument types $B_j$ has target atom $p$. Let
  $N_j:=\finrep{\seq{\Delta}{B_j}}{\Xi,Y:\rho}$ for all $j$. In order
  to have that $\finrep\sigma\Xi$ is deep, we need that $N$ is
  deep. We therefore assume that
  $\nbinfp{\emptyset}{\finrepempty{\rho}}$ and show that the summand
  $E:= y \fl j{N_j}$ is deep. We even show for all $j$ that $N_j$ is
  deep and that $\nbinfcan{N_j}$ holds. Since $\vec B$ is non-empty,
  this in particular yields a $j^*$ s.\,t.\ $N_{j^*}$ is deep. Fix
  some $j$.  $N_j$ is a sub-expression of $\finrep\sigma\Xi$, hence
  the induction hypothesis applies and gives that $N_j$ is deep
  provided $\seq\Delta{B_j}$ is a g.t.n., but this is obvious since
  the target atom of $B_j$ is still $p$, and the double negation at
  $p$ among the formulas of $\Delta$ is still available. It remains to
  show $\nbinfcan{N_j}$. From the assumption
  $\nbinfp{\emptyset}{\finrepempty{\rho}}$ and
  Lemma~\ref{lem:decide-fin-inhab}.\ref{lem:decide-fin-inhab.2}, we get
  an inhabitant of $\rho=\seq\Delta p$. By vacuous
  lambda-abstractions, this gives an inhabitant of $\seq\Delta{B_j}$
  (again because the target atom of $B_j$ is $p$). By virtue of the
  same theorem, this gives
  $\nbinfp{\emptyset}{\finrepempty{\seq\Delta{B_j}}}$. By monotonicity
  of $\nbinfsymb$ in its parameter, this can be weakened to
  $\nbinfcan{\finrepempty{\seq\Delta{B_j}}}$, and by the ``second
  consequence'' of the main result on $\nbinfcansymb$ mentioned after
  Lemma~\ref{lem:decide-fin-inhab}, this is equivalent to
  $\nbinfcan{N_j}$.
\end{proof}

\begin{theorem}[G.t.n.'s are i.o.n.]\label{thm:gtn}
	Let $A$ be a generalized triple negation.
	Then $A$ has either $0$ or infinitely many inhabitants.
\end{theorem}
\begin{proof} Immediate consequence of Theorem~\ref{thm:deep} and Lemmas~\ref{lem:ion} and \ref{lem:gtn}.
\end{proof}

We now obtain the theorem by Ben-Yelles \cite{benyellesthesis} (\cite[Theorem 8D9]{HindleyBasicSimple}) about monatomic types. The original proof and the textbook proof were as a consequence of a more difficult result called \emph{Stretching Lemma}. But here we see the theorem about monatomic types is just an instance of the
more general phenomenon captured by our Theorem~\ref{thm:gtn}.

\begin{corollary}[Monatomic inhabitation]\label{cor:monatomic}
	Let $A=\vec{A}\impl p$ be a monatomic type. If $A$ is flat, that is, each $A_i$ is $p$, then $A$ has exactly $n$ inhabitants where $n$ is the length of $\vec A$. Otherwise, $A$ has either $0$ or infinitely many inhabitants.
\end{corollary}
\begin{proof}
	The first case is immediate (this includes the case when $n=0$). The second case is an instance of Theorem~\ref{thm:gtn}: for monatomic types $A$,
	$A$ is a g.t.n.~iff $A$ is non-flat.
\end{proof}


\section{Final remarks}\label{sec:final}

\textbf{Contribution.} We are developing a comprehensive approach to reductive proof search that is naturally integrated with the Curry-Howard isomorphism: the lambda-terms used to represent proofs are seen co-inductively in order to capture (possibly infinite) solutions of search problems. But this Curry-Howard representation is just a convenient definition of the structures generated by proof search. An effective analysis
has to be conducted in an accompanying, equivalent, finitary representation, which may be seen as the main technical contribution. The role of formal sums also stands out, specially in connection with the new operation of decontraction. 
Also noteworthy is both the design of the finitary calculus (with its combination of formal sums, fixed points, and a relaxed form of fixed-point variable binding) and its typing system, which uses the relaxed form of binding to detect cycles in proof search, and which is sound and complete w.\,r.\,t.~a declarative semantics into coinductive forests.

This infrastructure was put to use in the study of proof search, as detailed elsewhere~\cite{EspiritoSantoMatthesPintoInhabitation,JESRMLP-FI2019}. A brief indication of the results there obtained was given in Section~\ref{sec:analysis}, together with a fresh example of the infrastructure at work in obtaining a generalization of a well-known theorem. Our approach has proved so far to be robust, comprehensive, and innovative. Robust because we could rely on it to obtain many results about proof search, including the benchmark results about decidability of inhabitation. Comprehensive because with the approach we were able to address a wide range of questions, from decision and counting problems to so-called coherence theorems, which is unusual if not
unprecedented in the literature. Innovative because we obtained new solutions for old problems, but we were also led to investigate and solve new problems, like those stemming from the consideration of solutions instead of just proofs, and to obtain new results when trying to prove old ones, like in the case of monatomic inhabitation. As detailed in Section~\ref{sec:analysis}, the innovative aspect of our applications and solutions can be summarized in these characteristics: (1) Separation of concerns; (2) Run the proof search only once; (3) Syntax-directedness; (4) Solutions and solution spaces as first-class citizens.

\smallskip
In order to test the comprehensiveness of our approach, we have already successfully  applied it to the case of full intuitionistic propositional logic as described in~\cite{DBLP:conf/types/SantoM020} (actually even via a more elaborate polarized intuitionistic logic~\cite{JESENTCS17}), developing coinductive and finitary representations of the full solution spaces, establishing their equivalence and obtaining decidability of inhabitation in a form that is analogous to the predicate  $\nbinfsymb_\emptyset\circ\finrepsymb$ of Lemma~\ref{lem:decide-fin-inhab} (and that thus factors through a recursive predicate on finitary expressions). As we anticipated, and similarly to this paper, the main theorem,
establishing the equivalence of representations, 
rests on the subformula property of the object logic. 
In the present paper, we preferred to explore a simple case study (proof search in $\LJT$) in order separate the complexities of the proposed approach for proof search from the complexities of the object logic. 

\medskip
\textbf{Related work.} 
In the context of
logic programming with classical first-order Horn clauses, the use of coinductive structures is seen in
\cite{KomendantskayaPowerCSL2011}, in order to provide a uniform algebraic semantics for both finite and infinite
derivations by SLD-resolution. Building on the the type-theoretical approach to resolution documented in \cite{FuKomendantskayaFAC2017}, an attempt to give semantics for a large class of infinite derivations is seen in \cite{FuKSP-FLOPS2016}: here, the central tool is a type system, called \emph{corecursive resolution} (CR), for a language of proof terms possibly containing fixpoints, and equipped with an operational semantics. A theorem about the operational equivalence with resolution is attempted, but it only captures a limited use of the fixed point operator (``simple loops''). In a later work \cite{BasoldKL19}, 
reductive proof search in a sequent calculus for uniform proofs is targeted, instead of resolution. A system called \emph{coinductive uniform proofs} (CUP) is proposed, for first-order and higher-order coinductive theories, again only allowing a limited used of fixpoints, this time without proof terms, but with a declarative semantics into Herbrand models. In our approach the coinductive forests representing the full search spaces are not built by a particular algorithm and constitute a semantic universe. Our finitary system, including the typing system, plays the role that CR or CUP play for the approaches above, the role of a ``logic of coinductive proofs''. 
But our finitary system, besides the already mentioned soundness and completeness,  has other distinctive properties:
no limitation in the nesting of fixpoints; the use of arbitrary formulas as ``coinductive hypotheses'', coming from the fact that no essential restriction is imposed on the sequents attached to these fixpoints (the atomic restriction being considered as inessential since the context can be freely expanded to accommodate hypotheses for the premisses $\vec A$ in a formula of form $\vec A\impl p$). These comparisons seem valuable, despite the fact that we are restricted to the implicational fragment of propositional logic, and the evidence that the first-order case brings unexpected results \cite{KomendantskayaRozplokhasBasoldTPLP20}.

In \cite{pym2004reductive} we find a comprehensive approach to proof search, where the generalization of proofs to searches (or ``reductions'') is accounted for semantically. Parigot's $\lambda\mu$-calculus is used to represent proofs in classical and intuitionistic sequent calculus, but no indication is given on how such terms could represent searches.

\smallskip
In Section~1.3.8 of \cite{lambdacalculuswithtypes} we find a list of types, for each of which the set of inhabitants is described through an ``inhabitation machine''. This list covers among others all our examples in Example~\ref{ex:types} with the exception of $\Inftyt$ and
$\DNPeircet$. We invite the reader to compare those descriptions in graphical presentation in the cited book with our succinct presentations of the solution
spaces worked out in Section~\ref{sec:coinductive} and Section~\ref{sec:decontraction} (see Examples~\ref{ex:boole}, \ref{ex:Church}, \ref{ex:Peirce}, and \ref{ex:Three-cont1}). While our expressions do not display the types of the subexpressions, they are explicit about when a variable
gets available for binding (in their example (vii), their variable $x$, that corresponds to our $y$ in Example~\ref{ex:Three-cont1}, looks as if it was available from the outset), and our expressions are even more explicit about the generation process for new names (the book speaks about ``new
incarnations'') using standard lambda abstractions and the decontraction operator.
While our presentations of solution spaces in Section~\ref{sec:coinductive} and Section~\ref{sec:decontraction} are still on the informal level of infinitary terms with meta-level fixed points, and for that reason may seem far from
a ``machine'' for the generation of inhabitants, the finitary expressions we obtained in Example~\ref{ex:dn-Peirce-cont2} and Example~\ref{ex:finitary} with the
machinery of Section~\ref{sec:finitary-calculus} compare in the same way with the inhabitation machines of \cite{lambdacalculuswithtypes} and are proper syntactic elements and can thus qualify as ``machine'' descriptions of the process of obtaining the inhabitants (and even the infinite solutions---notice that infinite solutions are not addressed at all in the description of inhabitation machines in \cite{lambdacalculuswithtypes}).

The work~\cite{SchubertDB15} also studies mathematical structures for representing proof search, and can partly be seen as offering a realisation of the intuitive description of the inhabitation machines in~\cite{lambdacalculuswithtypes}.
Similarly to our work, \cite{SchubertDB15} handles search for normal inhabitants in the simply-typed lambda-calculus. However, the methods of~\cite{SchubertDB15} are very different from ours. Their methods come from automata and language theory, and proof search is represented through automata with a non-standard form of register, as a way to avoid automata with infinite alphabets, and still meet the need for a supply of infinitely many bound variables in types like $\DNPeircet$ or the ``monster'' type (cf.~our discussion after Example~\ref{ex:types}). 
Unlike in our work, infinite solutions are not a concern of the approach in~\cite{SchubertDB15}, but this approach is concerned with computational complexity and is capable of obtaining the usual {\sf PSPACE} bound for the decision of the inhabitation problem.  

Besides the approach just mentioned, the literature offers a rich variety of other approaches to handle the search space determined by a type and its collection of inhabitants and to address  inhabitation, counting or enumeration problems in simply-typed lambda-calculus or extensions of it
\cite{benyellesthesis,HindleyBasicSimple,TakahashiAH96,DowekYing09,WellsYakobowski04,BrodaD05,BourreauS11}.
We briefly consider these approaches below. 

To the best of our knowledge, \cite{benyellesthesis} (nicely revisited in~\cite{HindleyBasicSimple}) is the first work to address the question of enumerating all inhabitants (in long normal form) of a simple type. (Next we refer to the presentation of the approach in~\cite[8C]{HindleyBasicSimple}).  The approach is very different from ours, since it does not explicitly build a structure representing the full collection of inhabitants of a type. Instead, it develops an iterative search algorithm,  that takes a type and may run forever, and at each stage produces a finite collection of \emph{normal form schemes} (lambda-terms with meta-variables), so that any inhabitant of the type can be extracted from one of these schemes. 

Context-free grammars are used in~\cite{TakahashiAH96} to represent the collection of inhabitants (in long normal form) of a simple type. Although (finite) context-free grammars suffice to capture inhabitants obeying the \emph{total discharge convention} (forbidding multiple variables with the same type, or, in the logical reading, forbidding multiple assumptions of the same formula), an infinite grammar is required to capture the full set of inhabitants
(due to the potential need for a supply of infinitely many bound variables, as alluded to above when relating to~\cite{SchubertDB15}). 
 Grammars are also used in~\cite{DowekYing09} to enumerate in two stages all inhabitants of a type in simply-typed lambda-calculus (and in certain fragments of system F). The first stage builds a context-free grammar description of the collection of \emph{schemes} of a given type. (Schemes are the proof terms of a so-called \emph{sequent calculus with brackets} LJB, where assumptions in the context are unnamed, thus following the total discharge convention.) In a second stage, an algorithm extracts the full collection of inhabitants of a type from its schemes.

The work~\cite{WellsYakobowski04} develops algorithms for counting and enumerating proofs in the context of full propositional intuitionistic sequent calculus $\LJT$. These algorithms are based on a direct representation of the search space of a sequent via directed graphs. (Roughly speaking, a sequent corresponds to a vertex that has outgoing edges to vertices with that sequent and a rule that applies to it (bottom-up), and the latter vertices have outgoing edges to the sequents resulting from the bottom-up application of the rule.) Even if the version of $\LJT$ considered there only allows proofs obeying the total discharge convention (context sequents are sets of formulas), and this is crucial to guarantee the finiteness of the graph representation, this is the only work we are aware of that addresses counting and enumeration of proofs for full intuitionistic propositional logic. As already mentioned, our recent work~\cite{DBLP:conf/types/SantoM020} shows that the coinductive approach developed in this paper is also applicable to (a system of focused proofs for) full intuitionistic propositional logic. \emph{Op.\ cit.} only treats the problem of type inhabitation and finite inhabitation (relative to focused proofs). However, we anticipate that also a simple counting function can be achieved analogously to what is briefly mentioned in Section~\ref{subsec:new-solutions} (for the implicational fragment), a direction we would like to explore in the future (alongside with other directions specified below). 

\cite{BrodaD05} is an extensive study of inhabitation in simply-typed lambda-calculus through the \emph{formula-tree proof method}, establishing new results and new proofs, in particular, in connection to uniqueness questions. The method relies on a representation of types as labelled trees called \emph{formula trees} (where each label identifies a \emph{primitive part} of the type), from which \emph{proof trees} are derived, and in turn allow the extraction of all inhabitants (in long normal form) of the type. This extraction also involves two stages: the first stage generates a context-free grammar representation of the inhabitants in so-called \emph{standard form} (imposing restrictions on the use of variables in the spirit of the total discharge convention); then, the second stage extracts finitely many inhabitants from each standard inhabitant (if any), producing the full collection of inhabitants of the type. The question of uniqueness of inhabitation in simply-typed lambda-calculus is also addressed in~\cite{BourreauS11}. This work uses yet a very different tool:
game semantics. Connecting typings with arenas and inhabitants with winning strategies for arenas, inhabitation questions can be recast in terms of game semantics. For example, \cite{BourreauS11} offers a new characterization of principal typings in simply-typed lambda-calculus through games. Actually, in~\cite{AlvesB15} one can find precise connections between this game semantics approach and the formula-tree proof method for addressing inhabitation in simply-typed lambda-calculus.

Since the above-mentioned work \cite{benyellesthesis,HindleyBasicSimple,TakahashiAH96,DowekYing09,WellsYakobowski04,BrodaD05,BourreauS11} is concerned with inhabitation (finite solutions) only, naturally they do not share with us the goal of having a mathematical representation of the full solution space, and a treatment of infinite solutions. Another distinctive feature of our work 
is that, as we stay within the lambda-calculus paradigm, we can profit from its binding mechanism, and avoid the need to restrict to inhabitants under total discharge convention, or the need for a two-stage process to capture the full collection of long normal forms.  Our representation of the entire space of solutions as a first-class citizen of a finitary lambda-calculus immediately offers the possibility of its structural analysis, and allows a new take on a wide range of questions related to inhabitation in simply-typed lambda-calculus, as explained in Section \ref{sec:analysis}.


\medskip
\textbf{Only seemingly related work.} Logics with fixed points or inductive definitions, as for example in~\cite{SantocanaleCircularProofs},
admit infinite or ``circular'' proofs, which are infinite ``pre-proofs'' enjoying an extra global, semantic condition to ensure that only valid conclusions are allowed. In addition, the proofs of these logics have alternative (sometimes equivalent) finite representations as graphs with cycles (e.\,g., trees with back edges). Despite superficial similarity, bear in mind the several differences relatively to what is done in the present paper: first, there is the conceptual difference between solution and proof; second, in our simple object logic $\LJT$, proofs are the finite solutions (hence trivially filtered amongst solutions), and therefore infinite solutions never correspond to globally correct reasoning; third, fixed points are not present in the object logic, but rather in the finitary calculus and its typing system: the latter, when seen as a ``logic of coinductive proofs'', is a meta-logic---a logic about the proof search process determined by $\LJT$.

\medskip
\textbf{Future work.} We would like to profit from the finitary representation of a (full) solution space to
extract individual solutions. 
As suggested in Section~\ref{subsec:reductive-proof-search}, this can be done by pruning the
solution space, an operation already studied in \cite{JESRMLP-FI2019} but only for coinductive representations (with the specific goal of obtaining the version of K\"onig's Lemma for simple types mentioned in Section~\ref{subsec:new-questions}).
We expect unfolding of fixed points
to play also a fundamental role in the process of extraction of individual solutions.
These ingredients
should provide a base for the accounting of
algorithmic control in proof search through rewriting.

\medskip
We would like to further test the comprehensiveness of our coinductive approach to proof search on other logical or type-theoretical settings. For example, it could be interesting to test our methodology on classical logic, for which there is already work in the context $\lambda\mu$-calculus \cite{DavidZaionc09}, or on the challenging  setting of intersection types, whose general inhabitation problem is undecidable, but where relevant decidable fragments have been identified \cite{BucciarelliKesnerRocca14,DudenhefnerRehof17}. 
Of course, it will be interesting and important  to test also our coinductive approach in the context of first-order logic. Recall that coinductive structures are already employed in \cite{BasoldKL19} to give a proof-theoretic account of   Horn clauses (even for coinductive theories), but the attainment of representations of solutions and of entire solution spaces with a rich collection of properties (like the one seen in this paper for intuitionistic implication) is likely to pose new questions.

\medskip
\textbf{Acknowledgements.} Jos\'{e} Esp\'{\i}rito Santo and Lu\'{\i}s Pinto  are funded by Portuguese Funds through FCT -- Funda\c{c}\~{a}o para a Ci\^{e}ncia e a Tecnologia, within the Projects UIDB/00013/2020 and UIDP/00013/2020. Ralph Matthes had been funded by the \emph{Climt} project (ANR-11-BS02-016 of the French Agence Nationale de la Recherche). All authors had been partially funded by COST action CA15123 EUTYPES.


\bibliographystyle{alpha}
\bibliography{proofsearch}

\appendix
\section{Technical appendix on regularity of finitary terms}\label{sec:app-regular}

In Section~\ref{sec:finitary-calculus}, we insisted that we do not
confine our investigation to trivially regular terms. This is directly imposed
by Definition~\ref{def:finrep}, as we will see next.

\begin{example}[A not trivially regular term]\label{ex:irregular}
  Assume three different atoms $p,q,r$, set $\Gamma:=y_1:q\impl
  p,y_2:(r\impl q)\impl p,x:r$ and $\Xi:=X:\seq\Gamma q$. Then
  Definition~\ref{def:finrep} yields
$$\finrep{\seq\Gamma p}{\Xi}=\gfp\,{Y^{\seq\Gamma p}}.y_1\tuple{X^{\seq\Gamma q}}+y_2\tuple{\lb z^r.X^{\seq{\Gamma,z:r}q}}$$
Fixed-point variable $X$ occurs free in this expression with two
different sequents as types, hence the expression is not trivially regular.
\end{example}

Definition~\ref{def:finrep} even leads us to consider
trivially regular terms with regular but not trivially regular
subterms, hidden under a greatest fixed-point construction:

\begin{example}[Hidden irregularity]\label{ex:hiddenirregular}
  Consider the following modification of the previous example: add the
  binding $y:p\impl q$ to $\Gamma$. Then, the above calculation of
  $\finrep{\seq\Gamma p}{\Xi}$ comes to the same result. And we calculate
 $$\finrepempty{\seq\Gamma q}=\gfp X^{\seq\Gamma q}.y\tuple{\finrep{\seq\Gamma p}{\Xi}}$$
 Hence, $X$ with two different sequents as types has to be bound by
 the outer fixed-point operator.
\end{example}

The following notion may be of further use:

\begin{definition}[Strong regularity in $\olfsfix$]
An expression $T$ in $\olfsfix$ is strongly regular, if all subexpressions of $T$ (including $T$) are regular.
\end{definition}

We can even strengthen Corollary~\ref{cor:finitary-regular}.

\begin{corollary}\label{cor:finitary-strongly-regular}
  $\finrep{\seq\Gamma C}{\Xi}$ is strongly regular.
\end{corollary}
\begin{proof}
  Regularity is already expressed in Corollary~\ref{cor:finitary-regular}.
  Concerning the regularity of the subexpressions,
  lambda-abstraction does not influence on regularity, and in the
  recursive case of the definition of $\finrep{\seq\Gamma C}{\Xi}$, the
  same $\xi_{\Xi,Y:\sigma}$ is admissible for all the occurring
  subterms, hence also for the summands that are bound by the $\gfp$
  operation.
\end{proof}


\end{document}